\documentclass[11pt,preprint2]{aastex}

\newcommand{\msun}{M_{\odot}}
\begin{document}

\title{A {\it Chandra} Study of the Rosette Star-Forming Complex. I. The Stellar Population and Structure of the Young Open Cluster NGC 2244}

\author{Junfeng Wang,\altaffilmark{1} Leisa
K. Townsley,\altaffilmark{1} Eric D. Feigelson,\altaffilmark{1}
Patrick S. Broos,\altaffilmark{1} Konstantin
V. Getman,\altaffilmark{1} Carlos G. Rom\'an-Z\'u\~niga,\altaffilmark{2,3} and
Elizabeth Lada \altaffilmark{3}}

\altaffiltext{1}{Department of Astronomy \& Astrophysics, The
Pennsylvania State University, 525 Davey Lab, University Park, PA
16802; {\tt jwang@astro.psu.edu}}

\altaffiltext{2}{Harvard-Smithsonian Center for Astrophysics, 60 Garden Street, Cambridge, MA 02138}

\altaffiltext{3}{Department of Astronomy, University of Florida, 211 Bryant Space Science Center, Gainesville, FL 32611}

\begin{abstract}

We present the first high spatial resolution X-ray study of NGC~2244,
the 2~Myr old stellar cluster immersed in the Rosette Nebula, using
the {\it Chandra X-ray Observatory}.  Over 900 X-ray sources are
detected; 77\% have optical or FLAMINGOS near-infrared (NIR) stellar
counterparts and are mostly previously uncatalogued young stellar
cluster members.  All known OB stars with spectral type earlier than
B1 are detected and the X-ray selected stellar population is estimated
to be nearly complete between 0.5 and 3~M$_{\odot}$. The X-ray
luminosity function (XLF) ranges from $29.4 < \log L_x < 32.0$ ergs
s$^{-1}$ in the hard ($2-8$~keV) band. By comparing the NGC~2244 and
Orion Nebula Cluster XLFs, we estimate a total population of $\sim
2000$ stars in NGC~2244. A number of further results emerge from our
analysis: 1) The XLF and the associated $K$-band luminosity function
indicate a normal Salpeter initial mass function (IMF) for
NGC~2244. This is inconsistent with the top-heavy IMF reported from
earlier optical studies that lacked a good census of $<4$M$_{\odot}$
stars. 2) The spatial distribution of X-ray stars is strongly
concentrated around the central O5 star, HD 46150. The other early O
star, HD 46223, has few companions. The cluster's stellar radial
density profile shows two distinctive structures: a power-law cusp
around HD~46150 that extends to $\sim 0.7$~pc, surrounded by an
isothermal sphere extending out to 4~pc with core radius 1.2~pc. This
double structure, combined with the absence of mass segregation,
indicates that this 2~Myr old cluster is not in dynamical
equilibrium. Our results will strongly constrain models of the cluster
formation process. The spatial distribution of X-ray selected
$K$-excess disk stars and embedded stars is asymmetric with an
apparent deficit towards the north. 3) The fraction of X-ray-selected
cluster members with $K$-band excesses caused by inner protoplanetary
disks is 6\%, slightly lower than the 10\% disk fraction estimated
from the FLAMINGOS study based on the NIR-selected sample. This is due
to the high efficiency of X-ray surveys in locating disk-free
weak-lined T Tauri stars. 4) X-ray luminosities for 24 stars earlier
than B4 confirm the long-standing $\log (L_x/L_{bol})\sim -7$
relation. The Rosette OB X-ray spectra are soft and consistent with
the standard model of small-scale shocks in the inner wind of a single
massive star. 5) About 50 intermediate-mass ($2 < M < 8$~M$_{\odot}$)
cluster members are identified; they exhibit a wide range of X-ray
luminosities consistent with previously studied samples of Herbig
Ae/Be stars.  Their $K$-excess disk fraction is $\sim$10\%, indicating
that the Herbig Ae/Be phenomenon is rare or short-lived.

\end{abstract}


\section{Introduction}

During their evolution from Class I protostars to
zero-age-main-sequence (ZAMS) stars, young stellar objects are readily
identified in X-rays due to their highly elevated X-ray emission
compared to the older Galactic stellar population \citep[see reviews
by][]{Feigelson99,Favata03,Feigelson07}. High spatial resolution {\em
Chandra} observations of well-known Galactic star forming regions
(e.g., the {\em Chandra} Orion Ultradeep Project, hereafter COUP, Getman
et al.\ 2005; RCW 38, Wolk et al.\ 2006; Cepheus B, Getman et al.\
2006), more distant molecular cloud and HII region complexes (e.g., NGC
6334, Ezoe et al.\ 2006; NGC 6357, J. Wang et al.\ 2007; M~16, Linsky et
al.\ 2007; M~17, Broos et al.\ 2007), the Galactic Center \citep[e.g.,
the  Arches and Quintuplet clusters,][]{Wangqd06,Muno06}, and extragalactic
star forming regions \citep[e.g., 30 Dor,][]{Townsley06,Townsley06b},
have greatly advanced our knowledge of star formation processes in these
regions.

Moreover, these studies demonstrate the unique power of studying star
formation in the X-ray band. Besides the high energy phenomena, other
new information about the young clusters such as population,
membership, and star formation environs can be obtained when combining
X-ray detections with knowledge obtained from longer wavelength
studies. For example, although X-ray-selected
samples contain a very small fraction of
extragalactic sources and Galactic field stars, X-ray sources that have optical and infrared (IR) counterparts are
mostly cluster members. This is in contrast to the high percentage of
non-members in the optical/IR images where, except for those sources with
massive dusty disks, membership for individual stars generally has to
be ascertained via spectroscopy. In initial mass function (IMF)
studies of more distant, high mass star forming regions, traditional
optical measurements are significantly encumbered by large reddening
and membership confusion. In comparison to the massive stars, the
lower mass pre-main-sequence (PMS) stellar populations are much less
accessible and thus much harder to evaluate. In the past, $H_{\alpha}$
emission was used in general as a youth and membership
indicator. However, in addition to the observational challenges due to
prevalent bright $H_{\alpha}$ nebulosity in HII regions, the emission
itself requires accretion activity in protoplanetary disks
\citep{Muzerolle98,Muzerolle01} and thus is susceptible to the disk
evolutionary stages. 

It has long been recognized that X-ray emission circumvents these
problems and is very effective for securing PMS membership of young
clusters \citep{Feigelson99}. Modern X-ray observatories like {\em
Chandra} and {\em XMM-Newton} enable identification of hundreds of
individual members from their X-ray emission. In this work and
subsequent papers, we report {\em Chandra} studies of a well-known
star forming complex, concentrating here on a new census of the low
mass cluster members and new knowledge of the IMF arising from the
X-ray perspective.

The Rosette star-forming complex, situated in a large star formation
site in the Perseus spiral arm, provides an ideal testbed for studying
sequential formation of clusters due to the favorable orientation of
its morphological components, consisting of an expanding blister HII
region on the edge of a giant molecular cloud oriented perpendicular
to the line-of-sight.  Thorough review of past and present research on
this popular and important star formation region appear in
\citet{TFM03} (hereafter TFM03) and \citet{RL07} (hereafter
RL07). Here we highlight what is most relevant to this work.

\begin{figure}
\centering
\epsscale{0.58}
\plotone{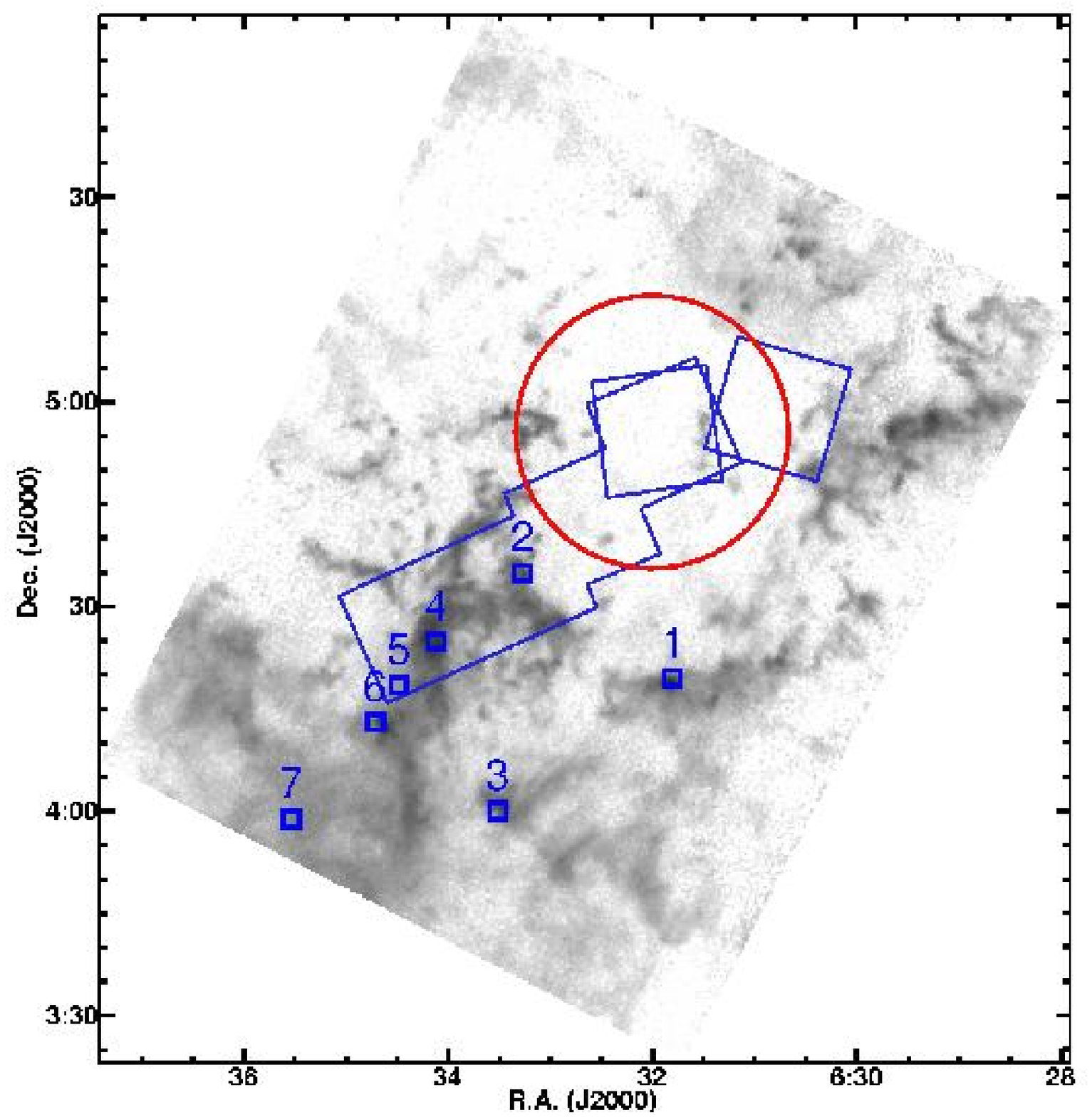}
\epsscale{0.61}
\plotone{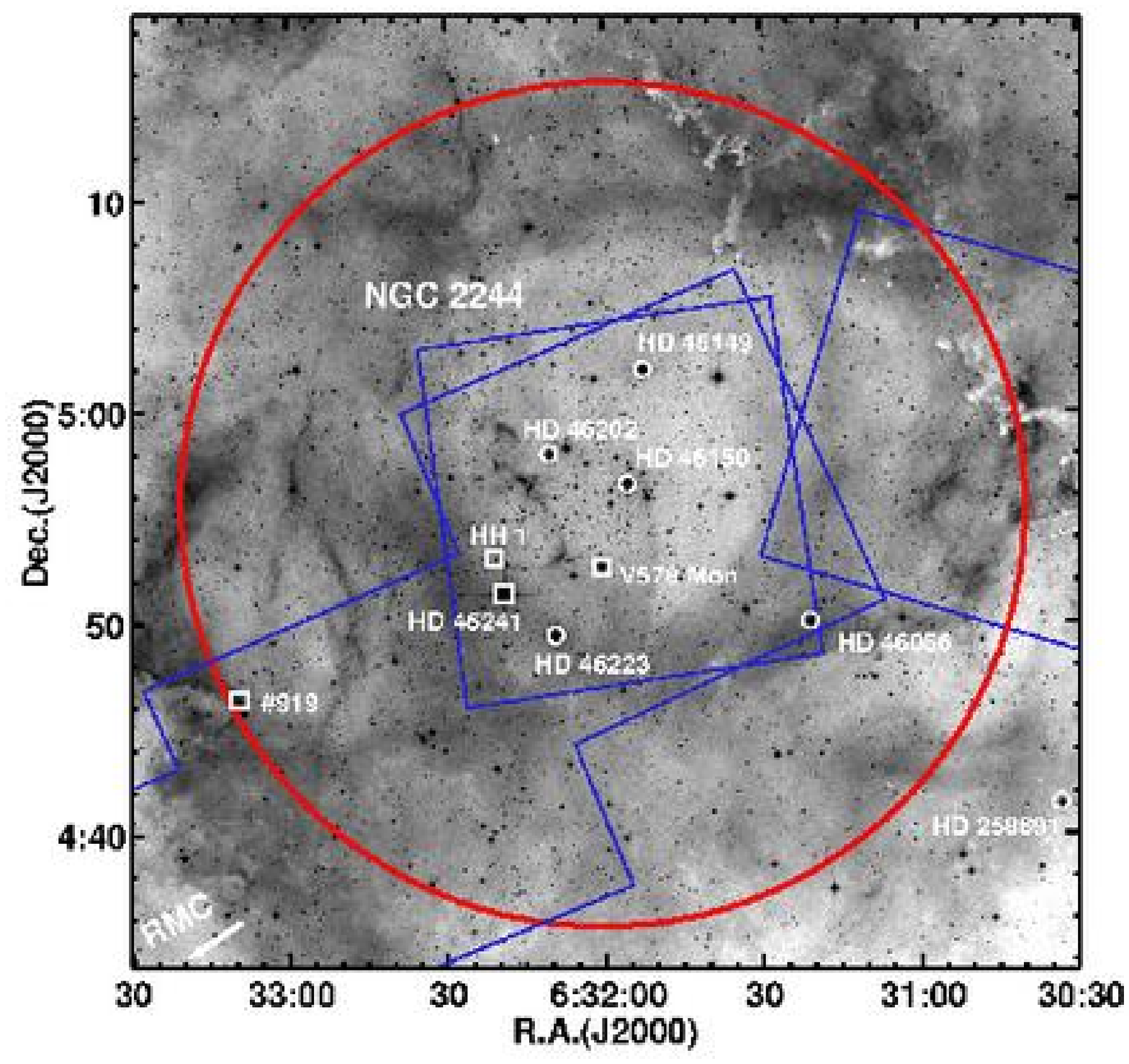}
\epsscale{1.0}

\caption{(a): A large scale ($\sim 2^{\circ}\times 1.5^{\circ}$) view
of the Rosette star forming complex in $^{12}$CO$J=1-0$ emission from
\citet{Heyer06}. The multiple ACIS FOVs (polygons) and the extent of the
NGC 2244 cluster (circle) are shown. Squares mark the embedded
clusters in the RMC with \citet{Phelps97} sequence numbers. (b):
A $45^{\prime} \times 45^{\prime}$ DSS2 $R$-band image of the Rosette
Nebula. All known O stars in the FOV that belong to the NGC 2244 cluster
(TFM03 Table 6) are labeled by circles. Four stars are marked as
squares: V578 Mon is an eclipsing binary; HH~1 is a stellar microjet;
ACIS \#919 is a candidate massive star; the visually brightest star HD
46241 (K0V) is foreground. These objects are described further in the
text.\label{fig:large_view}}

\end{figure}

The Rosette Nebula (= Sharpless 275 = W 16 = NGC 2237--2239, NGC 2244,
and NGC 2246) is a large HII region at the tip of the Rosette
Molecular Cloud (RMC). In both radio and optical images, it shows a
prominent ring-like morphology, with a cluster of ionizing young stars
located in the central hole \citep{Celnik85,TFM03}.  The large-scale
IRAS data \citep{Cox90} and CO emission map \citep{Heyer06} clearly
show a similar annular morphology extending into the molecular cloud
(Figure~\ref{fig:large_view}a).

Extending to the southeast of the Rosette Nebula, the RMC is an
elongated giant molecular cloud with $\sim 10^5 M_{\odot}$ of gas and
dust \citep{Blitz80}. Multiwavelength observations in mid-IR and radio
show clumpy structure in the RMC \citep{Williams98}.  Embedded star
clusters have been revealed in the densest parts through near-IR
imaging surveys \citep{Phelps97,Roman05,Roman07}.
Figure~\ref{fig:large_view}a demonstrates the association between
\citet{Phelps97} IR clusters and the molecular clumps where the CO
emission peaks. \citet{Lada03} have shown that embedded clusters are
physically associated with the most massive and dense cores in
molecular clouds, based on systematic and coordinated surveys
\citep[e.g.,][]{Lada92}.

In the X-ray band, previous imaging study of the Rosette Complex was
hampered by low spatial resolution. An early {\it Einstein
Observatory} study detected a few individual O stars and extended
$\sim 2$~keV X-ray emission at the center of the nebula
\citep{Leahy85}. \citet{BC02} considered the integrated contribution
from X-ray emitting low mass PMS stars in the cluster and concluded
that apparently diffuse emission seen by {\em ROSAT} could be
explained by unresolved point sources. \citet{Chen04} analyzed the
same {\em ROSAT} data set and attributed the brightest X-ray sources
to massive stars, active T Tauri stars, and foreground stars.
\citet{Gregorio98}, studying a {\em ROSAT} observation of the RMC,
reported faint X-ray point sources associated with T Tauri stars and
Herbig Ae/Be (HAeBe) stars and X-ray ``hot spots'' from unresolved
embedded low mass star clusters. In TFM03, the first high resolution
{\em Chandra} X-ray image mosaic of this high-mass star forming region
was presented and soft diffuse X-ray plasma ($kT\simeq 0.06$ and
$\simeq 0.8$ keV) with luminosity $L_x\simeq 6\times 10^{32}$ ergs
s$^{-1}$ was detected in the HII region. It was attributed to a
combination of fast O star winds and unresolved T Tauri stars.

The young stars powering the Rosette HII region are members
of the massive open cluster NGC 2244.  Figure~\ref{fig:large_view}b
shows a Digital Sky Survey image ($59^{\prime}\times 59^{\prime}$) of
this region with a few bright stars and interesting objects
labeled. Despite its apparent low concentration of stars in the
optical, NGC 2244 contains $\sim 30$ early-type stars between O4V and
B3V (Table 6 in TFM03), whose cluster memberships are secured from deep
photometric study along with proper motion data and spectroscopy
\citep{Verschueren91,PS02}. No other known massive young stellar
cluster within 2 kpc, other than RCW 38 \citep{Wolk06} and M 17 (Broos
et al.\ 2007), is comparably rich. The PMS members
were largely unknown from optical studies; only a handful of
$H_{\alpha}$ emission objects have been identified as young PMS
members \citep{PS02,Li04}. \citet{BC02} presented 138 sources selected
from {\it ROSAT} PSPC and HRI detections. Although incomplete, they
revealed previously unknown PMS stars to K spectral types.

The distance to NGC 2244 has been measured in many visual photometry
studies and ranges between 1.4 kpc and 1.7 kpc.  \citet{Hensberge00}
derive a distance of $1.39\pm 0.1$ kpc to an eclipsing binary member V578
Mon, using a novel Fourier spectral disentangling technique. The main
sequence (MS) turn-off age estimated by \citet{PS02} is 1.9 Myr, which
is consistent with the inferred age from V578 Mon, $2.3\pm 0.2$ Myr
\citep{Hensberge00}. This makes NGC~2244 the youngest cluster within
the larger Mon OB2 association \citep{Hensberge00}. We adopt a
distance of 1.4~kpc and a cluster age for NGC~2244 of 2~Myr throughout
our studies; these are consistent with TFM03. Note that a larger
distance value $d=1.6$ kpc has been used recently by other researchers
\citep{PS02,Roman07}. This discrepancy in distance affects our derived
X-ray luminosities ($\log L_x$) by only 0.1 dex. At the distance of
1.4~kpc, 1$^{\prime}$ corresponds to 0.4 pc.

Whereas the TFM03 paper was dedicated to study of the diffuse X-ray
emission in the Rosette HII region, we present here an X-ray point
source study of the Rosette complex based on TFM03 data and a new 75
ks {\em Chandra} observation centered on NGC 2244. The field of view of our
mosaic {\em Chandra} fields are outlined by the polygons in
Figure~\ref{fig:large_view}a. We separate our study into a series of
four papers with different astrophysical emphasis. In this work (Paper
I), we report {\em Chandra} observations of the NGC 2244 cluster
and the Rosette HII region and study the young stellar population in
detail.  In an upcoming paper (Wang et al., in preparation; Paper II),
we describe {\em Chandra} observations of the embedded clusters
in the RMC, aiming to investigate cluster formation in a sequential
manner and to test whether molecular clumps preferentially forming
embedded clusters of low-mass stars make up the fundamental building
blocks of star formation in molecular clouds. The westernmost {\em
Chandra} field, designed to study triggered star formation and the
X-ray detection of a twin cluster to NGC 2244 \citep{Li05,Roman07},
will be presented in Paper III (Wang et al., in preparation). A
detailed analysis of the diffuse X-ray emission in the HII region
will appear in Paper IV (Townsley et al., in preparation). 

New mid-IR observations are also contributing to our knowledge of the
Rosette complex.  A shallow {\em Spitzer Space Telescope} survey of
NGC 2244 has been reported \citep{Balog07} and a deep {\em Spitzer}
survey for disk emission from {\em Chandra} stars in NGC 2244 is also
underway (PI: Bouwman). Together with the {\em Spitzer} MIPS coverage
of the massive cores in the RMC (PI: Bonnell) and a new {\em Spitzer}
program (PI: Rieke), virtually the entire nebula and the molecular
cloud will be completely mapped.

This paper is organized as follows. First, we describe the {\em
Chandra} observations and data reduction in \S2. In \S3, we identify
the X-ray sources with optical and infrared counterparts, and evaluate
the fraction of contaminants through simulations utilizing stellar
population synthesis model and the $\log N$--$\log S$ distribution for
extragalactic X-ray sources. \S4 is devoted to global properties of
the NGC 2244 cluster such as the X-ray luminosity function, the
initial mass function, the $K$-band luminosity function, spatial
structures, mass segregation, and the $K$-band excess disk fraction
among the X-ray detected stars. We present collective properties of
interesting X-ray sources in \S5, ending with a summary in \S6.
 
\section{{\it Chandra} Observations and Data Reduction \label{obs.sec}}

\begin{figure}
\centering
\includegraphics[width=1.0\textwidth]{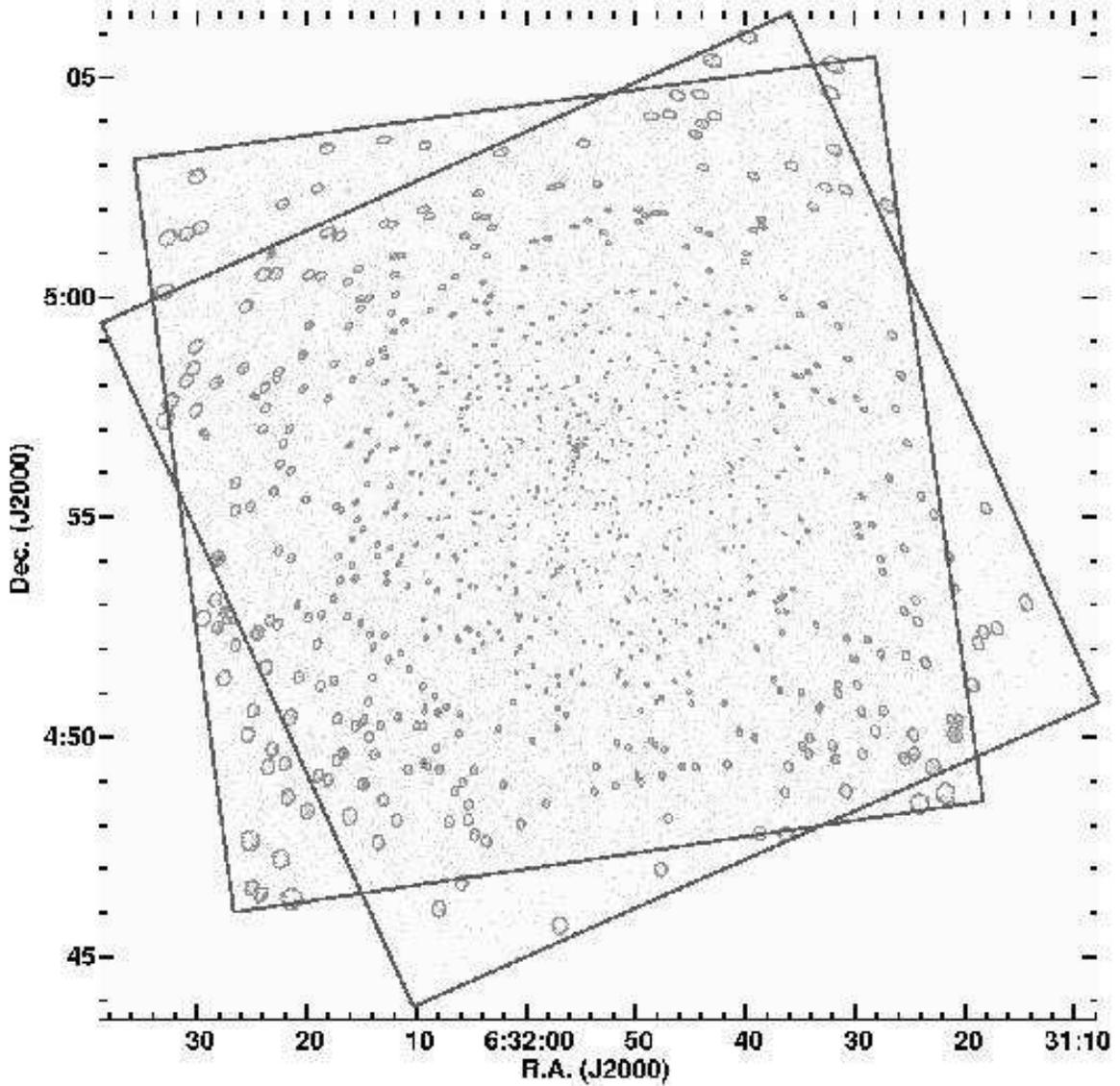}
\caption{(a): A merged 94~ks ACIS-I image of the NGC 2244 cluster from
ObsIDs 1874 and 3750 (outlined by two $17^{\prime}\times 17^{\prime}$
boxes) with reduced resolution (binned by 2 pixels). The two ObsIDs
have slightly different roll angles. (b): X-ray composite image
created from {\it csmooth} for the merged fields. Blue intensity is
scaled to the soft (0.5--2 keV) X-ray emission, green intensity is
scaled to the hard (2--7 keV) X-ray emission. (c): Same as (b) but the
scaling emphasizes soft diffuse emission and red intensity is scaled to
the DSS $R$-band optical emission. \label{fig:diffuse}}
\end{figure}

The Rosette complex was observed with the Imaging Array of the {\em
Chandra} Advanced CCD Imaging Spectrometer (ACIS-I). The ACIS-I field
of view is $17\arcmin \times 17\arcmin$ in a single pointing and a
mosaic observation was designed to best image the ionizing cluster,
capture the interface between the photoionized gas and the cold
neutral material, step into the dense molecular cloud, and study its
recently reported secondary cluster. As shown in Table~1, the entire
observation consisted of four $\sim$20 ks ACIS-I snapshots in January
2001 (TFM03, Figure~2), a deep 75 ks ACIS-I image in January 2004
centered on the O5 star HD~46150 in NGC 2244
(Figure~\ref{fig:diffuse}a), and one 20 ks ACIS-I pointing at the twin
cluster to NGC 2244 \citep{Li05} in 2007.  The image mosaic covers a
$\sim 1^{\circ}\times 0^{\circ}.25$ field of the Rosette Nebula and
RMC. All images were taken in standard ``Timed Event, Faint'' mode
with 3 pixel $\times$ 3 pixel event islands except ObsID 3750 and
ObsID 8454, which used the ``Very Faint'' mode (5 pixel $\times$ 5 pixel
event islands).

We follow the same customized data reduction and source extraction
described in TFM03, \citet{Wang07}, and \citet{Broos07}. The
processing of Level 1 data is presented in detail in Appendix B of
TFM03; the same reduced dataset used in that study (Rosette Field 1-4),
augmented by the deep observation (Rosette Nebula/NGC 2244) reduced following TFM03, were used here for further analysis. With slightly different
roll angles, we reprojected the ObsID 1874 data and merged them with
the ObsID 3750 field. Figure~\ref{fig:diffuse}a shows the merged 94 ks
ACIS image of NGC 2244 overlaid with source extraction regions (see
details below). Many point sources are visible; this is further
illustrated in the smoothed X-ray composite image
(Figure~\ref{fig:diffuse}b) for the merged fields created with the
CIAO tool {\it csmooth} \citep{Ebeling06}. In
Figure~\ref{fig:diffuse}c the existence of soft diffuse emission is
emphasized in the context of the DSS optical image, where the diffuse
X-ray emission nicely fills in the cavity of the HII region. This
component will be discussed in Paper IV.

\subsection{Source Finding and Photon Event Extraction}

For identifying X-ray point sources, first we assemble a large number
of candidate sources using a variety of techniques and criteria,
including image reconstruction and visual inspection.  The source
searching for NGC 2244 is performed on the merged fields from ObsID
1874 and ObsID 3750. For each of the ACIS fields, twelve different
images were created: soft (0.5--2~keV), hard (2--7~keV), and full
(0.5--7~keV) X-ray wavebands with four different pixel binning scales
(4$\times$, $\sim$2$\times$, 1$\times$, and 0.5$\times$ a sky
pixel). The {\it wavdetect} program \citep{Freeman02} was run with
wavelet scales from 1 to 16, 8, 4, and 2 pixels in steps of $\sqrt 2$
(for the four different binnings respectively) and a source
significance threshold of $1 \times 10^{-5}$ (which is very sensitive
but permits some false sources) on each of the images described above.
These twelve source lists were merged, with the source position from
the highest-resolution image retained, to generate a single list of
candidate sources.

\begin{figure}
\centering
\plotone{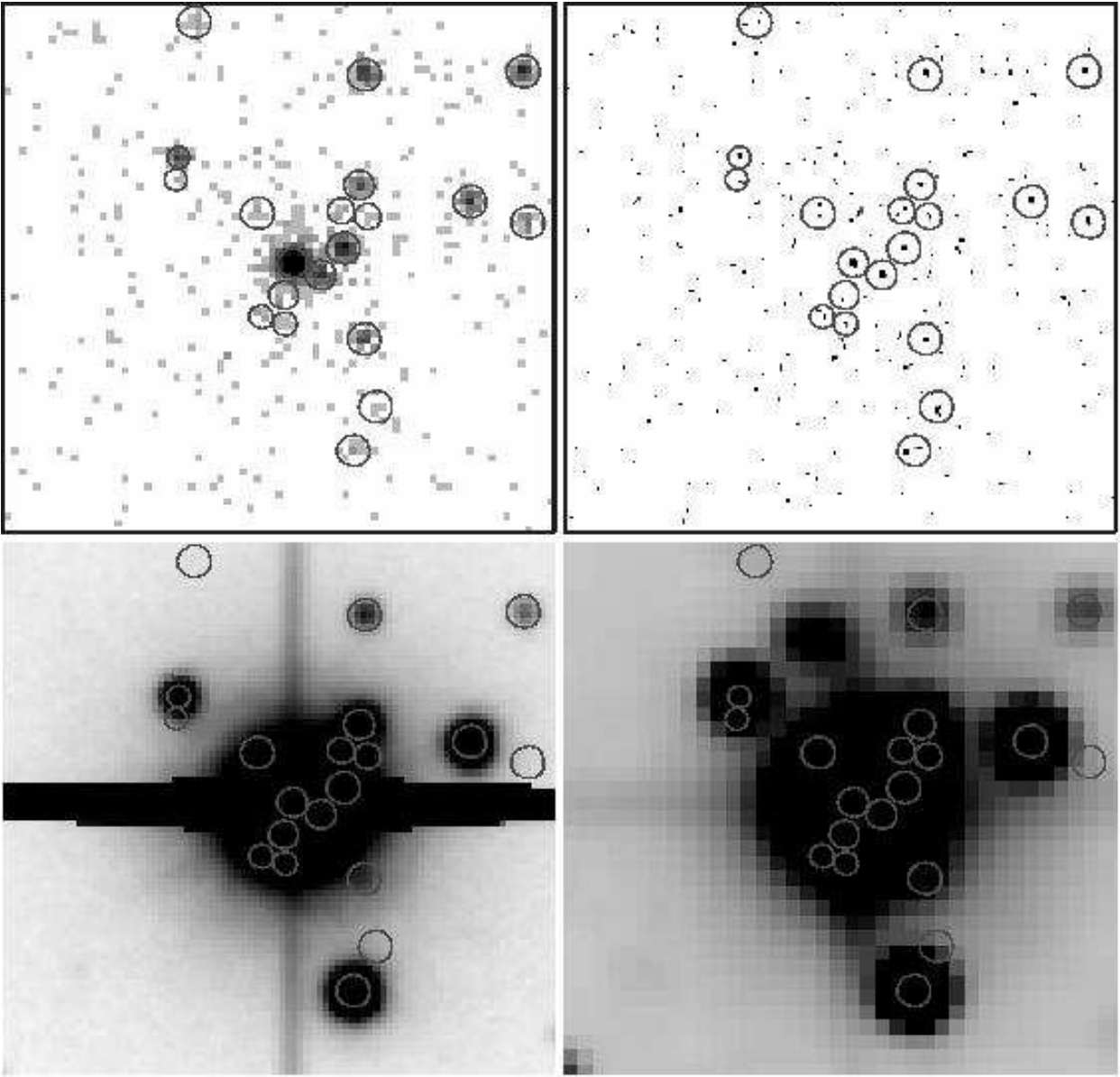}
\caption{$30^{\prime\prime}\times 30^{\prime\prime}$ ACIS unbinned
image (top left), reconstructed image (top right), the H$\alpha$ image
(bottom left), and the 2MASS-$Ks$ image (bottom right) of the central
region around the O5V star HD~46150. Several new sources are resolved
within 5\arcsec\/ of the dominant star by the {\em Chandra}
observation.\label{fig:O_cluster}}
\end{figure}

To take advantage of the sub-arcsecond point spread function (PSF) at
positions around the aimpoint, we applied a subpixel positioning code
\citep{Mori01} to improve spatial resolution in the inner part of the
field. An image reconstruction with the Lucy-Richardson maximum
likelihood algorithm \citep{Lucy74} was performed in the central
$50^{\prime\prime} \times 50^{\prime\prime}$ around HD 46150
(Figure~\ref{fig:O_cluster}) in ObsID 3750 \citep[examples of maximum
likelihood image reconstruction can be found
in][]{Townsley06,Wang07}. Eighteen additional candidate sources from
the image reconstruction were added to the source
list. Adaptive-kernel smoothed flux images in the three energy bands
were also created with {\it csmooth} to help visually identify
additional faint potential sources. Source lists from all ObsIDs
(except the recent observation, ObsID 8454) were then merged to form a
master candidate detection list.

The source finding procedure described above results in a total of
1452 potential sources identified for five ObsIDs (omitting the
westernmost field ObsID 8454). A preliminary event extraction for the
potential X-ray sources was made with our customized IDL script {\it
ACIS
Extract}\footnote{\url{http://www.astro.psu.edu/xray/docs/TARA/}}
\citep[version 3.98; hereafter {\it AE},][]{Broos02}.  Using the {\it
AE}-calculated probability $P_B$ that the extracted events are solely
due to Poisson fluctuations in the local background, source validity
can be statistically evaluated while taking into account the large
distorted PSFs at far off-axis locations and spatial variations in the
background.  After a careful review of the net counts distribution and
IR counterparts frequency for all candidate sources, we rejected
sources with $P_B > 0.014$, i.e. those with a 1.4\% or higher likelihood of being a background
fluctuation. The trimmed source list includes 1314 valid sources.

Since our data analysis involves multiple {\em Chandra} observations,
for convenience and to avoid repeating source designations we
divide the X-ray point sources into NGC 2244 sources and RMC sources
based on positions.  According to the stellar density distribution of
2MASS sources in NGC 2244 \citep{Li05}, the CO emission maps
\citep{Williams95,Heyer06}, the IRAS 60$\mu m$ emission \citep{Cox90},
and radio continuum \citep{Celnik85}, we assign all X-ray sources
within 20 arcmin of the cluster central position
(R.A.=06$^h$31$^m$59.$^s$9,
Dec.=+04$^\circ$55$^\prime$36$^{\prime\prime}$) as potential NGC 2244
sources. This is indicated by the large circle in
Figure~\ref{fig:large_view}. The resulting cluster extent is
consistent with the size of this massive open cluster determined by
systematic studies from the All-Sky Compiled Catalog of 2.5 Million
Stars \citep{Kharchenko05}.  We remind the reader that the dividing
line is not unique; ambiguity of exact physical associations certainly
exists for sources located in the interface between the HII region and
the molecular cloud to the east, and between the main NGC 2244 cluster
and secondary NGC 2237 cluster to the west. In this paper (Paper I) we
focus on a total of 919 sources located within the NGC 2244 cluster
region (as defined above); the rest are presented in Papers
II\footnote{A small number of sources that are in ObsID 1875 and
listed in Tables~2 and 3 here will be discussed scientifically in
Paper II with the RMC sources. To avoid confusion, they will not be
presented in RMC source list tables, but we will list these sources
separately in Paper II for consistency between the papers.} and III. 

The 919 valid sources are divided into a primary list of 805 highly
reliable sources ($P_B<0.001$; Table~\ref{tbl:primary}) and a
secondary list of 114 tentative sources with $P_B \ge 0.001$
likelihood of being spurious background
fluctuations. Table~\ref{tbl:primary} and Table~\ref{tbl:tentative}
have formats that are identical to Tables 1 and 2 in
\citet{Townsley06}, \citet{Wang07}, and \citet{Broos07}. A detailed
description of the table columns is given in the table footnotes.

\subsection{Source Variability}

\begin{figure}
\centering
\includegraphics[width=0.36\textwidth,angle=-90]{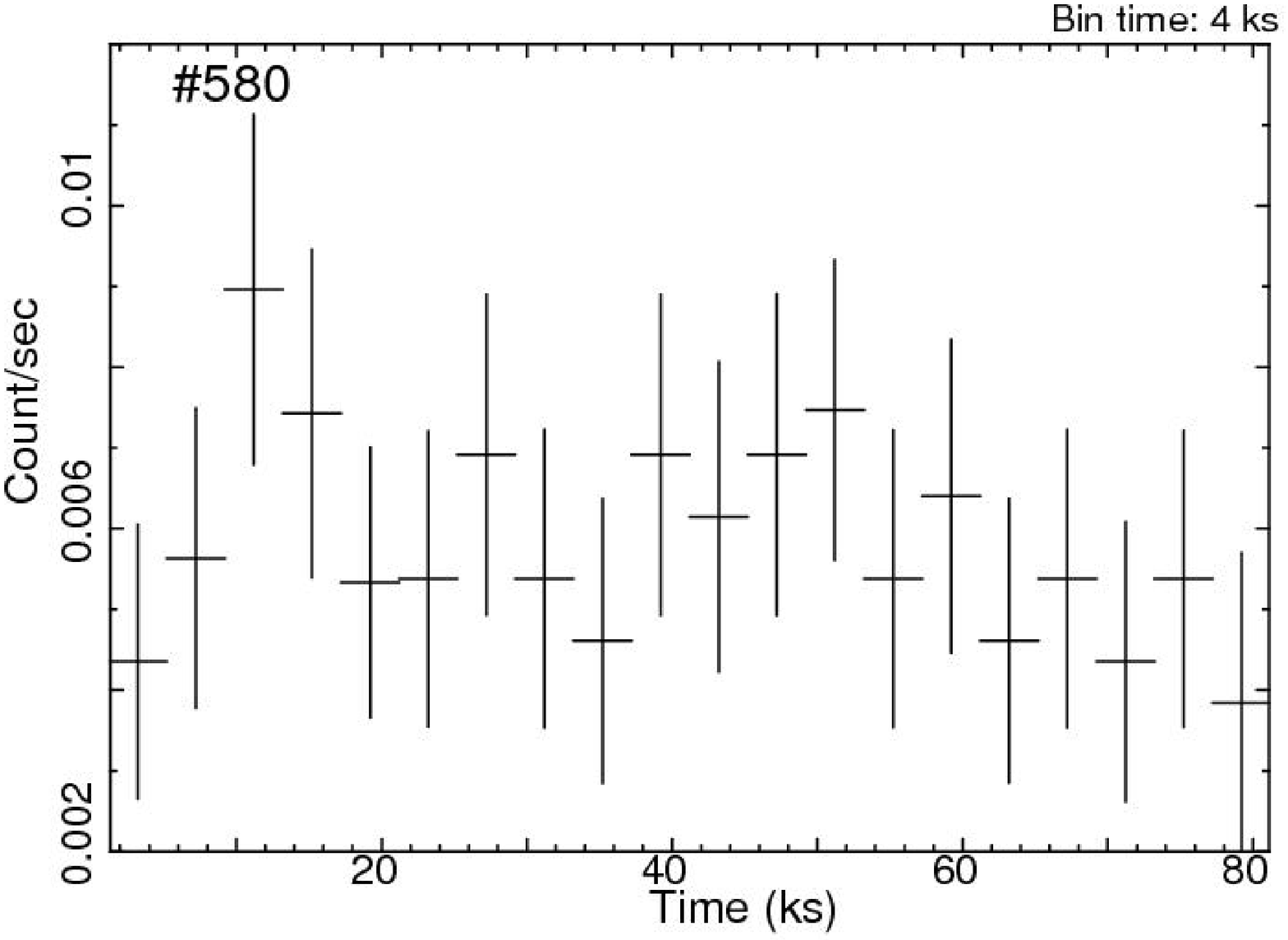}
\includegraphics[width=0.36\textwidth,angle=-90]{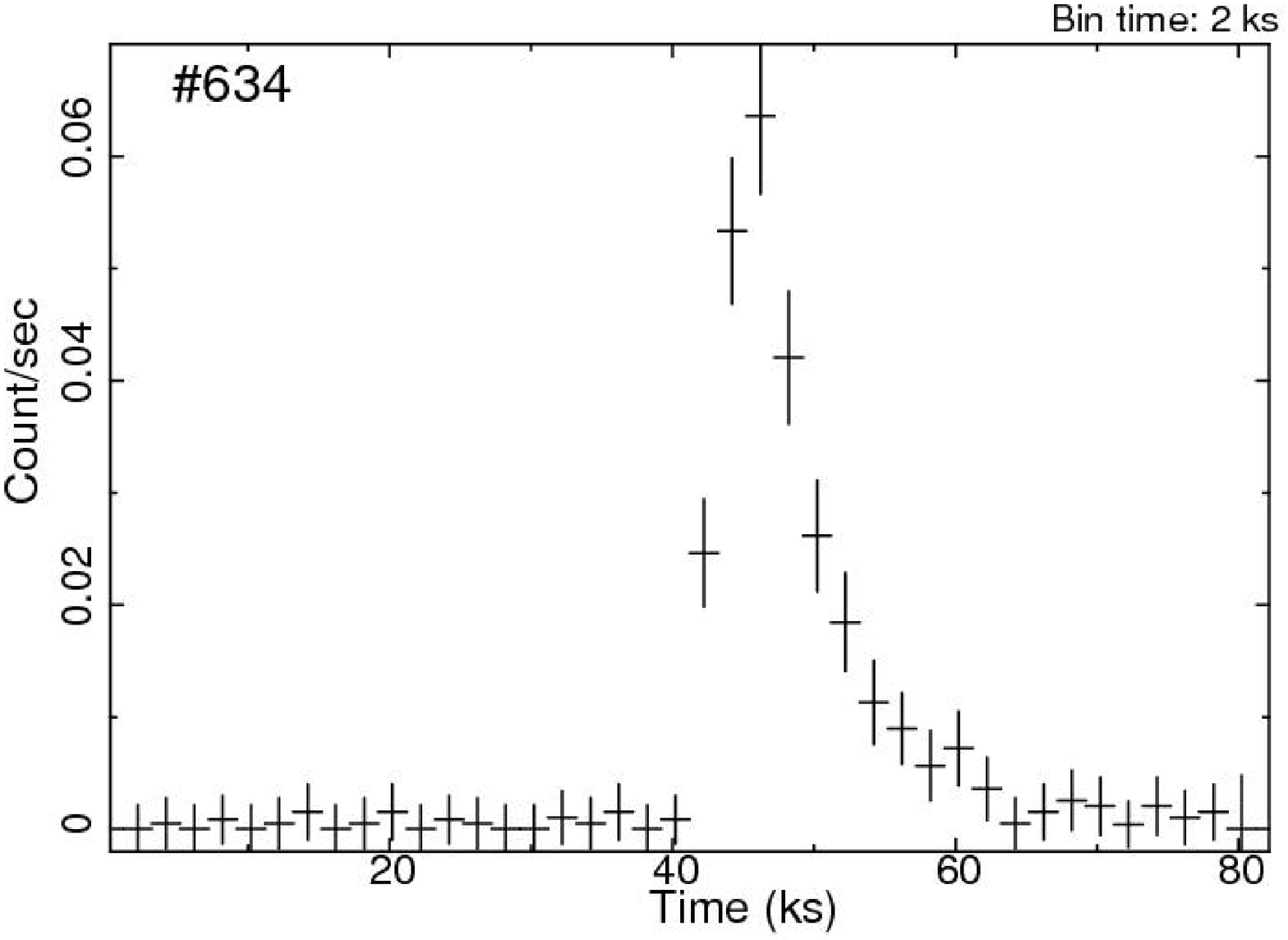}
\includegraphics[width=0.36\textwidth,angle=-90]{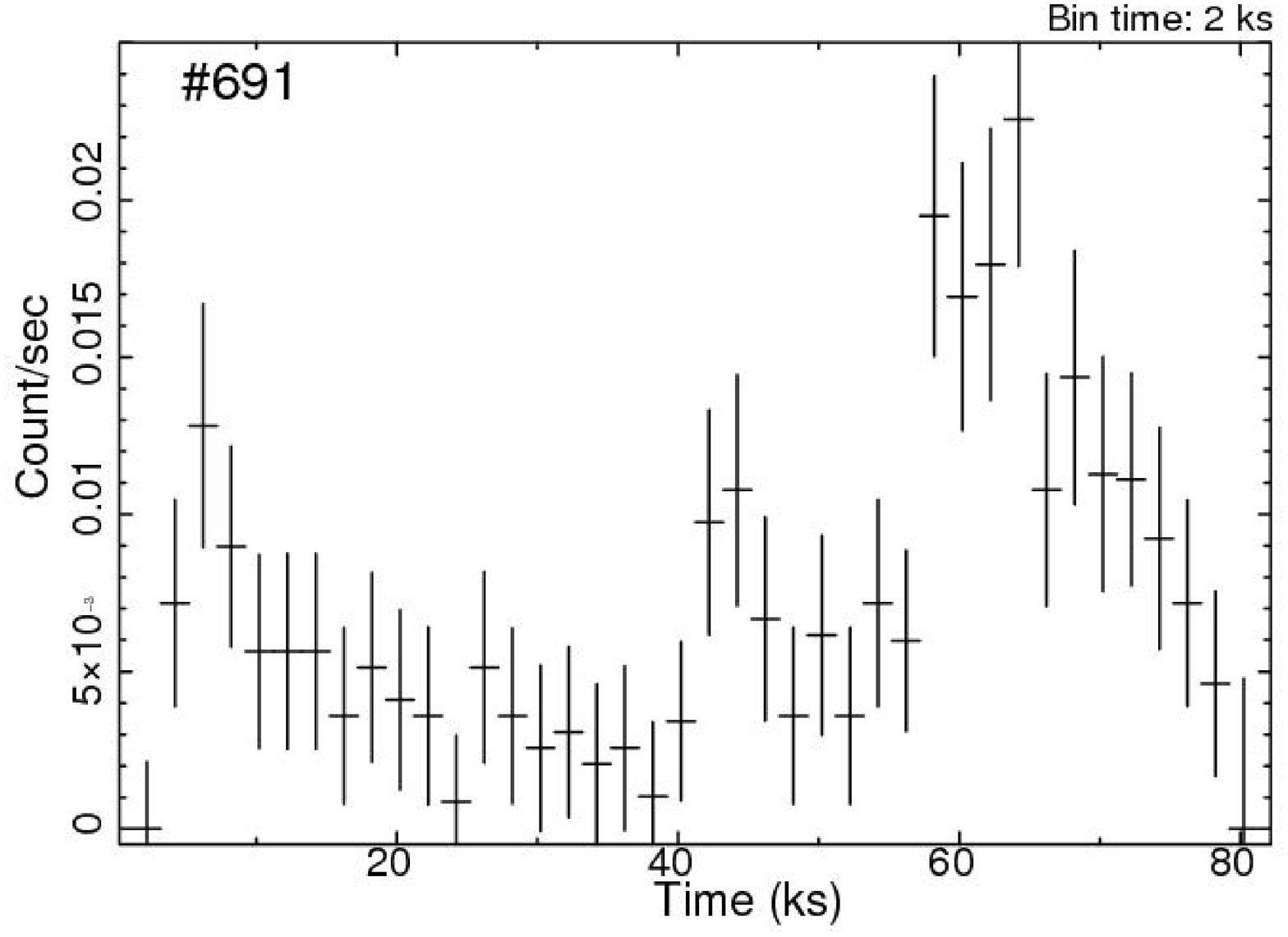}
\includegraphics[width=0.36\textwidth,angle=-90]{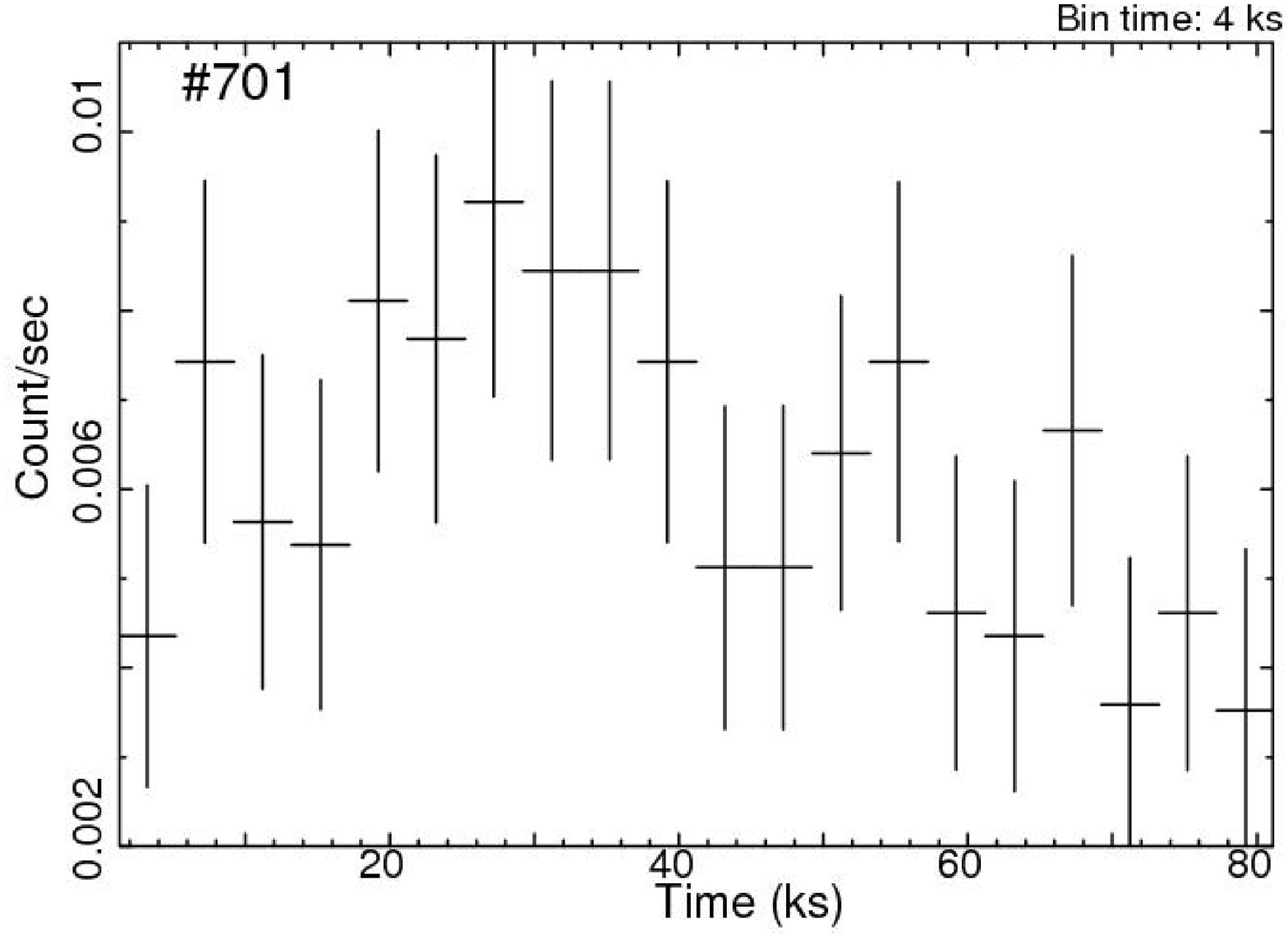}
\includegraphics[width=0.36\textwidth,angle=-90]{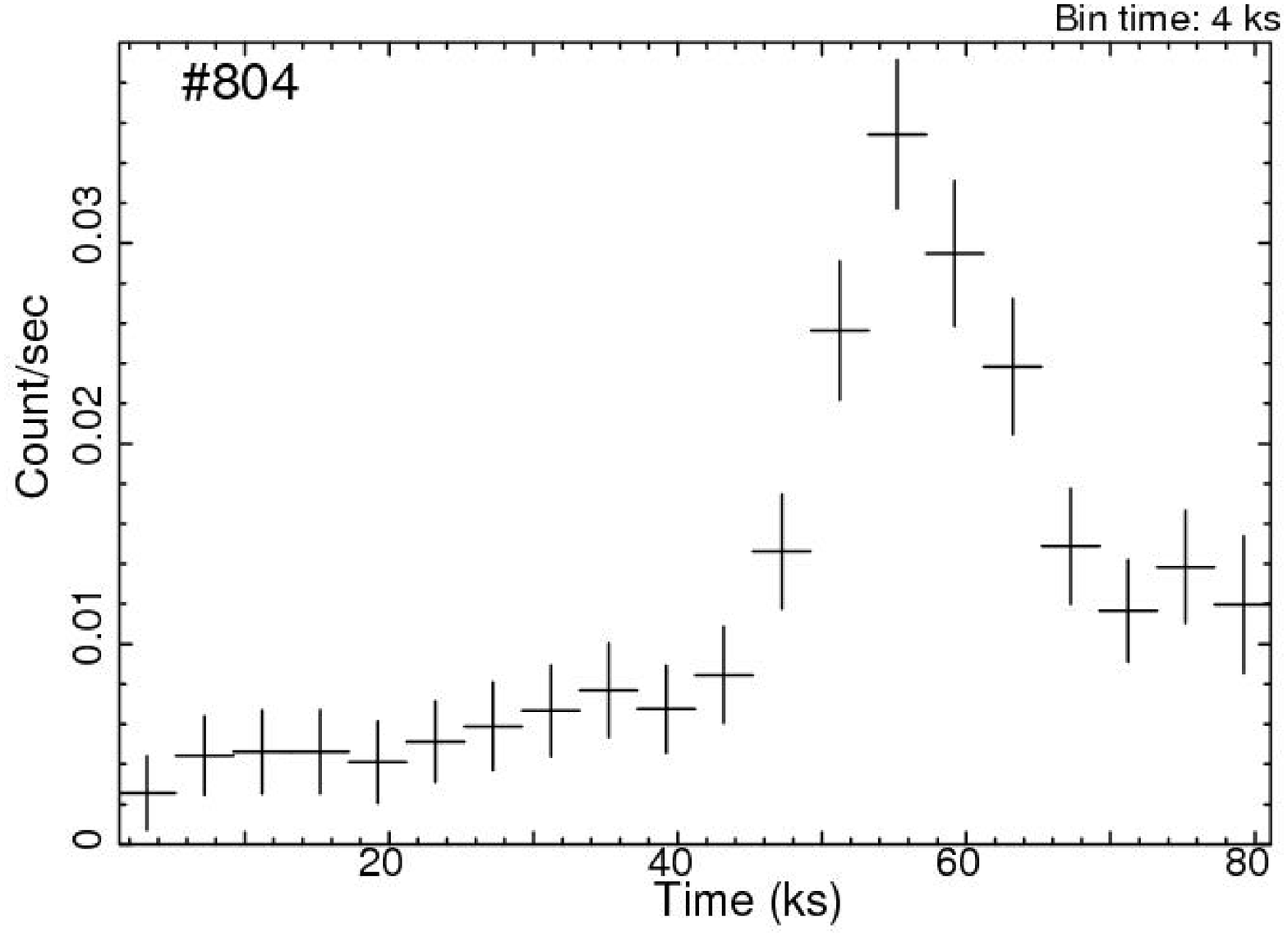}
\includegraphics[width=0.36\textwidth,angle=-90]{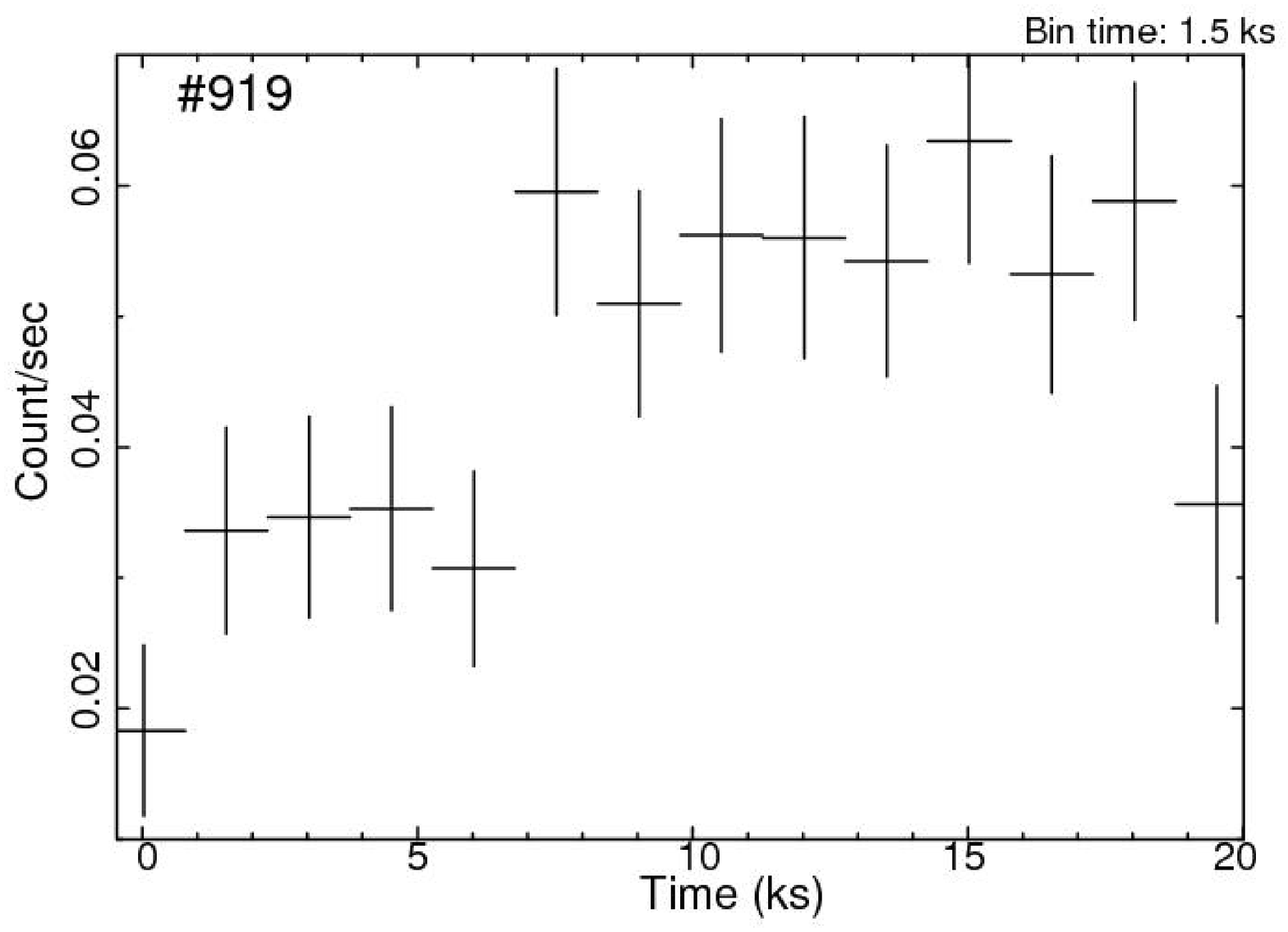}
\caption{Lightcurves of sources with more than 500 counts that are
significantly variable ($P_{KS}\le 0.005$). The ACIS sequence numbers
and binsizes are marked.\label{fig:lightcurve}}
\end{figure}

One of the most notable characteristics of PMS stars is flaring in the
X-ray band, and a few extraordinarily powerful X-ray flares have been
reported in several PMS stars \citep[e.g.,][]{Imanishi01, Grosso04,
Favata05, Getman06, Wang07, Broos07}.  A Kolmogorov-Smirnov (K-S) test
is performed by {\it AE} on each observation in order to evaluate
X-ray light curve variability within that observation, comparing the
source event arrival times to that of a uniform light curve model.
Seventy-eight sources display significant variability ($P_{KS} <
0.005$ in column 15 of Tables~\ref{tbl:primary} and
~\ref{tbl:tentative}) and 14 of them have more than 200 net
counts. Six of these highly variable light curves, for sources having
more than 500 counts, are shown in Figure~\ref{fig:lightcurve}. 

Due to the short duration (20~ks) of most of our Rosette observations,
few flares are observed in their entirety.  However ObsID 3750 is
75~ks long and completely captures a giant flare from source \#634
(average luminosity $\log L_{t,c}=31.3$~ergs~s$^{-1}$). The count rate
during the flare peak is $\sim$100 times that of the quiescent
level. The estimated peak luminosity is $\log
L_{t,c}=32.3$~ergs~s$^{-1}$. The shape of the flare is not symmetric,
with a $\sim$1.5 hr rising phase and a $\sim$4 hr decaying phase,
resembling the fast rise and slow decay X-ray flares commonly seen in
the COUP young stars \citep{Favata05}. Its IR counterpart
(\S~\ref{sec:counterpart_sec}) does not show $K$-band excess, and the
color and magnitude are consistent with a T Tauri star. The X-ray spectrum
is very hard ($kT\sim 7$ keV). Multiple flares are seen in \#691 with
different intensities. Source \#919 is only covered in the 20ks
observation, but it shows a ``flat-top'' flare, characterized by a
fast rise from low flux to high flux and a constant high flux level.

\subsection{Spectral Fitting}
\label{specfit.sec}

For brighter sources with photometric significance {\em Signif}$>2.0$
(column 12 in Tables~\ref{tbl:primary} and \ref{tbl:tentative}), the
extracted spectra were fit using single temperature and
two-temperature {\em apec} thermal plasmas \citep{Smith01} and power
law models subjected to an absorbing column ($N_H$) of interstellar
material with the {\it XSPEC}\footnote{
http://heasarc.gsfc.nasa.gov/docs/software/lheasoft/xanadu/xspec}
package \citep[version 12.2.1ap,][]{Arnaud96}, based on source
spectra, background spectra, ancillary response functions (ARFs) and
redistribution matrix functions (RMFs) from {\it ACIS Extract}.  The
best-fit model was achieved by the maximum likelihood method
\citep{Cash79}. Abundances of 0.3~$Z_{\odot}$ were assumed for the
automated fitting performed by {\it AE}. 

In general, we prefer unconstrained fits (including power law fits) to
constrained fits. The single temperature thermal plasma {\it apec}
model is the default model used for spectral fitting.  For sources
brighter than 100 counts, if a one-temperature thermal plasma model
did not fit the data well, a two-temperature thermal plasma model or
variable abundance {\em vapec} thermal plasma model was invoked. A
power law model was adopted if it represented the data more adequately
than the thermal model (visually or with improved statistics) or if
the thermal model required nonphysical parameters (e.g., $kT \gg 15$
keV). Note that the adopted model should not be used to infer the
nature of the source; a source best fit by a power law is not
necessarily an AGN. For a number of sources that required a very hard
thermal plasma and were identified with known stellar counterparts or
exhibited flaring light curves that were suggestive of PMS stars, we
truncated the plasma temperature at $kT = 15$ keV and adopt the
thermal plasma model.  When no model was acceptable, we froze the
parameter $kT = 2$~keV in the thermal model, a typical value for young
PMS stars \citep{Getman05b,Preibisch05a}, and then fit for the
absorbing column density $N_H$ and the normalization parameters.  The
brightest source in the field is the O star HD 46150 (ACIS \#373),
which has 3588 ACIS counts in 94 ks of exposure, or a count rate of
0.12 counts per CCD frame. This is not bright enough to cause multiple
photon events in a single frame that would corrupt our spectral
fitting, thus it does not warrant correction for photon pile-up
\citep{Townsley02}.

Spectral analysis results for the 630 sources with {\em Signif} $\ga
2.0$ are presented in Tables~\ref{tbl:thermal_spectroscopy} (588
sources; thermal plasma fits) ~and~\ref{tbl:powerlaw_spectroscopy} (42
sources; power-law fits). The table notes give detailed descriptions
of the columns. Best-fit absorbing column densities range from
negligible to $\log N_H \sim 23.3$~cm$^{-2}$, equivalent to a visual
absorption of $A_V \sim$~120~mag \citep{Vuong03}. Temperatures range
from $kT \sim 0.2$~keV to the hardest sources truncated at $kT = 15$
keV.  The range of observed total band ($0.5-8$~keV) absorption
corrected luminosities able to be derived from spectral modeling is
$29.1 \lesssim \log L_{t,c} \lesssim 32.4$ ergs~s$^{-1}$. Assuming a 2
keV plasma temperature and an average $A_V=1.5$~mag visual extinction
($\log N_H \sim 21.4$~cm$^{-2}$ absorbing column), PIMMS\footnote{The
  Portable, Interactive Multi-Mission Simulator is software for
  high-energy astrophysicists, written and maintained by Koji
  Mukai. See
  \url{http://heasarc.gsfc.nasa.gov/docs/software/tools/}}
gives an apparent total band luminosity $\log L_t \sim
28.7$~ergs~s$^{-1}$ estimated for the faintest on-axis detection in
Table~\ref{tbl:tentative}. A conservative estimate for limiting
sensitivity of the entire NGC 2244 observation is $\log L_t \sim 29.4$
erg s$^{-1}$ assuming an average extinction $A_V=1.5$ and a 10 counts
detection; the exact value depends on off-axis location and
absorption.

\section{Identification of Stellar Counterparts and Their Properties}\label{sec:counterpart_sec}

\subsection{Stellar Counterparts Matching and IR diagrams}\label{sec:IRcolor}

We associate ACIS X-ray sources with optical and near IR (ONIR)
sources using positional coincidence criteria, as described in the
Appendix of \citet{Broos07} \footnote{Software implementing the
matching algorithm is available in the TARA package at
\url{http://www.astro.psu.edu/xray/docs/TARA/} }.  The optical and
infrared catalogs from recent literature and observations that we
adopted for counterparts identification include: $UBV$ photometry of
NGC 2244/Mon OB2 (Massey et al. 1995 =MJD95), $UBVI$H$_{\alpha}$
photometry of NGC 2244 (Park \& Sung 2002 =PS02), $BVIR$H$_{\alpha}$
photometry of NGC 2244 (Bergh\"{o}fer \& Christian 2002 =BC02), the
Whole-Sky USNO-B1.0 Catalog (Monet et al. 2003 =USNO), 2MASS All-Sky
Catalog of Point Sources (Cutri et al. 2003 =2MASS), and the
University of Florida FLAMINGOS Survey of Giant Molecular
Clouds\footnote{PI: Elizabeth A. Lada; details about the FLAMINGOS
Survey can be found at
\url{http://flamingos.astro.ufl.edu/sfsurvey/sfsurvey.html}. The
instrument design and performance of FLAMINGOS are described in Elston
et al.\ (2003). An overview of the instrument is also available at
\url{http://www.gemini.edu/sciops/instruments/flamingos1}. The
FLAMINGOS observations of the Rosette Complex fields and IR data
reduction are described in Rom\'an-Z\'u\~niga (2006). Note that both
$K$ and $K$-short ($K_s$) filters are available in FLAMINGOS imaging
mode, and the observations here are taken in $K$ band.}. The reference
frame offsets between the ACIS fields (astrometrically aligned to the
Hipparcos frame using 2MASS sources in the data reduction) and the
catalogs are $0.4^{\prime\prime}$ to MJD95, $0.3^{\prime\prime}$ to
PS02, $0.3^{\prime\prime}$ to BC02, $0.2^{\prime\prime}$ to USNO, and
$0.2^{\prime\prime}$ to FLAMINGOS.  These offsets were applied before
matching sources.

Likely associations between ACIS sources and ONIR sources are reported
in Table~\ref{tbl:counterparts}; 712 of the 919 ACIS sources (77\%)
have an ONIR counterpart identified. Since \citet{PS02} provide the
full list of their high precision photometry, visual photometry is
reported in the priority order of PS02, BC02, MJD95, and USNO. USNO
photometry is photographic with $\sim 0.3$ magnitude photometric
accuracy \citep{Monet03}.  $JHK$ magnitudes from FLAMINGOS photometry
are reported if available for {\em Chandra} sources.  The FLAMINGOS
photometric data was zero-pointed to 2MASS.  (See Rom\'an-Z\'u\~niga et
al. 2007 for details on photometry.  A photometric catalog for optical
stars in NGC 2244 is presented in RL07 and the complete catalog from
the FLAMINGOS survey of the Rosette will be reported in
Rom\'an-Z\'u\~niga et al. 2007.)  For areas that were not covered by
the FLAMINGOS survey and for bright stars that were saturated ($H<11$
mag, Rom\'an-Z\'u\~niga 2006), 2MASS photometry is reported. The
SIMBAD and VizieR catalog services are used for complementary
information. Notes to other published characteristics of the selected
sources can be found in the table footnotes.

\begin{figure}
\centering
\epsscale{0.7}
\plotone{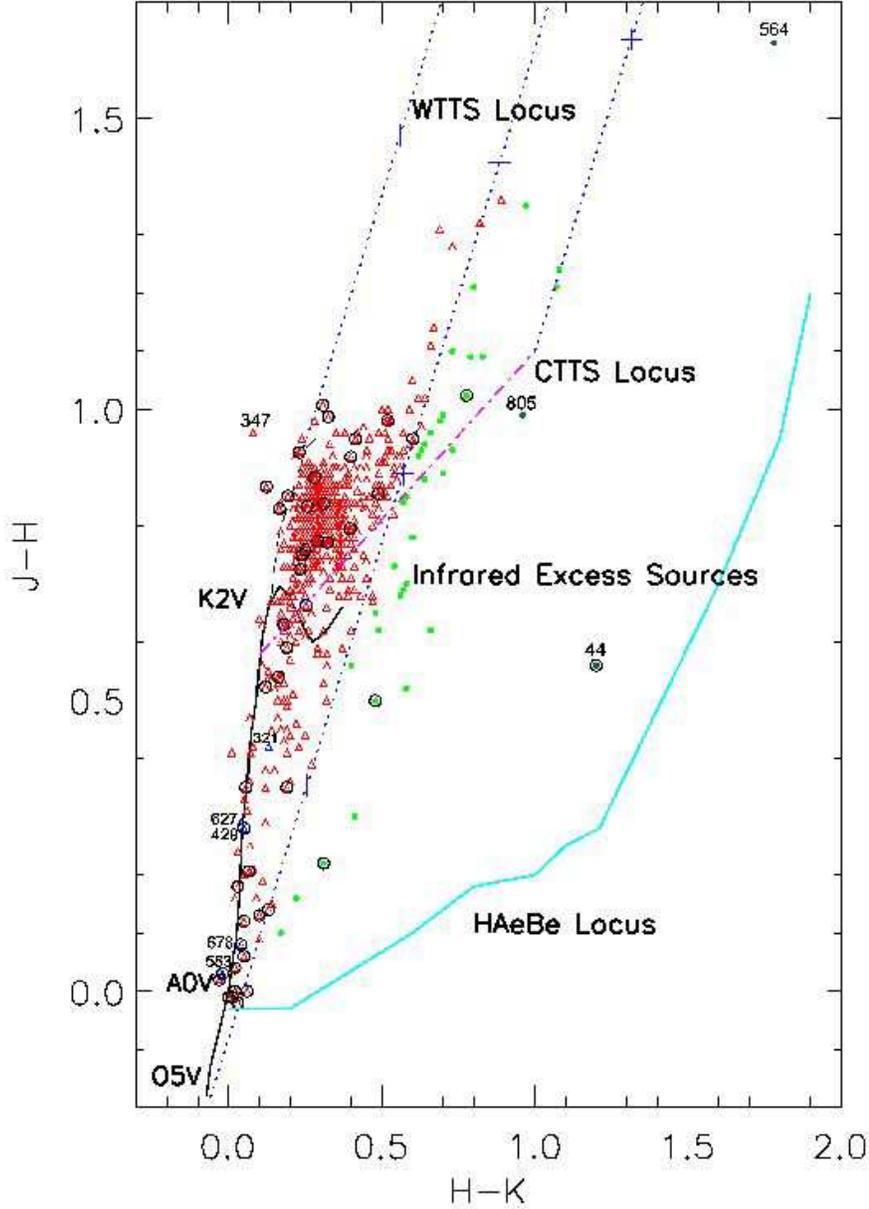}
\caption{NIR $J-H$ vs.\ $H-K$ color-color diagram for 617 {\em
Chandra} stars with high-quality photometry, from combined FLAMINGOS
and 2MASS data (error in both $J-H$ and $H-K$ colors $<0.1$ mag). The
(light and dark) green circles and red triangles represent sources
with significant $K$-band excess and sources without excess,
respectively. The dark green circles represent three Class I objects
and are labeled with their sequence numbers from Table~2. The five
blue triangles are foreground stars. Stars using 2MASS photometry are
indicated with black circles. The black solid and long-dash lines
denote the loci of MS stars and giants, respectively, from Bessell \&
Brett (1988). The purple dash dotted line is the locus for classical T
Tauri stars from Meyer et al. (1997), and the cyan solid line is the
locus for HAeBe stars from Lada \& Adams (1992). The blue dashed lines
represent the standard reddening vector with crosses marking every
$A_V=5$ mag. Most {\em Chandra} sources are located in the reddening
band defined by the left two dashed lines associated with Class III
objects (triangles). To the right of this reddened band are 38
IR-excess sources.\label{fig:ccd}}
\end{figure}

Figure~\ref{fig:ccd} shows the NIR $J-H$ vs.\ $H-K$ color-color
diagram for 617 out of 919 {\em Chandra} stars with high-quality $JHK$
photometry (error in both $J-H$ and $H-K$ colors $<0.1$ mag) listed in
Table~\ref{tbl:counterparts}. Most {\em Chandra} sources are located
between the left two dashed lines, the color space associated with
diskless young stars (Class III objects) that are reddened by
interstellar extinction. We emphasize that while this region is
usually filled with field star contaminants in NIR-only studies, here
nearly all of these stars are cluster members (\S3.2). A concentration
of cluster members subjected to $A_V\sim 1-2$~mag (assuming late type
stars) is apparent, centered at $J-H$=0.25, $H-K$=0.8. To the right of
this reddening band are 38 $K$-band excess sources\footnote{Note that
the definition of $K$-band excess slightly varies when considered by
different researchers. In \citet{Roman07}, a $K$-band excess star is
required to have colors that lie above $J-H=0.47(H-K)+0.46$ (locus of the
Classical T Tauri Stars; CTTS), in addition to requiring that the star
is located to the right of the reddening band for zero age main
sequence dwarfs. The region below the CTTS locus could include detections
of unresolved galaxies. But for the IR counterparts to our X-ray
selected sample, this color space has little contamination from
galaxies, and likely contains Herbig Ae/Be stars in the young
cluster. Therefore no further constraint is applied in defining the
$K$-excess.}, defined as stars that have colors
$(J-H)>1.7(H-K)+2\sigma(H-K)$. The presence of $K$-band excess is
frequently used to identify stars still possessing inner disks that
are relatively hot ($T\sim$1200K) and dusty. All except three of these stars
occupy the color space between the middle and right-most dashed lines;
they are likely PMS stars with circumstellar accretion disks (Class
II objects).

The three stars (\#44, \#564, and \#805) located beyond the right-most
dashed line show large color excess ($E(H-K)=0.87$, 0.82, and
0.38, respectively). Stars in this domain are likely surrounded by
extended envelopes, and hence are classified as candidate Class I
protostars \citep{Kenyon93,Strom95}. Sources \#44 and \#564 are
particularly interesting ($E(H-K)> 0.8$).  Source \#44 was
identified as a member of NGC 2244 \citep{OI81} and assigned spectral
type B7Ve \citep{Verschueren91}. It is further classified as a Herbig
Be star in \citet{Li02} based on the tenuous nebula seen in their KPNO
H$_{\alpha}$ image, large [V-25$\mu$m] color, and its confirmed late B
spectral type. Source \#805 is very faint ($J\sim 18.5$, close to the
FLAMINGOS imaging sensitivity limit) and may be a low mass protostar
or an embedded background object with large $H-K$ color
\citep{Froebrich05}.

\begin{figure}
\centering
\epsscale{0.8}
\plotone{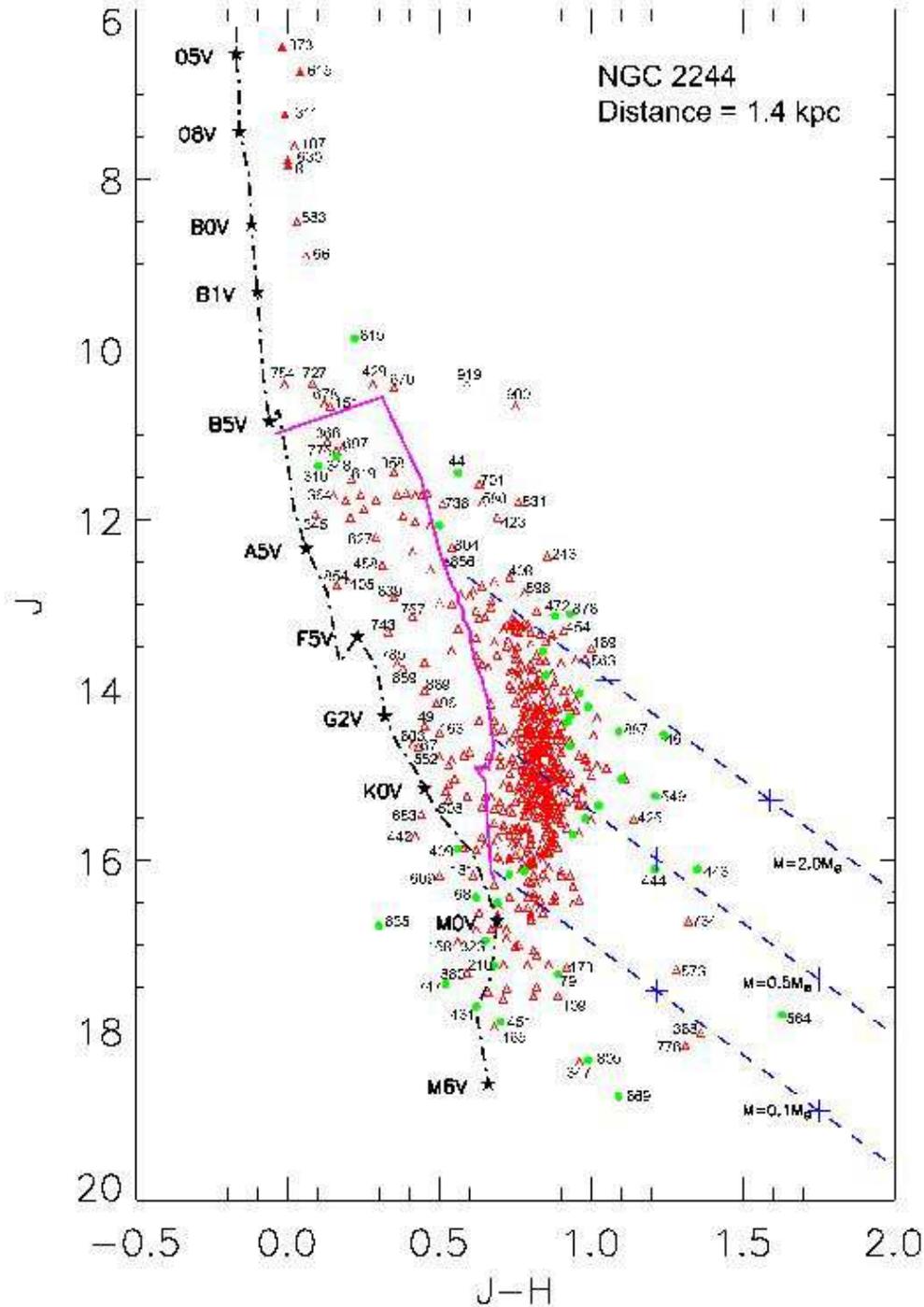}
\caption{NIR $J$ vs. $J-H$ color-magnitude diagram using the same
sample and symbols as Figure~\ref{fig:ccd}, except that known O stars
are denoted as the red filled triangles. ACIS source numbers are
marked for some stars.  The purple solid line is the 2~Myr isochrone
for PMS stars from \citet{Siess00}. The dash dotted line marks the
location of Zero Age Main Sequence (ZAMS) stars. The blue dashed lines
represent the standard reddening vector with asterisks marking every
$A_V=5$ mag and the corresponding stellar masses are
marked.\label{fig:cmd}}
\end{figure}

Figure~\ref{fig:cmd} shows the NIR $J$ vs.\ $J-H$ color-magnitude
diagram for the same stars shown in Figure~\ref{fig:ccd}. Known OB
stars are located at the top and reddened from the ZAMS with $A_V\sim
1$~mag. Unlike NGC 6357 \citep{Wang07} and other more heavily obscured
clusters \citep{Wolk06,Broos07}, we do not locate any new candidate
massive stars earlier than B0. However two likely new B0-B2 stars
(\#900 and 919) and a dozen new late B candidates are found.

Individual masses and reddening of the lower mass stars can be
estimated assuming that they are reddened from the 2~Myr
isochrone. Adopting a larger distance of $d=1.6$ kpc does not have a
significant effect here; the mass estimates for the low mass stars are
only affected by $\sim$0.1 M$_\odot$. The majority of ACIS stars
appear to be concentrated around $0.7\la J-H \la 1$ and $13 \la J \la
16$, consistent with F, G, K, and M stars ($0.1 \la M \la
2$~M$_\odot$) reddened by $1 \la A_V \la 2$~mag. Those showing
$K$-band excess seem to have larger reddening compared to the Class
III objects.  Sources to the left of the 2~Myr isochrone are probably
field stars older than the NGC~2244 population (see \S
\ref{sec:contamination}).  In addition, to the right of the ZAMS track,
around twenty stars have inferred stellar masses
$<0.1$~M$_{\odot}$. The IR photometry for these $J>17$ mag stars may
be less reliable, especially the fainter ones that are close to the
detection limit. These are likely a mixture of the lowest mass cluster
members, background stars, and a few IR luminous extragalactic
sources. The high percentage of NIR excess sources among the faintest
stars may not be real; \citet{Froebrich05} have shown that background
stars embedded in distant clouds have an overall larger $H-K$
color. Contamination by faint background IR sources is also discussed
in Rom\'an-Z\'u\~niga (2006).

\begin{figure}
\centering
\plotone{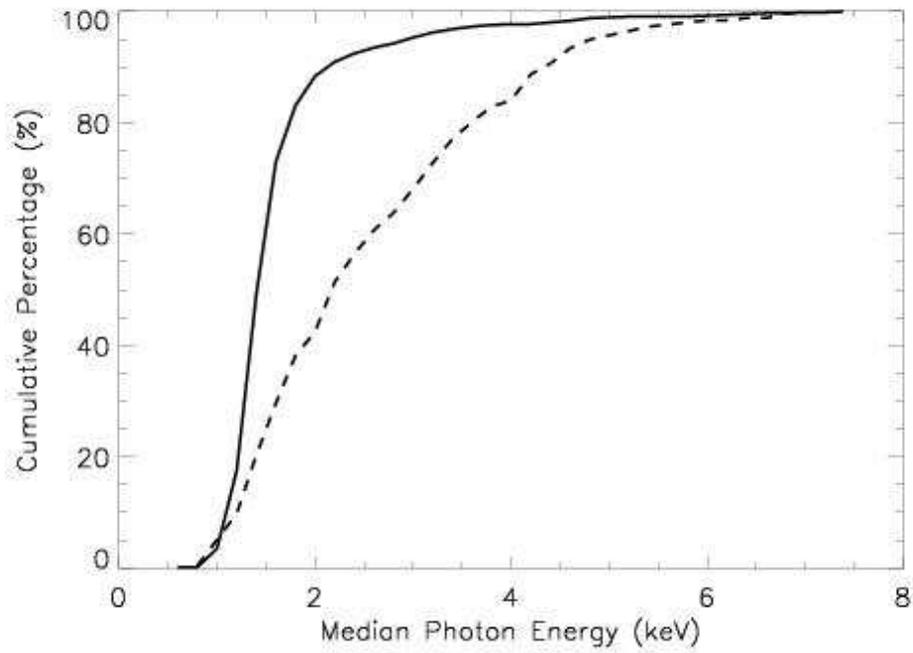}
\caption{The cumulative distribution of the source hardness indicator,
median photon energy, for {\em Chandra} sources with identified ONIR
counterparts (solid line) and those without counterparts (dashed
line). The sources with identified ONIR counterparts are considerably
softer than the latter group. \label{fig:cdf}}
\end{figure}

\begin{figure}
\centering
\plotone{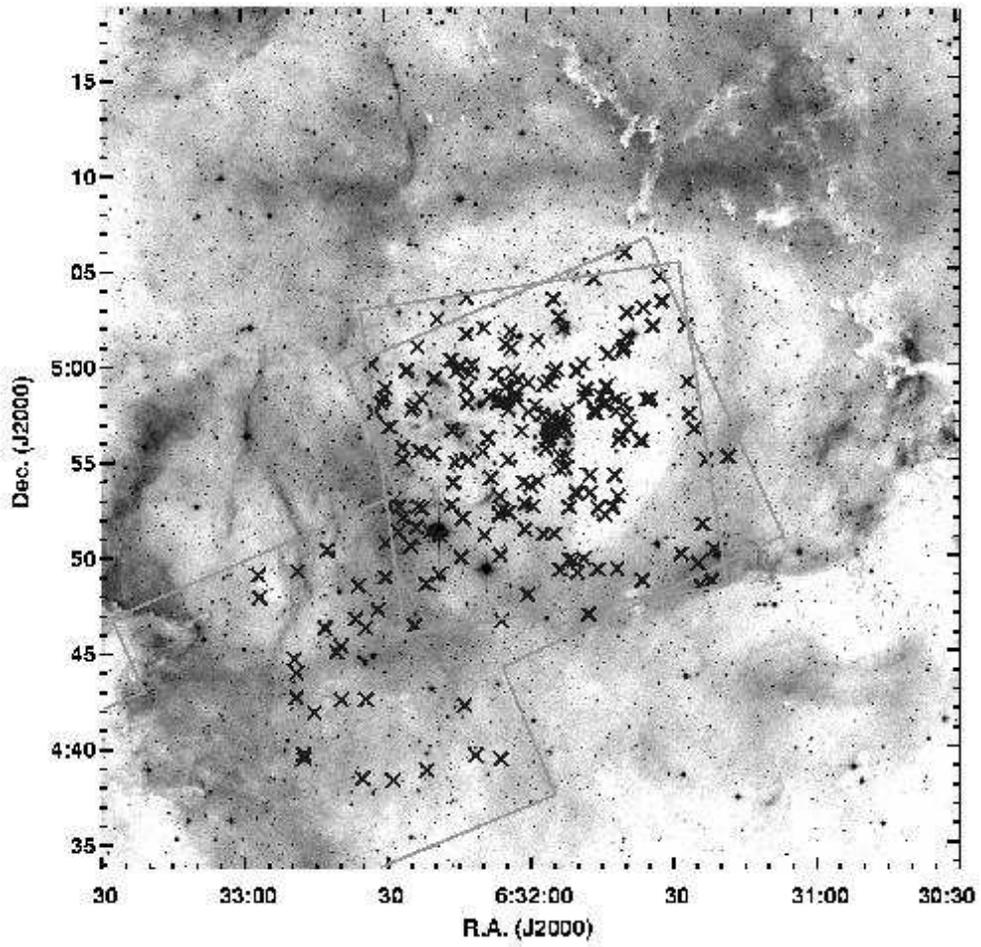}
\caption{The spatial distribution of {\em Chandra} sources without
identified ONIR counterparts. The background is the DSS $R$-band
image. ACIS-I FOVs are shown.\label{fig:no_counterpart}}
\end{figure}

The remaining $\sim$200 ACIS sources without matched counterparts are
likely to be a mixture of newly discovered embedded members of the
Rosette complex and distant background stars and extragalactic sources
(see \S \ref{sec:contamination} for estimated fractions). Cumulative
distribution functions of the median photon energy are shown in
Figure~\ref{fig:cdf} for sources that have matched counterparts and
those that do not have counterparts.  The harder median photon energy
of sources that do not have counterparts indicates that these sources
are deeply embedded or behind the cloud ($A_V \ga 10$~mag).  The new
low mass cluster stars reveal themselves because of strong X-ray
flares due to magnetic activity. Such X-ray discovered stars are
commonly found in the molecular clouds surrounding other clusters
\citep{Getman05a,Wang07,Broos07}.  Some of these may be very young
protostars with local absorption in an envelope or disk;
\citet{Getman07} found in IC 1396N that sources with $\log N_H\ga
23.0$ cm$^{-2}$ are protostars with dense envelopes.  Their spatial
distribution (Figure~\ref{fig:no_counterpart}) shows that they are
distributed around the bright stars and along the rims of the Rosette
nebula where infrared source identification may be challenging.

\subsection{Identification of Likely Contaminants} \label{sec:contamination}

As emphasized in \S 1, stellar X-ray emission decays rapidly only
after $\sim 100$ Myr and show only a small dependence of X-ray
luminosity on disk accretion
\citep{Preibisch05,Preibisch05a,Telleschi07}. Therefore X-ray surveys
have very high efficiency in detecting disk-free PMS stars over MS
stars, complementing traditional optical and infrared surveys of star
forming regions.  We evaluate the level of contamination by
extragalactic X-ray sources and Galactic disk stars following the
simulations described in \citet{Getman06} and \citet{Wang07}. The
calculations indicate that $\la 35$ sources are extragalactic
(\S~\ref{sec:hard_unidentified_section}), $\sim$20 are foreground and
$\sim$16 are background field stars (\S~\ref{sec:gal_con}). These
$\sim$70 contaminants constitute $\sim 8\%$ of the 919 ACIS sources.

\subsubsection{Extragalactic Contaminants}\label{sec:hard_unidentified_section}

Following Getman et al. (2006), we construct Monte Carlo simulations
of the extragalactic population by placing artificial sources randomly
across the detector.  We draw incident fluxes from the X-ray
background $\log N - \log S$ distribution \citep{Moretti03}, and power
law photon indices for the sources are assigned consistent with
flux-dependencies described by \citet{Brandt01}.  The spectrum of each
simulated source was passed through a uniform absorbing column density
$\log N_H \sim 21.9$ cm$^{-2}$, the H{\sc I} column density through
the entire Galactic disk in the direction of NGC 2244
\citep{Dickey90}.  After applying local background levels found in our
ACIS image, we calculate the photometric significance of each fake
source and then reject the weak extragalactic sources that would have
fallen below our source detection threshold. The simulations suggest
that $\sim 80$ extragalactic sources may be detected in our ACIS-I
field and $\sim 35$ may have photometric significance $Signif \ge 2.0$
(column 12 in Table~\ref{tbl:primary}). The true number is probably
smaller as we did not account for the patchy distribution of molecular
cloud material. The best candidates for the extragalactic contaminants
are those sources whose X-ray spectra are best fit by a power law, do
not have bright ONIR counterparts, and do not display the
characteristic PMS X-ray flares.

\subsubsection{Galactic Stellar Contamination}\label{sec:gal_con}

Monte Carlo simulations of the Galactic stellar population expected in
the direction of our ACIS fields ($l=206.3$, $b=-2.1$) were examined,
based on the stellar population synthesis model of Robin et al. (2003;
henceforth the Besan\c{c}on model)\footnote{These calculations are
made with the Web service provided by the \citet{Robin03} group at
Besan\c{c}on at \url{http://bison.obs-besancon.fr/modele}.}. In
addition to the smooth absorption component provided in the model, we
added a local absorption at $d=1.4$ kpc with a low $A_V \sim 1$ mag,
as inferred from the average NIR reddening of {\it Chandra} sources
shown in Figure~\ref{fig:cmd}.  Within the ACIS field-of-view, the
Besan\c{c}on model predicts $\sim 850$ foreground MS stars ($d < 1400$
pc) and $\sim 4300$ background MS stars, giants, and sub-giants within
the FLAMINGOS imaging sensitivity limit around $J < 19$ mag.

About $25\%$ of the foreground stars in the Besan\c{c}on simulation
have ages less than 1 Gyr, the younger population that could be
detected in X-ray surveys. Most reside at distances from 0.8 to 1.4
kpc; $\sim$54\% are M stars, $\sim$28\% are K stars, $\sim$13\% are G
stars, and 3\% are more massive. Following Getman et al. (2006), we
convolved the Besan\c{c}on model populations with the X-ray luminosity
functions of stars in the solar neighborhood measured from {\it ROSAT}
surveys \citep{Schmitt95, Schmitt97, Huensch99}. Luminosities were
adjusted to account for the different {\it ROSAT} and {\it Chandra}
spectral bands following the stellar hardness-luminosity relation
\citep{Gudel98}.  Following the same procedure we used for
extragalactic sources (\S \ref{sec:hard_unidentified_section}), we
applied our ACIS detection process to these simulated field stars.  A
typical Monte Carlo run predicts that $16-20$ foreground stars will be
detected in the ACIS-I exposures of NGC 2244.

Five optical counterparts to our X-ray stars are identified as
foreground stars based on their known spectral types and positions in
photometric diagrams. These are \#321, 429, 583, 627, and
678. Proper-motion data \citep{Marschall82,Dias06} also suggest that
they have low probability of being member stars of the cluster. In the
NIR color magnitude diagram, there are $\sim 30$ stars located between
2 Myr isochrone and the ZAMS track. Those close to the 2 Myr
isochrones are still consistent with being members due to uncertainty
in the age and distance of the cluster. About fifteen stars close to
the MS track are more likely to be unrelated foreground main sequence
stars \citep{Getman06}.  Available X-ray absorption columns derived
from spectral fits mostly require $\log N_H=20.0$ cm$^{-2}$,
suggestive of being foreground candidates\footnote{Spectral fits to
low counts sources can give $N_H$ values with large uncertainties,
hence we do not use low $N_H$ to identify foreground sources. Low
absorption columns are considered supporting evidence for foreground
stars suggested by their locations in the NIR color magnitude
diagram. The spatial distributions of the low $N_H$ sources with high
counts ($>30$ counts) and with low counts ($<30$ counts) are similarly
dispersed through the field, consistent with being part of the foreground
population.}. Their probabilities of being members derived from proper
motion data range between 0\% and 80\%, which are less
conclusive. With the above five confirmed foreground stars, altogether
twenty optical stars\footnote{ They are associated with {\em Chandra}
sources \#49, 67, 96, 105, 129, 163, 321, 429, 552, 583, 603, 627,
678, 743, 757, 785, 839, 854, 859, and 889.} are noted as the best
candidates for foreground stars in the Table~\ref{tbl:counterparts}
footnotes. This number is quite consistent with the Besan\c{c}on
simulated foreground population.

The Besan\c{c}on model is again convolved with X-ray luminosity
functions (XLFs) to simulate the number of stars behind the Rosette
star forming region that may enter our X-ray sample.  The model
predicts $\sim 18\%$ F dwarfs, $\sim 41\%$ G dwarfs, $\sim 33\%$ K
dwarfs, $\sim 1\%$M dwarfs, and $\sim 6\%$ giants. We use the dwarf
XLFs established for the solar neighborhood
\citep{Schmitt95,Schmitt97} and adopt the XLF for giants obtained from
Table 2 of \citet{Pizzolato00}. Typical runs of simulations result in
$\sim$11 dwarfs and $\sim$5 giants that are detectable in our {\em
Chandra} observation. To the left of the ZAMS track in the CMD and for $J>18$,
there are 16 stars\footnote{They are \#68, 157, 158, 181, 226, 253,
323, 325, 347, 409, 442, 583, 609, 747, 835, and 880.}  whose
locations match the Besan\c{c}on model predicted background population
of MS stars, subgiants, and giants, which we note as best candidates
for background stars in Table~\ref{tbl:counterparts}.

\section{Global Properties of the Stellar Cluster}

\subsection{X-ray Luminosity Function and Initial Mass Function}\label{sec:xlf}

As noted in \citet{Feigelson05}, the XLF (which is directly measured
here) can be considered to be the convolution of the IMF (which is
unknown) and the X-ray--Mass Luminosity ($L_x - M$) correlation (which
is measured in the COUP studies; Preibisch et al.\ 2005).  Using the
best-studied Orion Nebula Cluster XLF (COUP XLF) and IMF as a
calibrator, the NGC 2244 XLF can be used to probe the IMF of the
stellar cluster and to estimate the total X-ray emitting
population. Such a population analysis has been made for Cep OB3b, NGC
6357, and M17 \citep{Getman06,Wang07,Broos07}. In the following XLF
analysis, we use the hard band XLF rather than the total band XLFs,
since the unknown soft component of heavily absorbed X-ray sources can
introduce a large uncertainty in both the observed and absorption
corrected total band X-ray luminosity.

\begin{figure}
 \centering
\epsscale{0.7}
\plotone{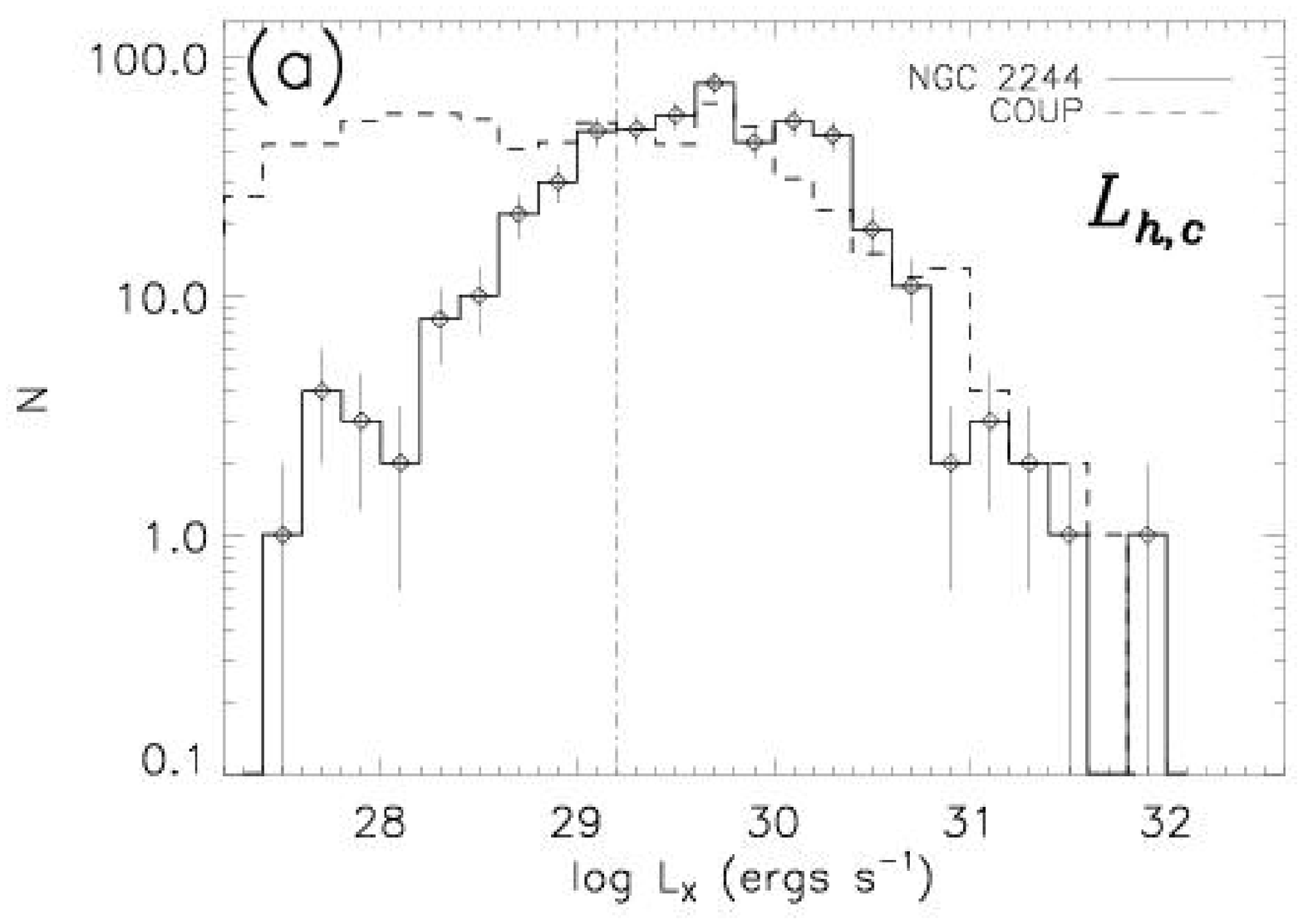}
\plotone{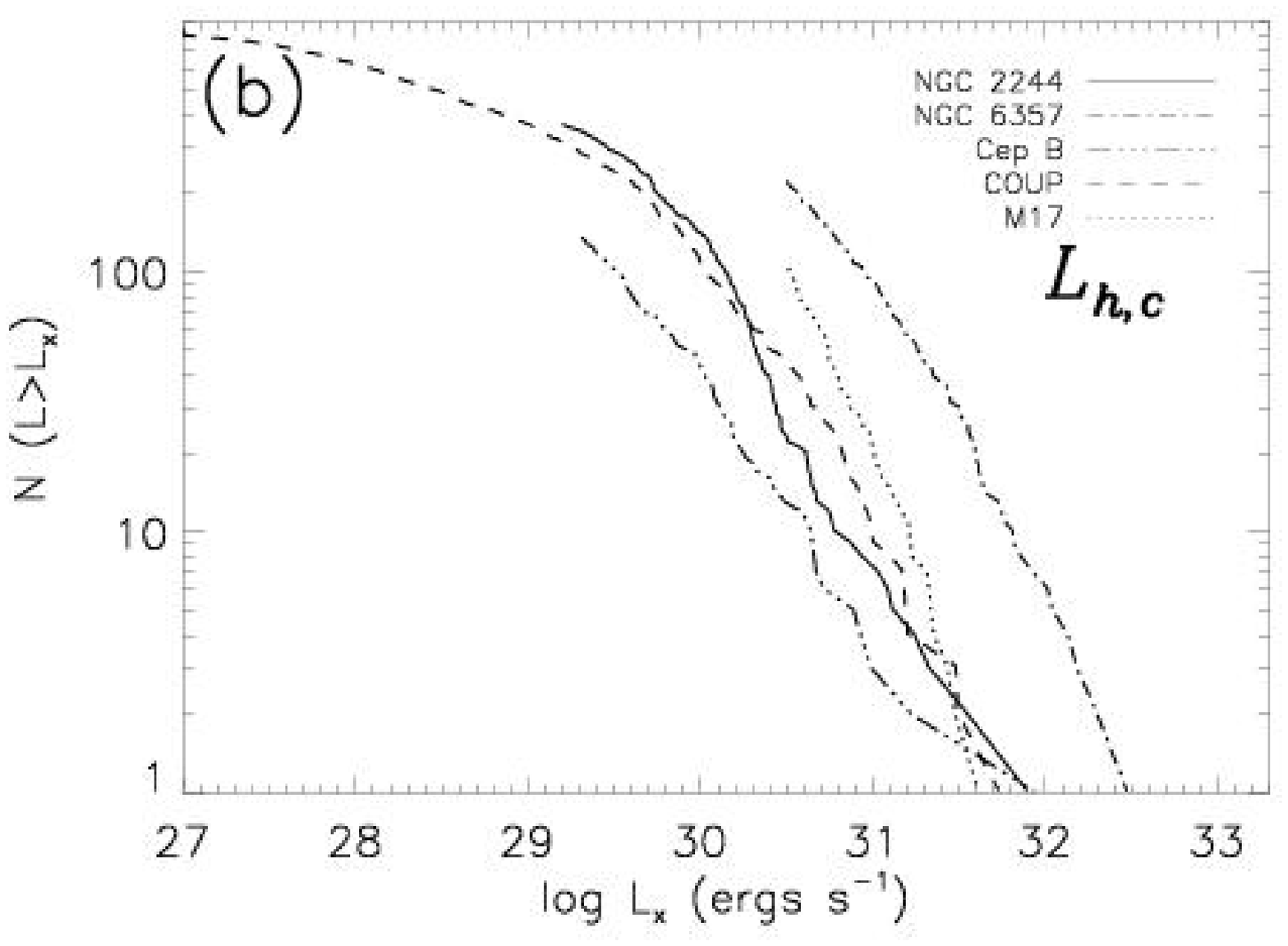}
\epsscale{1.0}
\caption{(a): X-ray luminosity function (XLF) constructed from the
absorption corrected hard band (2.0--8.0 keV) X-ray luminosity
$L_{h,c}$ for the unobscured NGC 2244 population (solid line), and the
COUP ONC unobscured cool stars population \citep[dashed
line,][]{Feigelson05}. The vertical line denotes the estimated
completeness limit for the NGC 2244 population. (b): The cumulative
distribution of X-ray luminosities for the unobscured NGC 2244 X-ray
sources as well as the unobscured populations in COUP, Cep B/OB3b
\citep[age$\sim$1-3 Myr,][]{Getman06}, NGC 6357 \citep[age$\sim$1
Myr,][]{Wang07}, and M17 \citep[age$\sim$1 Myr,][]{Broos07}. The
distributions for NGC 2244, NGC 6357, Cep B, and M17 are truncated at
their completeness limits, $\log L_{h,c}\sim 29.2$, 30.4, 29.3, and
30.4 ergs s$^{-1}$, respectively.\label{fig:XLF}}
\end{figure}

By counting the number of sources in different X-ray luminosity bins,
we construct the absorption corrected hard band (2--8 keV) XLF
($L_{h,c}$) for all unobscured NGC 2244 X-ray sources ($MedE\le
2.0$~keV) with derived X-ray luminosities in
Figure~\ref{fig:XLF}a. Here we exclude the five known foreground
stars and OB stars with spectral types earlier than B3 to be
consistent with the Orion cool stars sample. The absorption corrected
hard band fluxes ($F_{h,c}$) derived from XSPEC spectral fitting are
used to obtain luminosities assuming a distance of 1.4 kpc. As the
template, we also show the XLF of the COUP unobscured population (839
cool stars; Feigelson et al.\ 2005). 

The NGC 2244 XLF is largely consistent with the COUP XLF, suggesting
an NGC 2244 population comparable to the ONC. However, at $\log
L_{h,c} \ga 29.8$ ergs s$^{-1}$ the slope of the NGC 2244 XLF seems
steeper than the COUP XLF. At $\log L_{h,c} = 30.9$ ergs s$^{-1}$, the
NGC 2244 bin is short 9 stars compared to the ONC XLF; the luminosity
bins at $\log L_{h,c} = 29.5-29.7$ ergs s$^{-1}$ are $\sim 40$ stars
higher than the ONC XLF.  Admittedly these deviations from the ONC XLF
are not significant ($2\sigma$), more closely following the shape of
the ONC than the Cep B or the M17 XLF does (see
Figure~\ref{fig:XLF}b). The apparent steeper slope is not an artifact
of our distance estimate or detection completeness limit. We
comfortably detect 15 count sources at any off-axis location; the
X-ray luminosity of the detected sources is roughly $\log L_t \sim
29.4$ ergs s$^{-1}$ (corresponding to $\sim 0.5 M_{\odot}$
from the $L_x-M$ relation) and $L_{h,c}\sim 29.2$ ergs s$^{-1}$. The
luminosity bins with the steep slope seen in the NGC 2244 XLF are much
brighter than the completeness limit.

A few possible extrinsic reasons may account for the slope deviation
from the ONC XLF. First, while the log-normal COUP XLF represents the
best data and provides a good observational template, the underlying
physics and the variations from it among clusters remain to be
explored. There is evidence that the XLF may not be identical in all
regions \citep[see review by][]{Feigelson07}. Figure~\ref{fig:XLF}b
suggests that the XLFs of Cep B and M17 may vary from the ONC XLF in
different ways. The $L_x - M$ correlation is derived from the very
young ONC, but may vary for the older clusters like NGC 2244.  Second,
the exclusion of some hot stars to construct an XLF comparable to the
COUP XLF could implicitly affect the resulting XLF. We examined
luminosities ($\log L_{h,c}$) of these OB stars, which mostly
contribute to the luminosity bins around $\log L_{h,c}\sim 30.0$ ergs
s$^{-1}$; an XLF including them still shows the observed deviation
from the ONC.

There are two possibilities intrinsic to the NGC 2244 cluster that may
be responsible for the apparent steeper slope: (i) The NGC 2244
population is the same as the COUP population, but there is an excess
of $\sim 50$ stars in the luminosity range $30.0\la \log L_{h,c} \la
30.4$ (solar mass stars as inferred from the $L_x-M$ relation); or
(ii) The NGC 2244 population is $\sim 1.2$ times larger than the COUP
population, but NGC 2244 is $\sim 20$ stars deficient in stars with
$30.4\la \log L_{h,c} \la 31.0$ (intermediate mass stars). In either
case, the NGC 2244 XLF deviates from a scaled version of the COUP XLF
in a manner similar to that seen in the Cep B/OB3b field studied by
\citet{Getman06}.

To improve the statistics, remove possible binning effects, and
further investigate how the NGC 2244 XLF compares to that of the ONC
and other clusters, we derive the cumulative distribution of X-ray
luminosities for the unobscured NGC 2244 X-ray sources as well as the
unobscured COUP, Cep B/OB3b \citep[age$\sim$1-3 Myr,][]{Getman06}, NGC
6357 \citep[age$\sim$1 Myr,][]{Wang07}, and M17 \citep[age$\sim$1
Myr,][]{Broos07} populations. The resulting cumulative XLFs are shown
in Figure~\ref{fig:XLF}b. To avoid confusion caused by incompleteness
in detections, we cut off the cumulative XLFs at the corresponding
completeness limit for each region (see Figure~\ref{fig:XLF}).  At a
given X-ray luminosity above the completeness limit, the ratio between
the cumulative numbers of sources from two populations reflects the
relative scaling between the two populations that are more luminous
than this limit. As a result, the unobscured population of NGC 6357,
M17, and Cep B, is $\sim 5$ times, $\sim 3$ times, and $\sim 0.4$
times of the size of unobscured ONC population, respectively. These
are consistent with previously reported values.\footnote{In NGC 6357,
$\log L_{h,c}$ is $\sim$0.5 dex higher than the other clusters because
candidate OB stars are included in the sample \citep{Wang07}. Note
that, although the unobscured population of NGC 6357 is larger than
that of M17, the obscured population in M17 is significantly larger
than that of NGC 6357 \citep{Broos07}, which makes the estimated total
populations of the two clusters comparable.} The cumulative X-ray
luminosity function of NGC 2244 closely follows the COUP XLF, although
the deviation found in Figure~8 (deficit at $30.4\la \log L_{h,c} \la
31.0$ and excess at $\log L_{h,c} \la 30.4$) can still be clearly
seen. Depending on the treatment of this deviation, the NGC 2244
population ranges from 1.0 to 1.2 times the ONC population.

Further investigations were made to examine the possible excess of sources
with $30.0 \la \log L_{h,c} \la 30.4$ ergs s$^{-1}$. To identify a
previously unknown cluster in the field that may contribute extra
stars to NGC 2244, we inspected the spatial distribution of sources that
have luminosities in the excess bins, but did not find any apparent
clustering. To test whether this excess comes from contamination of
X-ray bright non-members, we removed candidate contaminants (foreground
stars, background stars, extragalactic sources) as suggested in
\S~\ref{sec:contamination} and reconstructed the XLF. The number drop
mainly appears in low luminosity bins while the excess is still
significant at $30.0\la L_{h,c} \la 30.4$ ergs s$^{-1}$. Therefore we
conclude that excluding candidate contaminants would not alter the
XLF, since they do not contribute much to the high luminosity bins
that characterize the X-ray emitting population.

\subsection{Initial Mass Function and K-band Luminosity Function}

\begin{figure}
\centering
\plotone{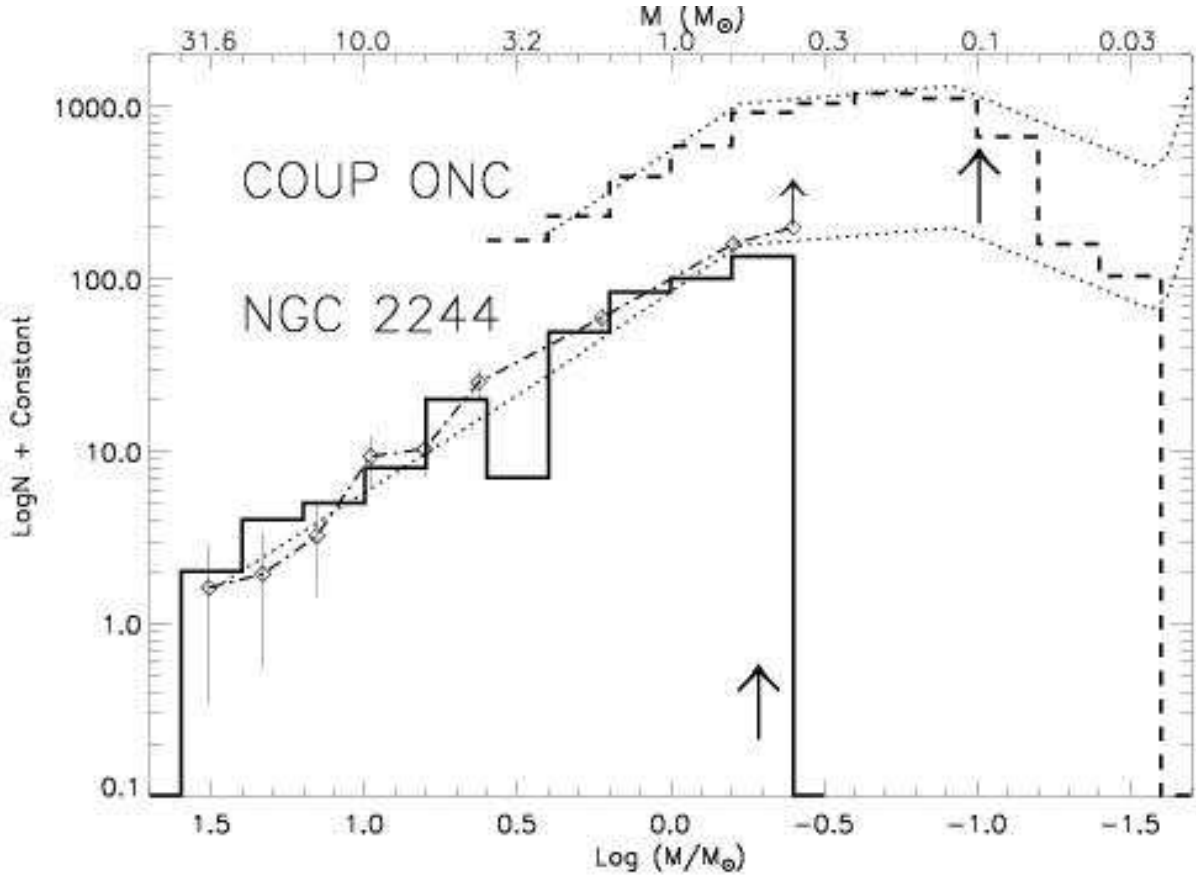}
\caption{Comparison between IMFs of the NGC 2244 X-ray stars (solid
line) and COUP ONC stars (dashed line) using NIR photometry derived
stellar masses. The dotted lines show the ONC IMF \citep{Muench02} and
its scaled version to match the NGC 2244 IMF. The dot-dashed line is
the NGC 2244 IMF estimated using the KLF derived from 2MASS data. The
arrows indicate the approximate mass completeness limits for the NGC
2244 and ONC X-ray stars, and the mass completeness limit for the KLF
derived from 2MASS data. Note that the 2MASS completeness limit is the
same as our {\em Chandra} completeness limit in stellar mass. The
incomplete bins in the NGC 2244 KLF/IMF are omitted.\label{fig:IMF}}
\end{figure}

To examine whether the deviation in the XLF is a reflection of an
intrinsic anomalous IMF in the NGC 2244 cluster, we perform two tests
on the IMFs using NIR data. One experiment is to use the location of
X-ray stars in the IR color magnitude diagram to derive their masses
and construct an approximate IMF. The exact mass will not be as
accurate as measured from spectral types and an HR diagram, but
their statistical distribution should be sufficient for our interest
here. If there is indeed an excess of X-ray sources with $\log L_{h,c}
\sim 30.0-30.4$ ergs s$^{-1}$ (option i in \S4.1), from the empirical
$L_x-M$ relation \citep{Preibisch05a}, we would expect to see an
excess of stars around a solar mass.  Figure~\ref{fig:IMF} shows the
IMF constructed from NIR estimated masses for unobscured COUP stars
and for NGC 2244 stars. No excess is apparent for stars in the solar
mass range, after the ONC IMF \citep{Muench02} is scaled to match the
NGC 2244 IMF. Instead, a deficit of intermediate mass stars around
$2-3$ M$_{\odot}$ in the X-ray selected sample is apparent (option ii
in \S4.1).

A second test is to obtain a statistical sample of cluster members and
construct a KLF \citep{Lada95} following the Cep B/OB3b KLF analysis
in \citet{Getman06}. We used 2MASS $K_s$ data to constuct the KLF
since it was spatially complete toward NGC 2244 and it is
photometrically complete to a similar mass limit as our {\em Chandra}
data for the Rosette. A control field from 2MASS centered at
($\alpha,\delta$) =($6^h30^m00^s$, $+3^{\circ}45^{\prime}00^{\prime
\prime}$) [J2000] is used for background population subtraction
(extinction toward this control field and in the foreground of NGC
2244 is low; see Li 2005 for its 2MASS color-color diagram). The
resulting KLF is similar to that derived by Li (2005) with a power law
slope $(d\log N(K_s)/ d K_s)\sim 0.3$. To get the IMF, we convert
$K_s$ magnitude to $\log M$ using the 2~Myr theoretical isochrone. The
relation between $K_s$ and $\log M$ from the theoretical isochrone
\citep{Siess00} can be approximated well with a power law as
demonstrated by \citet{Getman06}.  The resulting IMF derived from
$K$-band star counting (dot-dashed line in Figure~\ref{fig:IMF}) is
consistent with the IMF estimated from NIR properties of {\em Chandra}
sources (solid histogram in Figure~\ref{fig:IMF}). Note that the mass
derived using K magnitude as a proxy for photospheric luminosity can
be biased towards higher mass for K-excess sources.  However, our IMF
estimated from KLF is justified; the fraction of $K$-excess sources in
our sample is low and the IMF inferred from the KLF is coarsely binned
($\Delta\log (M/M_{\odot})\sim 0.2$) to account for the effect of
K-excess on the estimated mass.

This KLF suggests that our X-ray sample is largely complete down to
$\sim 0.5 M_{\odot}$, consistent with the mass limit estimated from
the X-ray completeness limit. It also suggests that our X-ray selected
sample is missing $\sim 20$ intermediate-mass stars around 3
M$_{\odot}$. It is not surprising that the detection efficiency of
intermediate-mass stars in the X-ray band is not as deep as in optical
or IR \citep{Schmitt85}. The X-ray production mechanism is not well
understood for stars in the intermediate-mass range
(\S~\ref{sec:HAeBes}). In many cases of X-ray detections of
intermediate-mass stars, the X-ray emission in fact comes from a low
mass companion to the intermediate mass star. Compared to the ONC,
some of the intermediate-mass stars in NGC 2244 may be X-ray quiet
because of the absence of low-mass companions. This is reflected as
the deficit of $\log L_{h,c} \sim 30.4-31.0$ ergs s$^{-1}$ sources in
the XLF. The dynamical evolution history of the young cluster might
result in a lower fraction of intermediate-mass stars with
secondaries.  Surveys of intermediate-mass member stars in NGC~2244 to
measure their multiplicity, such as the binarity study of Sco OB2
intermediate-mass stars by \citet{Kouwenhoven07}, are needed. Under
this assumption, the ``excess'' of $\log L_{h,c} \sim 30.0-30.4$ ergs
s$^{-1}$ sources would then be the result of a slightly larger
population of NGC 2244 stars compared to that of the ONC.

Based on all the above analysis, we conclude that the unobscured X-ray
emitting NGC 2244 population is about 1.2 times larger than the known
unobscured population in ONC, or $\sim$1000 stars with $\log L_t >
27.0$ ergs s$^{-1}$ (the COUP detection limit). The obscured
population, estimated from similar XLF scaling, is $\sim$ 500
stars. Given that the total COUP sample accounts for $\sim 75\%$ of the
ONIR sample of the ONC in \citet{Hillenbrand98}, we estimate that the
total size of the NGC 2244 stellar population is around
2000. \citet{Li05} gives a census of $\sim$1900 NGC 2244 members
estimated from the spatially complete 2MASS analysis.

Although the massive end of the stellar complement in NGC 2244 was
well-known from early studies \citep[e.g.,][]{OI81}, the low-mass
populations were poorly identified. \citet{Perez91} suggested that
there may not be $M<4$ M$_{\odot}$ stars in the NGC 2244
cluster. Perez (1991), Massey et al.\ (1995), and Park \& Sung (2002)
investigated the IMF for NGC 2244 in optical, and reported a top heavy
IMF with a flat power law slope $\Gamma$ $(d\log N(\log M)/d \log
M)\sim -0.7$, although Park \& Sung (2002) are cautious due to the
incompleteness of their intermediate- and low-mass
population. Indeed, their optical sample becomes incomplete for stars
with masses lower than $\sim 3M_{\odot}$. In comparison to the
traditional IMF studies, our high sensitivity X-ray sample of stars
largely benefits from our robust membership criteria and our ability
to identify the low-mass members in a reliable manner. The IMF slope
obtained from the X-ray selected membership, which is nearly complete to
0.5$M_{\odot}$, gives $\Gamma \sim -1.1$ instead of $\Gamma\sim -0.7$,
consistent with the Orion IMF.

\subsection{Spatial Structure of the Stellar Cluster}
\label{sec:spatial.sec}

The morphology of young clusters, including dependency on stellar
mass, provides clues for cluster formation and dynamical evolution.
As morphological studies based on optical or infrared samples are
complicated by patchy extinction, nebular contamination, confusion
with field stars, and bias towards stars retaining protoplanetary
disks, the spatial distributions of X-ray identified stars in populous
young clusters should be excellent laboratories to explore their
origins and dynamical evolution.  For example, an aspherical shape or
clumpy distribution would reflect unequilibrated initial conditions
while a spherical shape with mass segregation would indicate
well-developed virialization \citep{Clarke00}.  In the Orion A cloud,
the ONC, NGC~2024, and associated molecular filaments have flattened
shapes \citep{Lada91,Feigelson05} which have been attributed to global
gravitational collapse of an elongated cloud \citep{Hartmann07}.  In
contrast, the rich NGC~6357 and M~17 clusters appear spherical but
with subclusters that may reflect distinct (perhaps triggered)
subcluster formation \citep{Wang07,Broos07}. The absence of mass
segregation can either reflect a young stellar system that has not yet
achieved dynamical relaxation, or a mature system where many of its
massive members have been ejected by few-body interactions in the core
\citep{Pflamm06}.

It was recognized in the 2MASS study by Li (2005) that the apparent
center of the large-scale annulus (see Figure~1b) defining the optical
Rosette Nebula at ($\alpha,\delta$) = ($6^h31^m56^s$,
$+04^\circ59^\prime56^{\prime\prime}$) \citep{OI81} is offset from the
center of the IR surface density distribution at ($\alpha,\delta$) =
($6^h31^m59.9^s,+04^\circ55^\prime36^{\prime\prime}$). He interprets
this offset as a projection effect where the Rosette Nebula resembles
a tilted cylindrical cavity in the molecular cloud \citep[see models
in][]{Celnik86}. The distribution of the Class II sources in a recent
{\em Spitzer}/IRAC-MIPS survey of NGC 2244 \citep{Balog07} shows good
agreement with the 2MASS data. Here our {\em Chandra} data show that
the stellar concentration around HD 46150 is offset from the larger
stellar distribution, irrespective of its relation to interstellar
matter. The off-center massive star HD 46223 will be further discussed
in \S~\ref{sec:mass_seg}.

A `center' must be defined for structural analysis.  In NGC~2244, the
highest concentration of X-ray stars appears around the second most
massive star, HD~46150 (O5V), at ($\alpha,\delta$) =
($6^h31^m55^s,+04^\circ56^\prime34^{\prime\prime}$). We therefore
treat HD~46150 to be the center for our radial profile analysis of
this high density region in \S~\ref{sec:radial_pro}, although we are
aware of the asymmetrical distribution of X-ray stars as noted above:
HD~46150 lies about 2\arcmin\/ northwest of the IR center defined in
the 2MASS study (Li 2005).

\subsubsection{Morphology and Substructures}

\begin{figure}
\centering
\plotone{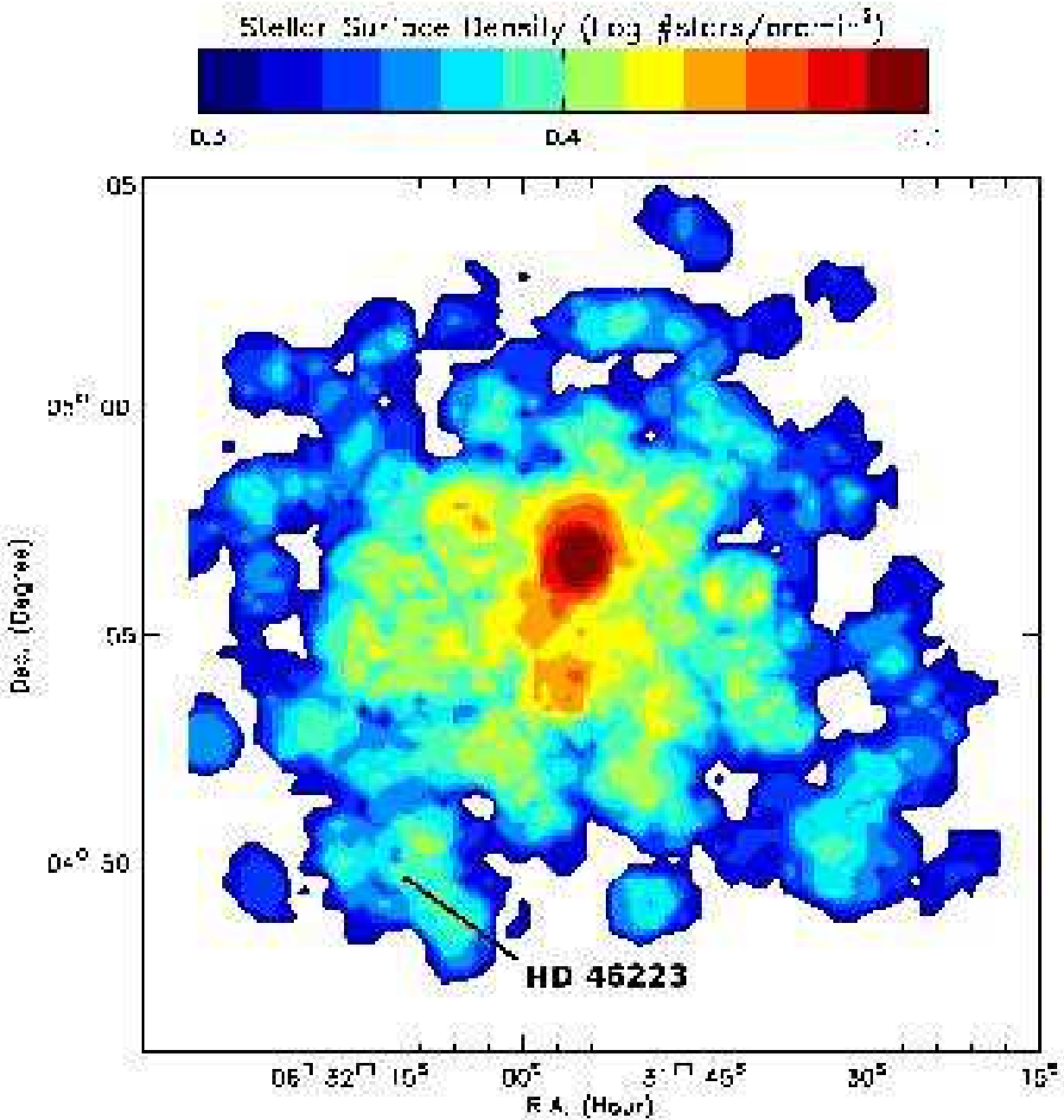}
\caption{The stellar surface density map for the unobscured stellar
population in NGC 2244. The cluster shows a spherical structure that
extends 8 arcmin in diameter. The highest concentration of stars is
around R.A.=06$^h$31$^m$55$^s$,
Dec.=$+04^{\circ}56^{\prime}34^{\prime\prime}$, the location of HD
46150. A secondary density enhancement is seen centered at
R.A.=06$^h$31$^m$56$^s$, Dec.=04$^{\circ}$54$^{\prime}$10$^{\prime
\prime}$. The O4 star HD 46223 is mostly isolated.
\label{fig:ssd}}
\end{figure}

Figure~\ref{fig:ssd} shows a smoothed map of the stellar surface
density for 572 lightly-obscured (median photon energy $\la 2$keV)
NGC~2244 X-ray stars. This map of smoothed spatial distribution is
constructed following \citet{Wang07} and \citet{Broos07}. A similar
smoothing technique has been applied to 2MASS data to identify
large-scale structure of the entire Rosette Complex
\citep{Li05,Li05a,Li05b}. A $\sim 20^\prime \times 20^\prime$ grid is
created to cover the stellar positions, and at each position the total
number of sources within a 0.5 arcmin radius sampling kernel is
counted to estimate the smoothed stellar density. Only sources covered
by both ObsID 1874 and ObsID 3750 are considered to guarantee roughly
equal X-ray sensitivity throughout the field. The heavily-obscured
sources are omitted because many of them are expected to be background
AGNs.

The cluster shows an approximately spherical structure that extends
8\arcmin\/ (3.2 pc) in diameter, centered at ($\alpha,\delta$)=($06^h
31^m 59^s, 04^\circ 55^\prime 30^{\prime\prime}$). This center, as
well as the large scale structure and substructure seen in the
X-ray-sampled cluster (Figure~\ref{fig:ssd}), are in good agreement
with the results derived from the surface density of $K$-excess stars
in the FLAMINGOS study \citep{Roman07}. It also matches the center
defined from the 2MASS star-count \citep{Li05} and the center defined
from the Spitzer Class II sources \citep{Balog07}. The large-scale
north-south asymmetry can be attributed to the off-center placement of
HD~46150.  The primary concentration is seen around HD~46150. Five of
these stars were noted by Sharpless (1954) as a visual compact
subcluster, but we find $\sim$50 stars extending to a radius of
1\arcmin\/ around this massive star.  The central stellar surface
density here is $\sim$700 stars per pc$^2$; recall that this value is
restricted to stars with masses above $\sim 0.5$~M$_\odot$ due to
X-ray sensitivity limits.

\begin{figure}
\centering
\plotone{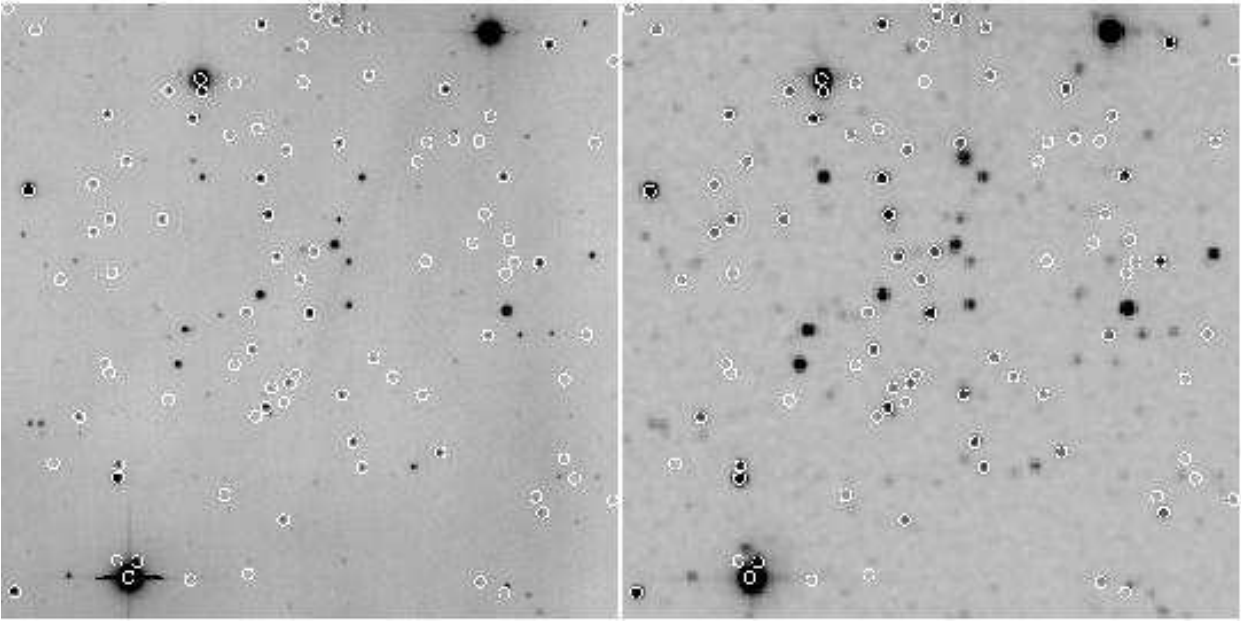}
\caption{The KPNO H$\alpha$ image \citep{Li04} and 2MASS $Ks$ image of
the secondary overdensity of stars seen in Figure~\ref{fig:ssd}. The
overlaid circles mark the ACIS sources. The 3.5\arcmin $\times$
3.5\arcmin\ images are both centered at
RA=06$^h$31$^m$56$^s$,Dec.=04$^{\circ}$54$^{\prime}$16$^{\prime\prime}$.
\label{fig:subcluster}}
\end{figure}

A secondary density enhancement of $\sim 15$ X-ray sources (about a
3$\sigma$ enhancement) is seen 3\arcmin\ south of HD~46150 at
($\alpha,\delta$)=($6^h31^m56^s$,+04$^\circ$54$^\prime
10^{\prime\prime}$). Different sampling scales are tested and this
substructure persists for smoothing kernels with radii
$<1^{\prime}$. The local density peak here has six stars tightly
clustered within 20$^{\prime \prime}$; it is also apparent in the
optical $H_{\alpha}$ image and the 2MASS-$Ks$ image
\citep[Figure~\ref{fig:subcluster} and][]{Li05}. Assuming they are
lightly-obscured late-type stars, the reddening indicated by their NIR
colors is $A_V\sim 1$ mag, similar to that of NGC~2244 cluster. Their
NIR estimated spectral types range from F to M if they are
located at the same distance as NGC~2244.

The existence of both substructures, around HD~46150 and 3\arcmin\ to
the south, is direct evidence that the NGC 2244 cluster has not
attained dynamical equilibrium. But perhaps most remarkable is the
absence of companions around the most massive cluster member, HD 46223
(O4V). Nine X-ray sources lie within 1\arcmin\/ of HD 46223 compared
to $\sim$50 around HD 46150. One possible explanation for the isolation of
HD 46223 is that it was ejected by dynamical interactions within the
HD 46150 subcluster. However, it does not exhibit high proper
motion (\S4.4) and it seems unlikely that such a massive member,
rather than less massive members, would be ejected at high
velocity. We note, however, that an O4 supergiant has been reported to
be probably ejected from Cyg OB2 (Comeron \& Pasquali 2007). Radial
velocity measurement of HD 46223 will be valuable to evaluate the
ejection scenario.

The spatial distribution of the NGC 2244 sources exhibits some common
characteristics and notable differences when compared to M17
\citep[see \S 3.1 in ][]{Broos07}. They both show the highest density
of stars close to massive O stars and the concentration is largely
spherical. However, the concentration in M17 is around the known early
O stars in NGC 6618, while in NGC 2244 the concentration is not around
its massive star with the earliest spectral type (HD 46223) but with
another O star, HD 46150. Both clusters show distinct substructures:
in M17 an obscured small cluster is found (M17-X), which is seen as an
elongation of the central cluster in M17; an unobscured substructure
is also found in NGC 2244. M17 shows a triggered stellar population
along the shock front and the eastern edge of the M17-SW molecular
core (south bar in Jiang et al. 2002). To the southeast of NGC 2244, a
sequence of triggered embedded clusters also exists along the midplane
of the RMC (Phelps \& Lada 1997). The different appearances of the
triggered populations are related to the geometric configuration of
the dense giant molecular clouds relative to the HII regions. The
``V''-shaped M17 is edge-on blister HII region emerging from the
surrounding molecular materials, while NGC 2244 is located in an
expanding HII bubble at the tip of the elongated RMC.

\subsubsection{Radial Density Profile}\label{sec:radial_pro}

\begin{figure}
\centering
\plotone{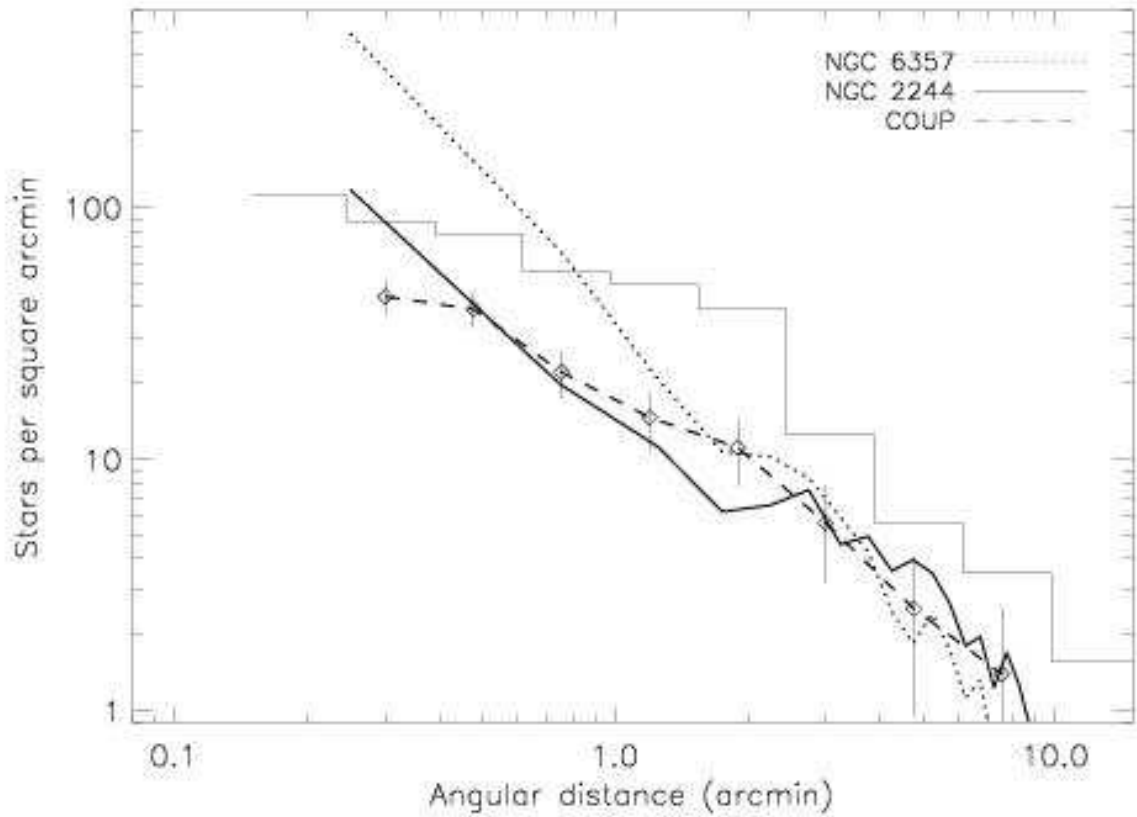}
\caption{The observed radial density profiles of the NGC 2244 cluster,
the ONC from COUP studies \citep{Feigelson05}, and the NGC 6357 region
from our {\em Chandra}/ACIS observation \citep{Wang07}. The histogram
shows the radial density profile of the ONC from ONIR studies
\citep{Hillenbrand98}. $1\sigma$ Poisson error is shown for the COUP
ONC. \label{fig:radial_a}}
\end{figure}

The radial density profile for the NGC~2244 cluster, centered at the
stellar density peak around HD~46150, is shown in
Figure~\ref{fig:radial_a} with comparison profiles from optical/NIR
and X-ray studies of the ONC \citep{Hillenbrand98, Feigelson05} and
from our X-ray study of Pismis~24 in NGC~6357 \citep{Wang07}. The
radial profile of NGC~2244 has two distinctive components: a powerlaw
structure around HD~46150 extending 1.5\arcmin\/, and a structure with
a flat core and steeper dropoff extending from 1.5\arcmin\/ to
8\arcmin. The power law structure is centered on, but is much more
extended than, the $\sim 20$\arcsec\/ X-ray resolved subcluster around
HD~46150 shown in Figure 3. NGC~6357, and perhaps the ONC, have a
similar radial profile with approximately the same powerlaw slope
(Figure~\ref{fig:radial_a}).

\begin{figure}
\centering
\plotone{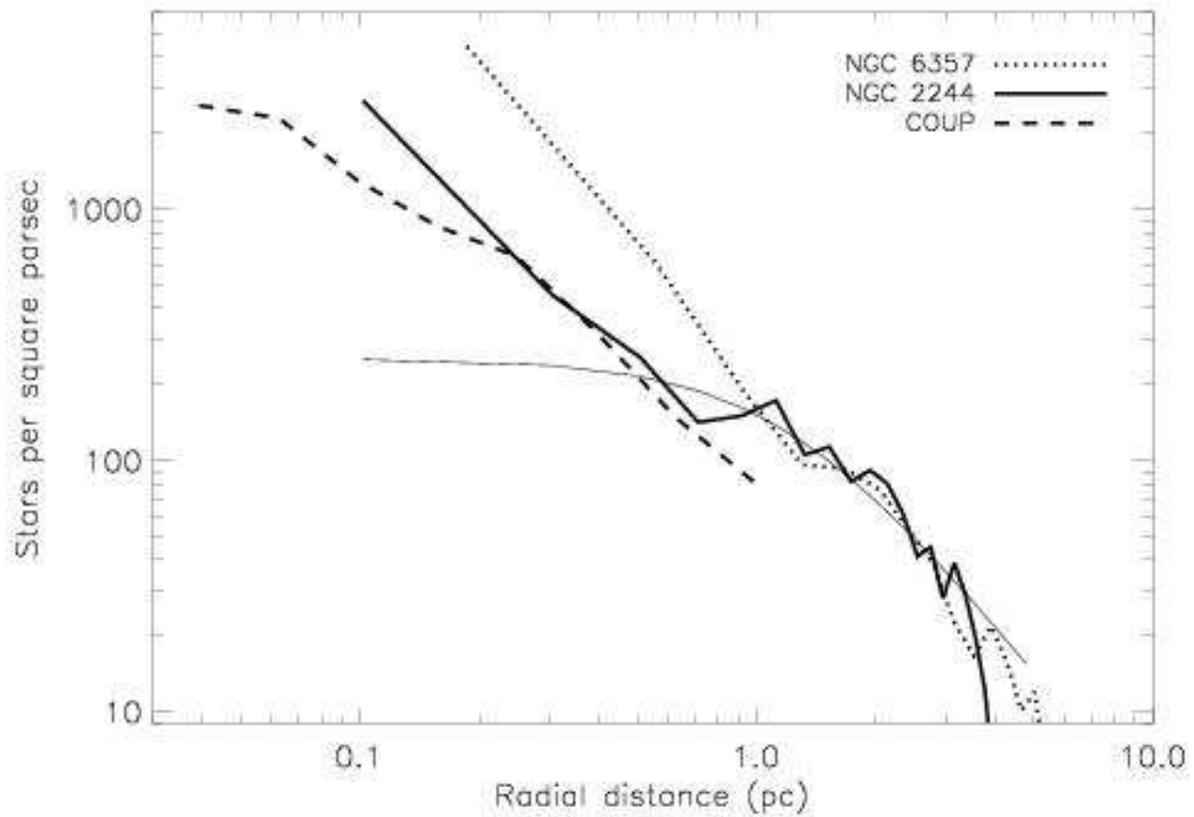}
\caption{The radial density profiles (thick lines) in physical scales
for the same three clusters, where the stellar densities of NGC 2244 and
NGC 6357 have been scaled to 1.2 times and 5 times the ONC population,
respectively, based on the XLF analysis. The thin line represents a
King model profile for the outer portion of NGC 2244. \label{fig:radial_b}}
\end{figure}

Figure~\ref{fig:radial_b} shows the inferred radial density profiles
in physical size units (parsecs) for the three clusters, where the
star densities are scaled to their estimated true densities based on
the comparison of the XLFs shown in Figure 9b.  The stellar density of NGC
2244 has been scaled to 1.2 times the ONC population
(\S~\ref{sec:xlf}), and NGC~6357 to 5 times the ONC population
\citep{Wang07}. Omitting the central $r^{-2}$ powerlaw structures, the
profiles of these two clusters can be fit as isothermal spheres with
core radii $r_c=1.2$pc and $r_c=1.4$ pc, respectively.

\subsection{Mass Segregation}\label{sec:mass_seg}

The concentration of massive cluster members at the center and lower
mass members at larger radii from the cluster center is commonly
observed in rich young star clusters
\citep[e.g.][]{Carpenter97,Hillenbrand98,Adams01}.
\citet{Schilbach06} investigate mass segregation in over 600 open
clusters with a wide range of ages.  For their youngest clusters
with ages $\sim$5 Myr, some show mass segregation whereas others do
not.

Mass segregation can occur as a natural consequence of dynamical
relaxation. Details of the process have been debated. For the ONC
Trapezium, some researchers argue that the dense collection of $\sim
10$ OB stars is an imprint of initial conditions
\citep{Binney87,Bonnell98}, while others argue that the core has
collapsed and many OB stars have been ejected
\citep{Pflamm06}. \citet{McMillan07} suggest a model of sequential
mergers of mass segregated subclusters.  \citet{Bonatto06} found that
the M~16 cluster, with age $\sim$1.3 Myr, has an overall relaxation
timescale around $\sim 20$ Myr, yet shows some degree of mass
segregation at this young age.

\begin{figure}
\centering
\includegraphics[width=0.7\textwidth]{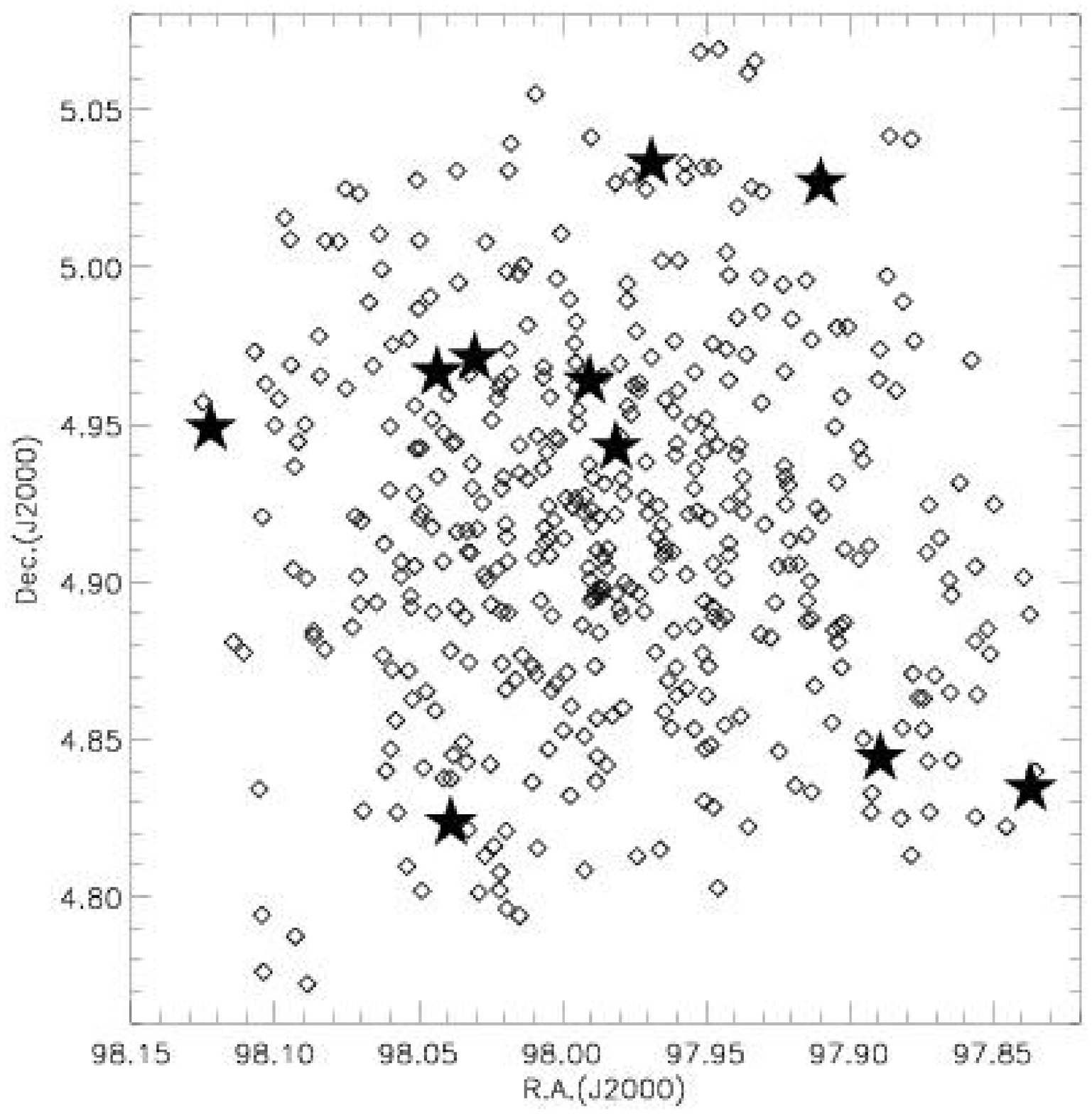}
\includegraphics[width=0.7\textwidth]{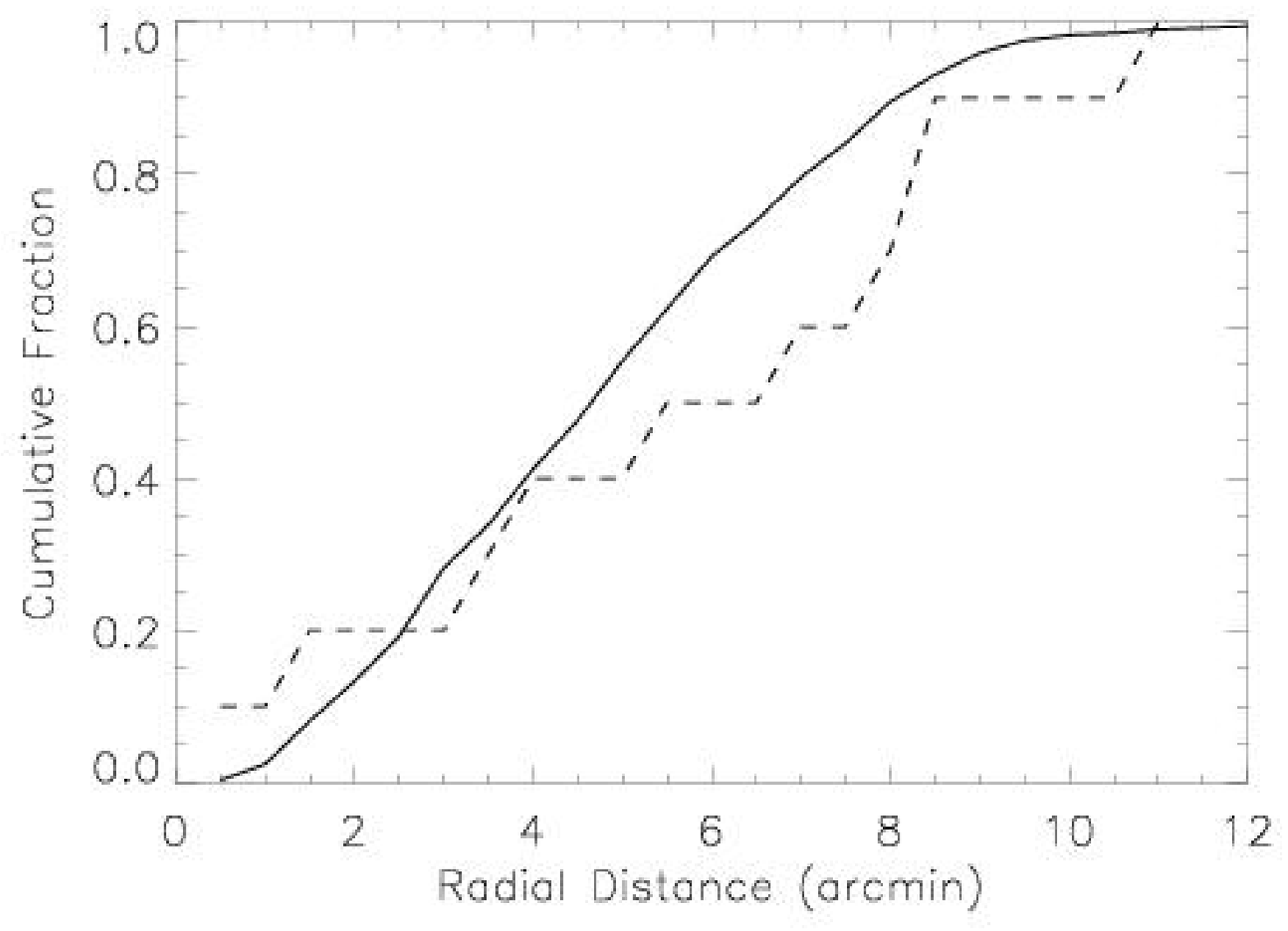}
\caption{(a): Spatial distribution of the massive stars (NIR estimated
mass $M\ga 8 \msun$, filled stars) and the low mass stars ($M\la 2
\msun$, open diamonds) in NGC 2244, using our X-ray-selected
sample. (b): The cumulative radial distributions for the massive stars
(NIR estimated mass $M\ga 8 \msun$, dashed line) and the low mass
stars ($M\la 2 \msun$, solid line).\label{fig:mass_seg}}
\end{figure}

In NGC~2244 in the Rosette Nebula, the O stars are not highly
concentrated, as shown in Figure \ref{fig:mass_seg}a (see also Figure
1a). The earliest O-type (O4V) star in the cluster, HD~46223, has a
rather puzzling location in the cluster near the southeast boundary of
the nebula.  Its proper motion is fairly small
\citep[$\mu_{\alpha}=-0.2$ mas yr$^{-1}$, $\mu_{\delta}=0.4$ mas
yr$^{-1}$;][]{Zacharias04} which does not suggest ejection from the
cluster center.  The issue of mass segregation has not been
extensively investigated for this 2~Myr old cluster in the literature,
mainly because the low-mass population was not adequately
identified. With the proper motions and membership probabilities of
stars in the NGC 2244 region derived from photographic plate data,
\citet{Chen07} suggest that the cluster shows evidence of mass
segregation.

Figure~\ref{fig:mass_seg}b shows the cumulative radial distributions
for the massive stars with NIR-estimated masses $M\ga 8 \msun$ and for
the X-ray identified low mass stars ($M\la 2 \msun$).  The
distributions appear very similar, and a Kolmogorov-Smirnov test of
the two distributions does not show significant difference.  Thus,
mass segregation is not present in NGC~2244 within our FOV.

We estimate a two-body dynamical relaxation time $t_{relax}$ for the
NGC 2244 cluster \citep[e.g.,][]{Bonatto06}: $t_{relax}\approx (N/8
\ln N)\times t_{cross}$ where $t_{cross}=2R/v_{disp}$ is the
characteristic crossing time for a star to travel through the cluster
with radius $R$ and velocity dispersion $v_{disp}$. Adopting $R\sim 4$
pc from the full cluster extent in Figure~\ref{fig:radial_b}, a rough
estimate for the unmeasured velocity dispersion $v_{disp}\sim 3$ km
s$^{-1}$ \citep{Binney87}, and $N\sim 1900$ stars (\S~\ref{sec:xlf}),
we obtain $t_{relax}\sim 30$ Myr for NGC~2244.  As the cluster age is
$<10$\% of this relaxation time, no significant mass segregation is
expected from two-body dynamical interactions.  If we consider only
stars within the estimated core radius $r_c=1.2$ pc, then
$t_{relax}\sim 9$ Myr which is still considerably larger than the age
of the cluster.

The absence of mass segregation is thus consistent with standard
dynamical theory, and implies that NGC 2244 (unlike some other
clusters) was not formed with a central concentration of massive
stars. The main challenge for explaining the dynamical state of NGC
2244 is the difference between the dominant member HD 46150, which has
a rich compact subcluster, and HD 46223, which is mostly isolated.

Similar to NGC 2244, \citet{Broos07} show that many other massive
stars are scattered all over the ACIS-I field in M17, in addition to
the concentration in NGC 6618. ACIS source \#51 is one of the most
massive stars in the field, yet it sits $>6$\arcmin from the center of
the cluster. For comparison, HD 46223 is about $\sim 7$\arcmin south
of HD 46150 and $\sim 5$\arcmin south of the cluster center in NGC
2244. Not all massive stars are participating in the mass segregation
seen in NGC 6618. One possible explanation is that the O stars are not
all co-eval. Indeed, several massive protostars have been found in M17
\citep[e.g.,][]{Nielbock01,Chini04,Chini05}. Together with the
presence of an UCHII region there, these young massive stars establish
that the massive populations are not all the same age in M17. Massive
stars to the east might be older and belong to a wider OB association,
while source \#51, for example, might be younger. This could also be
the case in NGC 2244.  The late-O stars are scattered as in M17. HD
46223 may be younger and not part of the same population as NGC 2244's
central cluster.

\subsection{X-ray Stars with Infrared Excess Disks}

X-ray selected samples have several advantages over optical and IR
samples (see review by Feigelson et al. 2007). X-ray emission arises
from stellar magnetic activity which is enhanced $10^1 - 10^4$ above
main-sequence levels for stars during the entire age range of interest
($<0.1$ to $>10$ Myr), thus X-ray surveys suffer only a small number
of field and extragalactic contaminants ($\sim8$\% in observations of
rich massive clusters), which are usually identifiable (\S3.2). They
naturally deliver a nearly disk-unbiased sample of young
stars\footnote{There may be additional complications: X-ray selected
PMS samples suffer a small bias against accreting stars in the $0.5-8$
keV band because Class II systems are on average $\sim 2$ times
fainter than Class III systems (Preibisch et al. 2005; Telleschi et
al. 2007), and there may also be a small bias toward accretion systems
in the soft $<1$ keV band due to emission at the accretion
shock. However accretion variations do not cause X-ray variations in
the {\em Chandra} band \citep{Stassun06}. These are minor effects
considering that the X-ray luminosity function spans $28 < \log L_x <
32$ ergs s$^{-1}$ and it is dominated by flare emission. See
discussion in Feigelson et al. (2007).}. The main disadvantage of
typical X-ray surveys is their incompleteness in detecting the lowest
mass objects, which can be identified in high-sensitivity IR
images. The complementary nature of the {\em Chandra} and IR data will
provide the best census to date for the young stellar population of
NGC~2244. It is worth noting that because the different disk
indicators in the IR trace thermal emission from circumstellar
materials of different temperatures, the inferred disk fraction of a
young cluster generally increases when measured in longer IR
wavelengths \citep[e.g.,][]{Lada04}. For example, \citet{Haisch01a}
show that $JHK$ observations alone are not sensitive enough to detect
circumstellar disks in a complete and unambiguous manner. They find that
21\% of the sources in the IC 348 cluster have $K$-band excess disks;
the disk fraction rises to 65\% when using $JHKL$ IR-excess
emission. In this subsection we focus on the X-ray selected sample of
stars with $K$-band excesses.

\begin{figure}
\centering
\includegraphics[width=0.8\textwidth]{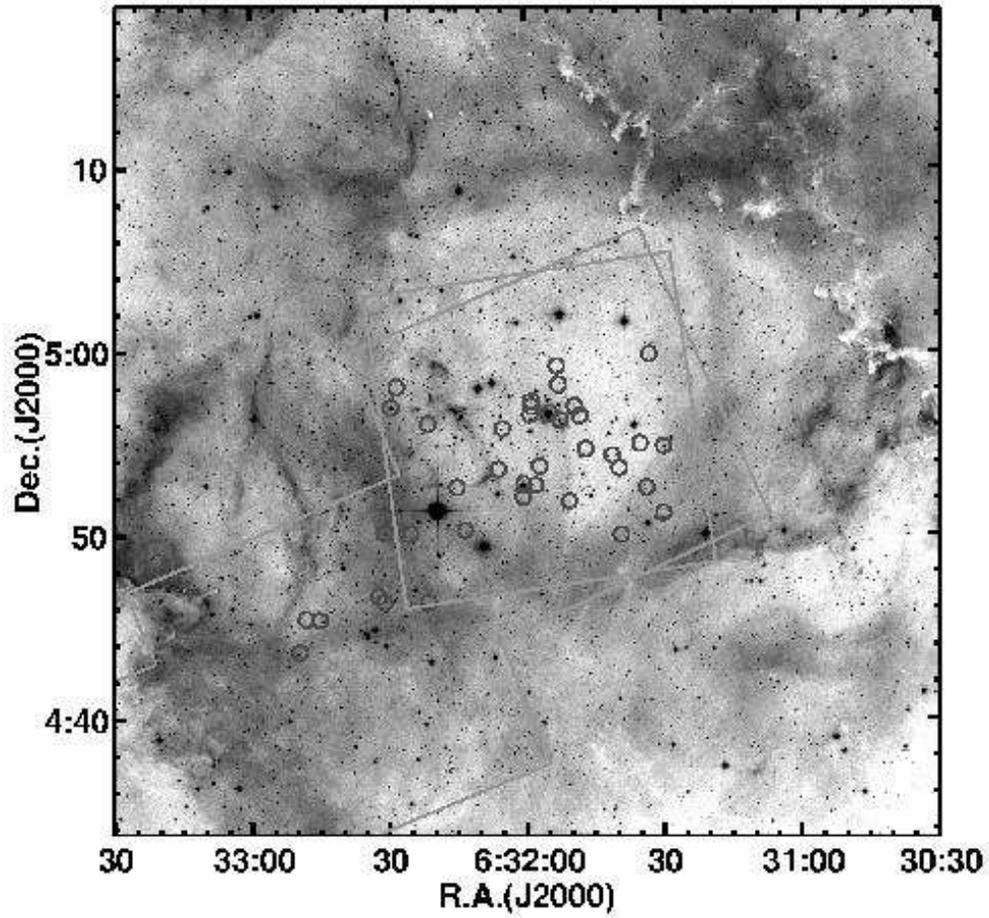}
\caption{The spatial distribution of 38 sources with significant NIR
color-excess (circles). The northern part of the nebula seems
deficient in NIR excess sources. The boxes outline the FOVs of the
multiple ObsIDs.\label{fig:excess}}
\end{figure}

The spatial distribution of 38 X-ray stars with $K$-band excesses
attributed to inner protoplanetary disks (\S~\ref{sec:IRcolor}) is
shown in Figure~\ref{fig:excess}.  Ten of these young stars cluster
around HD~46150 (O5V), and a few are around HD~259135 (B0.5V).

A deficit of color-excess stars in the northern part of the nebula is
seen: none of them are located in the northern part of the cluster
where HD~46149 (O8.5V) is located.  This region is also devoid of dust
emission in the IRAS image \citep{Cox90}.  The deficit of disky stars
cannot be attributed to photoevaporation by OB stellar UV radiation
and winds since a grouping of them is found around HD~46150.  We
consider two explanations for this asymmetry.  First, \citet{Li05}
suggested that the Rosette Nebula is open $30^{\circ}$ north of the
line of sight.  As gas and dust stream away from the HII region
\citep[as in M~17,][]{TFM03}, the stars in this region may have
undergone a faster inner disk dissipation so that their disks no
longer show K-band excess. Second, the star formation in NGC~2244 may
have proceeded over a considerable time span along the north-south
direction, with the older population in the northern region.  Similar
spatial-age patterns have been found in other young clusters
\citep[e.g., Cep OB3b,][]{Burningham05}.

For the $\sim$2~Myr old NGC 2244 cluster, using our X-ray selected
sample with high $JHK$ photometric quality (largely complete to 0.5
$M_{\odot}$), we derive an overall disk frequency of $\sim$6\% for
{\em Chandra} stars with mass $M \ga 0.5M_{\odot}$ (assuming a
presence of $\sim$20 foreground field stars). The disk fraction is
$\sim$10\% for stars with mass $M \ga 2.0M_{\odot}$ and $\sim$5\% for
stars with mass $0.5M_{\odot} \la M \la 2.0M_{\odot}$. Using a large
{\em Chandra} sample similar to that studied here, \citet{Wang07}
reported a low fraction of $K_s$-band excess among intermediate-mass
stars in the young massive star forming region NGC 6357 ($\sim$1 Myr
old) and a similar result is reported for M~17 ($\sim$1 Myr old) in
\citet{Broos07}. These are consistent with the findings that optically
thick circumstellar disks are already rare among the intermediate-mass
PMS stars with ages less than a few Myr and suggest that the disks are
short lived for the massive stars
\citep[e.g.,][]{Hillenbrand93,Natta00}. It has been suggested that the
disks around earlier type stars may evolve faster than around later
type stars based on studies of the IR-excess fraction as a function of
spectral type in a few clusters \citep{Lada00,Haisch01b,Lada06}.

\citet{Li05} used the 2MASS NIR sample after background population
subtraction to derive a $K_s$-excess disk fraction of $\sim 20.5$\%
above mass $\sim 0.8 M_{\odot}$ for NGC 2244. The discrepancy between
our results reflects the different criteria in selecting excess
sources: their $K_s$-excess sources included 2MASS sources with large
photometric error. With high quality FLAMINGOS $JHK$ photometry data,
\citet{Roman07} derive a lower $K$-excess fraction of $\sim 10\%$ for
$K<15.75$ stars, although it remains slightly higher than the
IR-excess fraction in our X-ray-selected sample. Their sample covers a
larger mass range, probably down to 0.1-0.2 $M_{\odot}$ depending on
extinction \citep{Roman07}.

Balog et al.(2007) present a {\em Spitzer} survey of NGC 2244
covering 3.6~$\mu$m to 24~$\mu$m and estimate that the overall disk
fraction in the cluster is 44.5\% using the IRAC data (and MIPS data
if available). The total cluster population in their field of view
(0.5 square degree), including the cluster members without disks, is
estimated by subtracting the average background population, whereas
our sample identifies the diskless stars from their elevated X-ray
emission. But the discrepancy between our results is mainly due to the
fact that {\em Spitzer} mid-IR observations are much more sensitive to
circumstellar dust than $K$-band excesses. The overall IR-excess
fraction among NGC 2244 {\em Chandra}-selected stars will be studied
with a deep {\em Spitzer} survey (PI: Bouwman) and compared to the IR-only
determined excess fraction, aiming to evaluate the different samples
of young stars selected through IR colors and those identified in
X-rays. A similar study has already taken place in the Serpens cloud
core \citep{Winston07}, which analyzed combined {\em Spitzer} and {\em
Chandra} observations of the embedded stellar cluster.

The overall low $K$-excess disk frequency seen here ($\sim 6\%$ for
{\em Chandra} stars with mass $M \ga 0.5M_{\odot}$; $\sim 10\%$ for
FLAMINGOS stars with mass $M \ga 0.1M_{\odot}$), at a cluster age of
2~Myr, may imply a faster disk dissipation while the cluster is
immersed in the hostile environment of UV radiation and strong stellar
winds of many massive stars. In clusters of similar age but without
the presence of O stars, the $K$-band excess fractions appear
higher. For example, in the 2.3 Myr old IC 348 cluster \citet{Lada95}
find that 20\% of the stars ($M\ga 0.08M_{\odot}$, estimated from the
$K$-magnitude completeness limit) have $K$-excess disks. The recent
{\em Spitzer} observations of IC 348 find a total disk frequency of
$\sim50$\% among stars with mass $M\ga 0.1M_{\odot}$ based on infrared
excess between 3.6 and 8.0~$\mu$m \citep{Lada06,Muench07}, not
significantly higher than the $\sim45$\% disk fraction in NGC 2244
reported in \citet{Balog07}.  This implies that although the inner
disks around the NGC 2244 stars may dissipate faster, there seems no
difference in the overall disk dissipation. 

However, \citet{Balog07} did notice a lower disk fraction (27\%) among
stars close to the NGC 2244 O stars (separation $d<0.5$~pc), hinting a
faster disk evolution near the massive stars. A number of
observational and theoretical studies have already demonstrated the
photoevaporation of disks by external radiation
\citep[e.g.,][]{Odell96,Johnstone98,Hollenbach00,Throop05}. \citet{Balog06}
presented 24$\mu$m images of three protoplanetary disks being
photoevaporated around high mass O stars, including one disk close to
the O5 star HD 46150 in NGC 2244 with an estimated mass loss rate
$10^{-10}-10^{-8}\msun$ yr$^{-1}$. We do not detect X-ray emission
from the IR point source in this cometary structure, likely due to its
low mass. Based on the {\em Spitzer} identified Class II and Class I
sources, it is further suggested that the effect of massive stars on
the circumstellar disks is significant in the immediate vicinity of
the hot stars \citep{Balog07}. \citet{Guarcello07} reported evidence
that the spatial distribution of the stars with a circumstellar disk
in NGC 6611 is anti-correlated with the distribution of OB
stars. Their findings also suggest that UV radiation from OB stars
does have an impact on the evolution of the disks close to massive
stars.

\section{X-rays across the Mass Spectrum}
\label{ximf.sec}

\subsection{X-rays from Massive Stars}

One of the important discoveries of early {\em Einstein} observations
was the soft X-ray emission from individual early type O stars
\citep{Harnden79}. Most O-type stars were found to be soft X-ray
emitters ($kT<1$ keV) with X-ray luminosities $L_x \sim
10^{31}-10^{33}$ erg s$^{-1}$.  A canonical relation between X-ray
luminosity and bolometric luminosity of $L_x/L_{bol} \sim 10^{-7}$ was
proposed and confirmed from {\em Einstein} and {\em ROSAT}
observations \citep{Pallavicini81,Chlebowski89}. \citet{Berghoefer97}
extended the same relation down to stars of later spectral type
(B1--B1.5). Recent {\em Chandra} studies of O7--B3 stars in Orion
(COUP) found both a soft wind-emission component and a hard flaring
component in many OB stars, and a larger dispersion was found for late
O and early B stars \citep[$\log (L_x/L_{bol})\sim$ $-4$ to
$-8$;][]{Stelzer05}. However, when only considering X-ray emission in
the 0.5--2.5 keV band, \citet{Sana06} derived a tight scaling law
$\log (L_x/L_{bol})=-6.91\pm 0.15$ for O-type stars with a deep {\em
XMM-Newton} observation of NGC 6231.

Wind-shock models were developed to explain the X-ray emission from
massive stars, where small-scale instabilities in radiatively-driven
stellar winds from massive stars produce shocks
\citep{Lucy80,Owocki88,Owocki99}. To account for the observed X-ray
emission line profiles and hard, variable continuum emission
\citep[e.g.,][]{Corcoran94,Evans04,Waldron04,Stelzer05}, more complex
models were invoked such as the magnetically channeled wind shock
(MCWS) model \citep{Babel97a,Babel97b,ud-Doula02}. \citet{Gagne05}
shows that the MCWS model with strong line-driven winds can adequately
reproduce both the soft and the hard components in {\it Chandra}
gratings spectra of $\theta ^1$ Ori C (O6V). In some cases, the
anomalously hard and luminous X-ray component ($kT>10$ keV and
$L_h\sim 10^{33}$ ergs s$^{-1}$) implies close binarity, as powerful
winds in two massive components collide and shock to produce very high
energy X-rays
\citep[e.g.,][]{Pollock05,Skinner06,Broos07}. \citet{Schulz06}
observed a large X-ray outburst in $\theta^2$ Ori A, which can be
attributed to reconnection events from magnetic interactions between
the binary stars.

\begin{figure}
\centering
\includegraphics[width=0.36\textwidth,angle=-90]{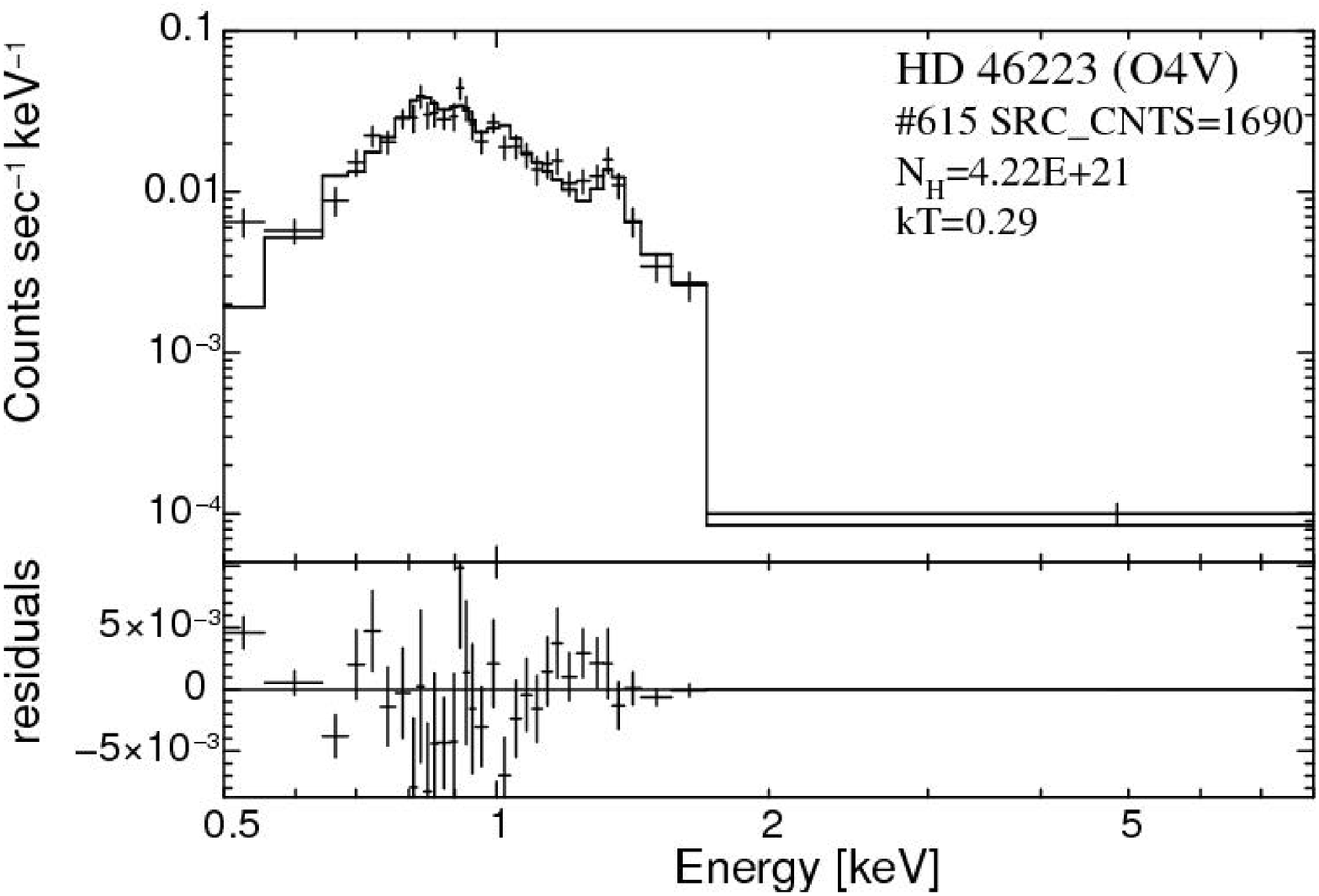}
\includegraphics[width=0.36\textwidth,angle=-90]{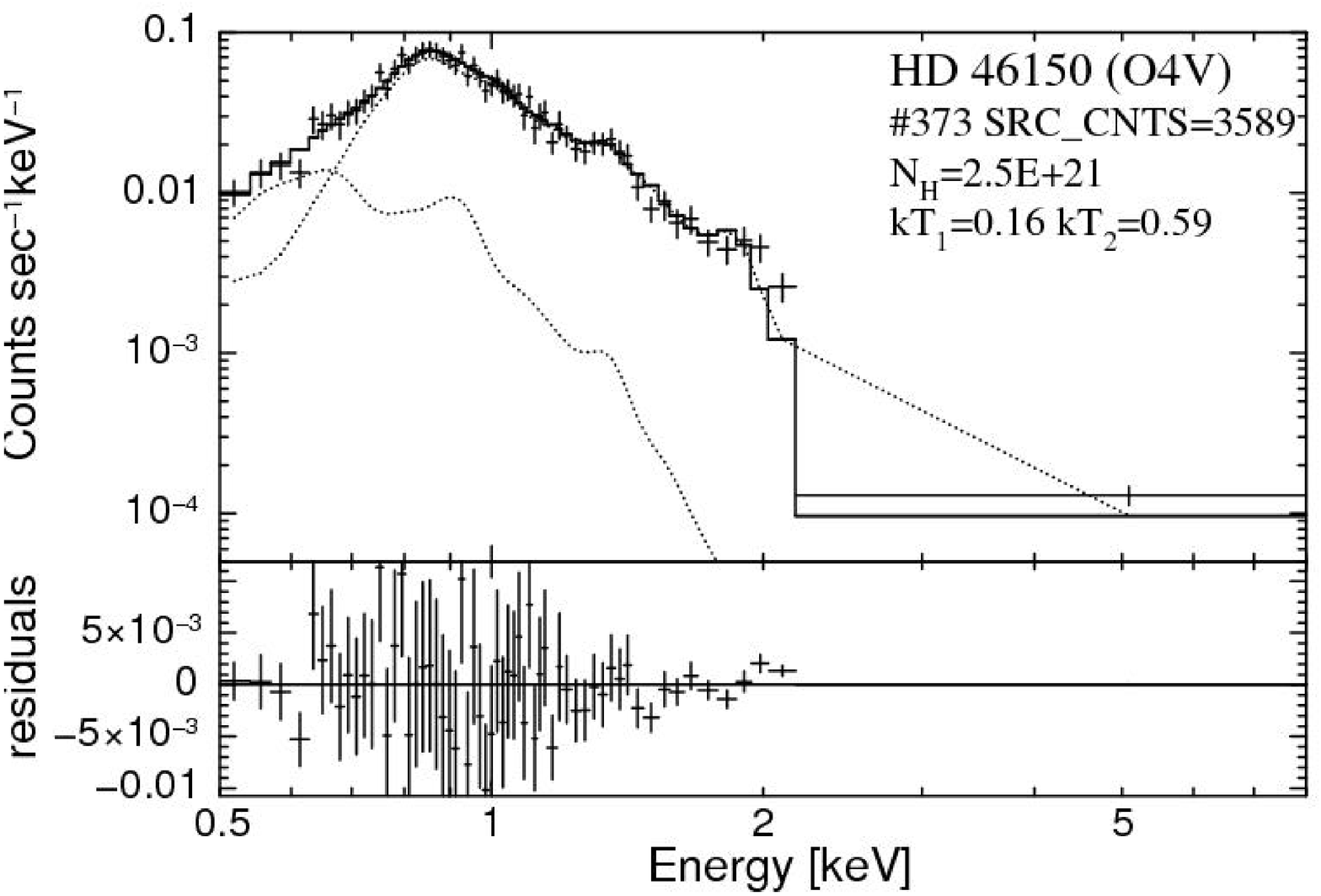}
\includegraphics[width=0.36\textwidth,angle=-90]{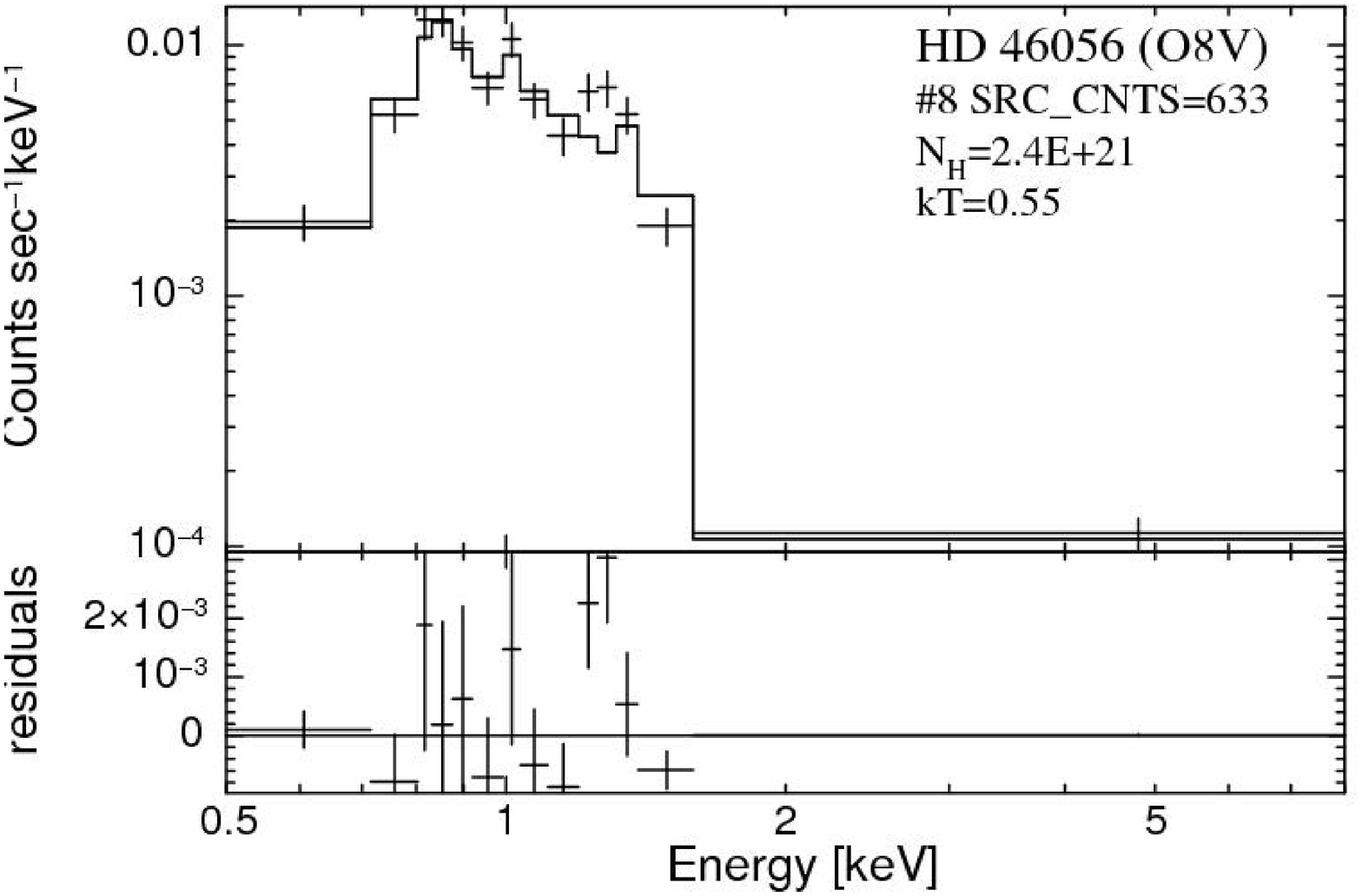}
\includegraphics[width=0.36\textwidth,angle=-90]{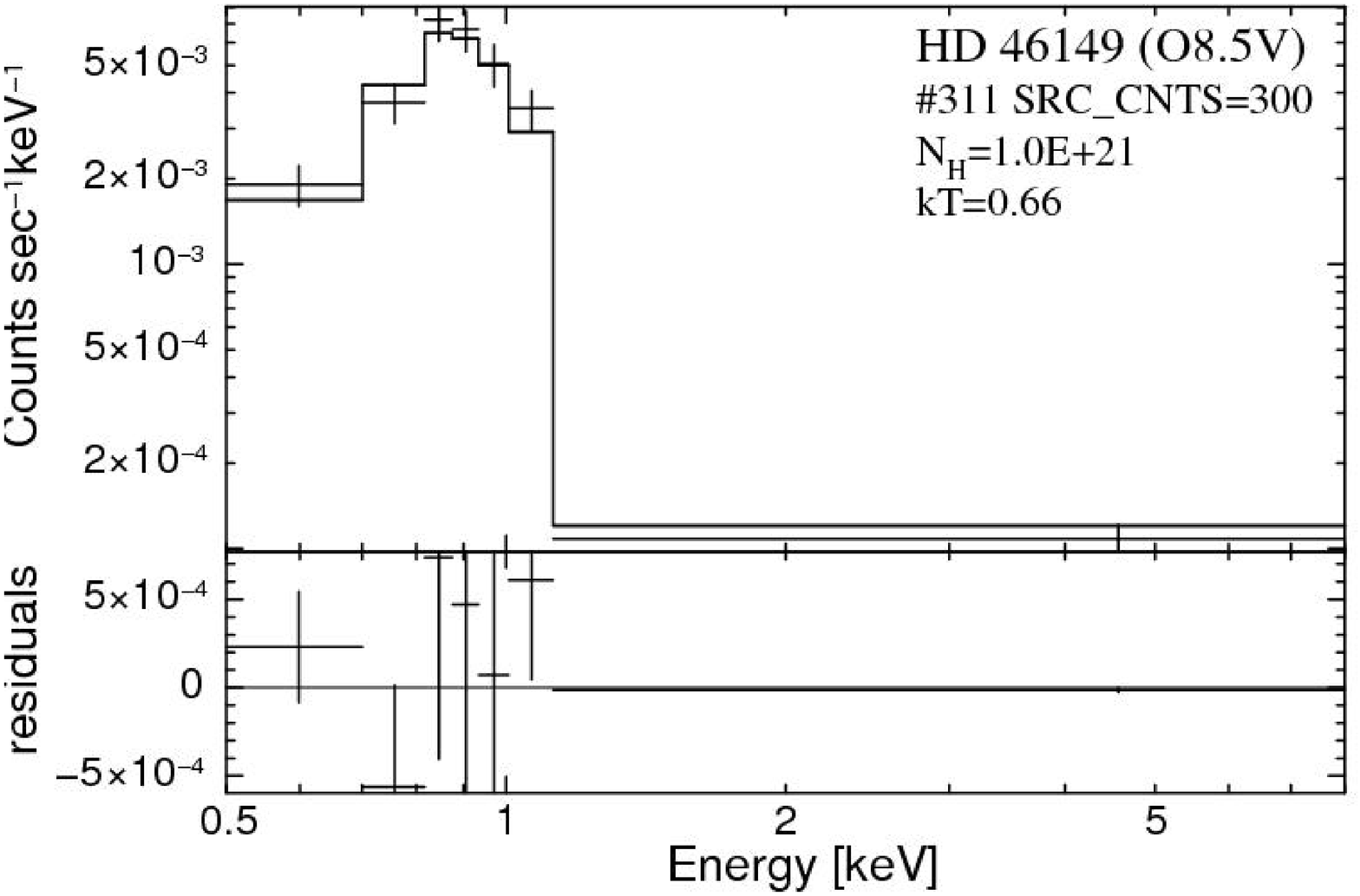}
\includegraphics[width=0.36\textwidth,angle=-90]{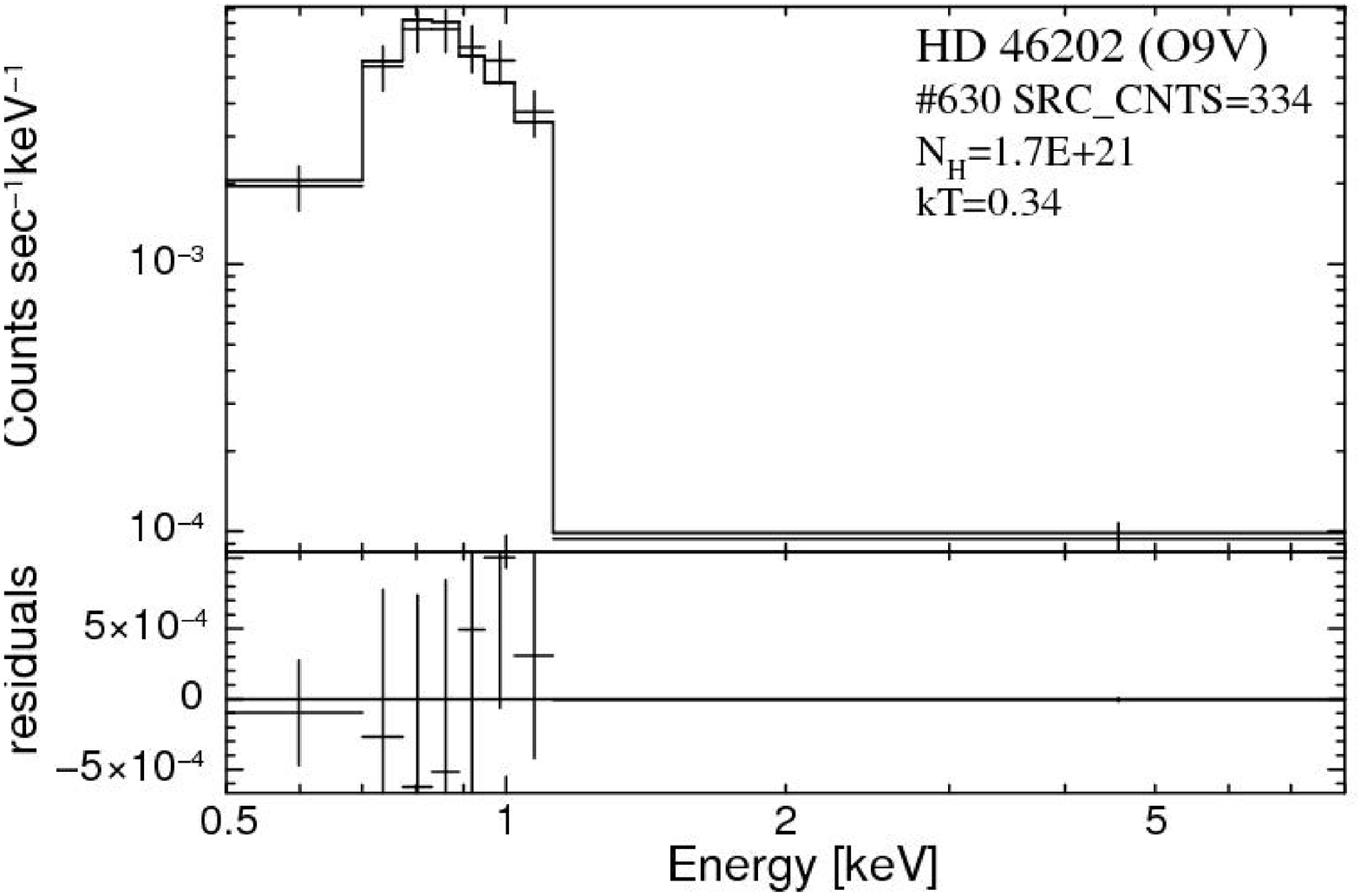}
\includegraphics[width=0.36\textwidth,angle=-90]{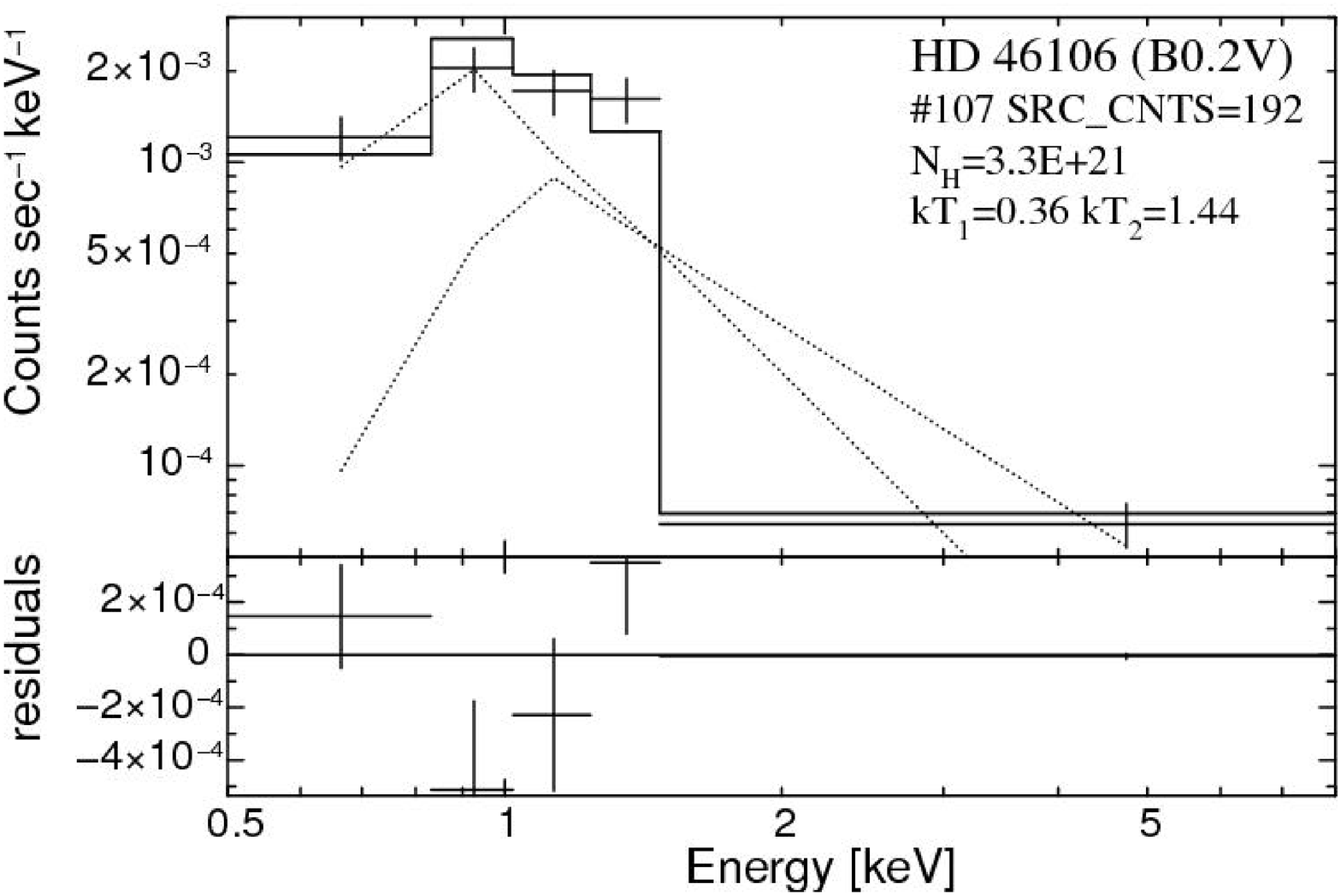}
\caption{Spectral fits to X-ray spectra of six O and early B
stars. Source name, source counts, and fit parameters are marked in
each panel. The two model components are shown as the dotted lines for
spectra that are best fit with two-temperature thermal plasma
models.\label{fig:OBs}}
\end{figure}

Due to its richness in population, NGC 2244 offers an excellent
opportunity to study X-ray emission in OB
stars. Table~\ref{tbl:OB.tab} summarizes the detection/non-detection
of O and early B-type stars in our observation, along with their
optical, IR, and X-ray properties. {\em Chandra} spectra are shown in
Figure~\ref{fig:OBs}. We detected all 9 OB stars with spectral types
B0.5 or earlier that were in the field. The two early O stars in NGC
2244 exhibit soft ($kT<1$ keV) and strong ($L_x \sim 10^{32}$ erg
s$^{-1}$) X-ray emission as expected in the classical wind-microshock
regime. All of the 2.3 Myr old NGC 2244 O stars show soft X-ray
emission ($kT <1$ keV), which is similar to the soft thermal spectra
seen in most O-type stars in other star forming regions (e.g., Rauw et
al.\ 2002, Rho et al.\ 2004, Skinner et al.\ 2005, Sana et al.\ 2006,
Albacete-Colombo et al.\ 2007). Nevertheless unusual cases of O stars
characterized by a hard spectrum have been reported in a number of
observations (e.g., Gagn{\'e} et al. 2005, Broos et al. 2007,
Tsujimoto et al. 2007). Based on analysis of X-ray grating spectra of
the Orion Trapezium stars, Schulz et al. (2003) proposed that the
presence of hard X-ray emission indicative of hot coronal plasma
($kT>1$ keV) indicates the presence of coronal magnetic fields. Linsky
et al. (2007) compared the X-ray properties of young massive stars
with known magnetic fields (e.g., $theta^1$ Ori C, O6V, $t\sim 0.3$
Myr; M16 ES1) with older stars with weak or no magnetic fields ($\tau$
CMa, O9.5I, $t<12$ Myr; $\zeta$ Ori, O9I, $t\sim 3-5$ Myr). The very
young stars with magnetic fields show high coronal plasma temperatures
($kT\sim 2-5$ keV), which may be heated as in the MCWS model
(Gagn{\'e} et al. 2005), and the total X-ray luminosities are far
larger than the expected X-ray luminosities from micro-shocks in the
stellar winds. The absence of such a hot component in the X-ray
spectra of the older massive stars (including our NGC 2244 O stars)
might imply a timescale of the presence of coronal magnetic fields in
the massive stars; the hot component may be restricted to stars much
younger than 2~Myrs. A detailed analysis of X-ray properties of a
large sample of O stars will be required to link the magnetic fields
in massive stars to the cluster ages.

We detected 6 out of 14 B stars with spectral types B1--B3. This low
X-ray detection rate among B stars is consistent with other recent
{\em Chandra} and {\em XMM-Newton} observations of massive star
forming regions \citep{Wang07,Broos07,Sana06}. The X-ray emission from
early B stars is consistently harder than that from the O stars
(higher $kT$ in Table~\ref{tbl:OB.tab}), which suggests that unseen
late-type companion stars rather than the B star itself is responsible
\citep{Stelzer05}.

The earliest exciting star in this complex, HD 46223, is of spectral
type O4V \citep{Walborn02}. Its X-ray spectrum is adequately fit by a
soft $kT=0.3$ keV single temperature plasma subjected to $N_H=4\times
10^{21}$ cm$^{-2}$ absorption (Figure~\ref{fig:OBs}). HD 46150 (O5V)
is the visually brightest early type star in the cluster. A two
temperature plasma model fit ($kT_1=0.2$ keV, $kT_2=0.6$ keV) is
needed to describe the X-ray spectrum, with an absorption column
$N_H=2.5\times 10^{21}$ cm$^{-2}$. Their X-ray luminosities are
similar, $L_{t,c}\sim 2.5\times 10^{32}$ ergs s$^{-1}$. The light
curves of the O-stars are examined and no variability is suggested by
the K-S statistics. In contrast, both O4V stars in M17 show a very
hard plasma component ($kT > 10$ keV) and are nearly an order of
magnitude brighter in their intrinsic full-band X-ray emission
\citep{Broos07}.  If this hard emission is caused by close binarity or
fossil magnetic fields, as suggested by \citet{Broos07}, then it is
likely that the early O stars in NGC 2244 lack at least one of these
features; their soft X-ray emission suggests that these are single O
stars without close, massive companions and/or that they do not
possess strong magnetic fields.

\begin{figure}
\centering
\includegraphics[width=0.6\textwidth,angle=90]{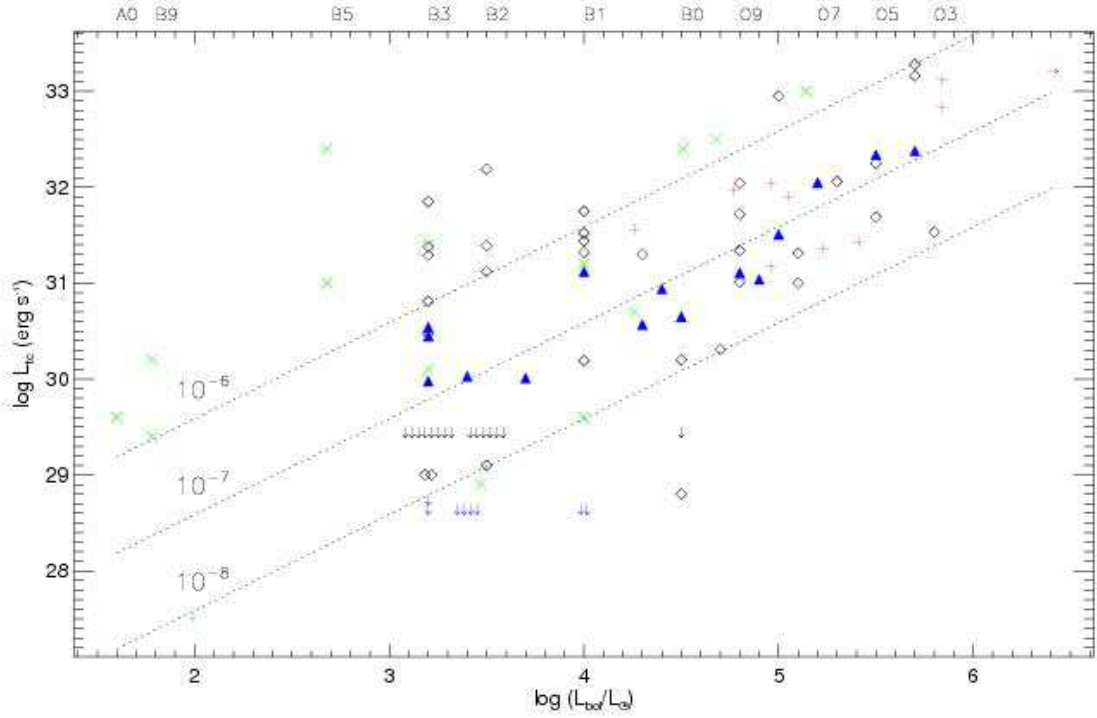}
\caption{The $L_x$ vs. $L_{bol}$ relation for X-ray detected O and early B
stars. The samples are from the NGC 2244 cluster (filled triangles;
this work, Table~\ref{tbl:OB.tab}), the ONC
\citep[crosses,][]{Stelzer05}, and the massive star forming regions
M17 \citep[diamonds,][]{Broos07} and NGC 6357
\citep[pluses,][]{Wang07}. Upper limits are marked as arrows. The
bolometric luminosities are adopted from \citet{Broos07} to be
consistent. \label{fig:lxlbol}}

\end{figure}

\citet{BC02} reported the X-ray and optical luminosities of NGC 2244
early type stars from {\em ROSAT} PSPC and HRI observations and $BVI$
photometry, and concluded that they are consistent with the canonical
relation $L_x/L_{bol} \sim 10^{-7}$. The $L_x/L_{bol}$ relation for OB
stars in NGC 2244 determined from our {\em Chandra} data is shown in Figure~\ref{fig:lxlbol}. Statistical
tests for the correlation between $L_x$ and $L_{bol}$ were performed
using the ASURV survival analysis package \citep{Isobe86}. To take
into account the available upper limits, the generalized Kendall's tau
correlation test for censored data is adopted. The null hypothesis (a
correlation is not present) probability is $P<0.01\%$, supporting a
significant $L_x-L_{bol}$ correlation. As shown in
Figure~\ref{fig:lxlbol} as well as Table~\ref{tbl:OB.tab}, the NGC
2244 O stars closely follow the X-ray to bolometric luminosities
$L_x/L_{bol} \sim 10^{-7}$ ratio, although the B spectral type stars
show larger (yet still $<0.5$~dex) scatter. For comparison, previously
reported $L_x$ vs. $L_{bol}$ values for additional OB stars from the
massive star forming regions Orion \citep[1 Myr;][]{Stelzer05}, NGC
6357 \citep[1 Myr;][]{Wang07}, and M17 \citep[1 Myr;][]{Broos07} are
also shown in Figure~\ref{fig:lxlbol}. The overall scatter in
$L_x/L_{bol}$ is considerably smaller for O stars from different
clusters ($\sim$2 orders of magnitude) than for B stars ($\sim$4
orders of magnitude).

\subsection{X-rays from Intermediate Mass Stars \label{sec:HAeBes}}

X-ray emission from intermediate-mass stars with spectral types mid-B
to A is unexpected since no X-ray production mechanism is known; they
lack strong stellar winds and convective surfaces
\citep{Berghoefer94,Berghoefer97,Stelzer03,Stelzer06a}. However X-ray
detections of PMS intermediate-mass stars known as Herbig Ae/Be stars
are widely reported
\citep[e.g.,][]{Zinnecker94,Damiani94,Berghoefer96,Stelzer05,Hamaguchi05}. A
systematic {\em Chandra} archival study \citep{Stelzer06b} rules out
radiative winds as the origin for X-ray emission in HAeBes based on
the observed high X-ray temperatures. Thus an X-ray generating
mechanism from magnetic flares similar to late type stars or emission
from an unknown/unresolved companion is favored, although the role of
accretion for the production of X-rays remains unclear
\citep{Hamaguchi05,Stelzer06b}.

Using masses estimated from the NIR color magnitude diagram (Figure~6), we
detect around 50 stars in the intermediate mass range $2M_{\odot} \la
M \la 8M_{\odot}$. The absorption-corrected X-ray luminosities of the
detected sources in the 0.5-8.0 keV band are in the range of $\log
L_{t,c} \sim 29.5-31.8$ ergs s$^{-1}$.  This is fully consistent with
the level of X-ray emission detected in HAeBes from
\citet{Hamaguchi05} and \citet{Stelzer06b}, but also with the level
expected from late-type companions. Optical spectral classification of
these stars may further clarify the link between the X-ray emission
properties and the spectral types \citep[e.g.,][]{Li02}. Eight of
these stars show significant temporal variability ($P_{KS}\le 0.005$),
which strongly supports emission from flaring, possibly from
unresolved low mass companions. For example, source \#804 shows a big
flare during our long observation (Figure~\ref{fig:lightcurve}). The
rising and decay times during the flare are rather symmetric. The count
rate during the peak of the flare is 18 times higher than that of the
quiescent level. As noted in \S~\ref{sec:IRcolor}, source \#44 is a
previously identified Herbig Be star in NGC 2244 \citep{Li02}. We
detected 40 net counts at its optical position in 94 ks. The fit to
its X-ray spectrum indicates a low absorption and a hard ($kT=2.8$ keV)
plasma with a nonvariable lightcurve.

\subsection{X-rays from Other Interesting Sources}

{\bf Herbig-Haro Jets and Knots--} Two optical jet systems, namely
Rosette HH~1 (R.A.=06$^h$32$^m$20.$^s$76,
Dec=04$^{\circ}$53$^{\prime}$02.9$^{\prime \prime}$) and HH~2
(R.A.=06$^h$32$^m$14.$^s$14,
Dec.=05$^{\circ}$02$^{\prime}$17.95$^{\prime \prime}$ [J2000]), have
been discovered in the Rosette Nebula
\citep{Li03,Li04,Li05,Meaburn05,Li06}. HH 1 consists of a collimated
jet originating from a faint optical star with a mass-loss rate
$\dot{M}\sim 10^{-8}M_{\odot}$.  We have detected X-ray emission
(ACIS source \#743) coincident with the location of the exciting source of HH 1, a
weak-lined T Tauri star (F8Ve, Li et al.\ 2005). At the location of
the base, knot, or terminal shock of the jet, the distribution of
X-ray photons is consistent with background counts. No concentration
of photons is coincident with the shock structure seen in optical near
the end of the collimated jet. The X-ray spectrum of the star is soft,
with $kT\sim 0.8$ keV and a negligible absorption column. It has been
suggested that the combination of low extinction and high Lyman photon
flux inside the Rosette Nebula makes the jet optically visible
\citep{Meaburn05}.

A group of bright ionized knots in the Rosette Nebula were proposed to
be collisionally ionized, either by bow shocks formed around globules
by the strong winds from O stars which are then overrun by an
expanding shell, or collimated flows of shocked gas driven by the wind
\citep{Meaburn86}. \citet{Chen04} found an X-ray source in the
nebulous region D with high speed knots and identified two stars as
counterparts within {\em ROSAT} positional errors. We examined our
{\em Chandra} image with much higher resolution at all knot
locations. No X-ray emission was found to be associated with these
high-speed knots, but we detected point sources coincident with
optical stars embedded in the nebulous region. ACIS source \#678 (= Chen et
al. source 30) matches the position of HD 259210, a likely foreground
star with spectral type A1V.

{\bf Binaries--} The eclipsing binary V578 Mon, used to determine the
age and distance to NGC 2244, is detected in our X-ray image with 186
net counts (\#476). It consists of two early B-type stars, one of the
very few massive eclipsing systems known \citep{Harries98}. Its
orbital period is precisely determined in the optical to be $P=2.40848
\pm 0.00001$ days \citep{Hensberge00}. Using the eclipse ephemeris and
the date at the beginning of our observation, we examined the X-ray
light curve together with the optical light curve with orbital phases
(Figure~\ref{fig:eclipse}). No significant variability is suggested by
a K-S test for the X-ray light curve, although dips might be seen in
the X-ray light curve around phase $\phi=0.85$ and $\phi=1.0$, where
the primary eclipse is expected. This could be one of the rare cases
where X-ray eclipses can be used to constrain emitting geometry
\citep[e.g.,][]{Schmitt99}. However this could simply be a statistical
fluctuation given the limited number of counts. The spectral fit gives
a plasma temperature of $kT\sim 1.6$~keV with low absorption, $\log
N_H=20.6$ cm$^{-2}$.

\citet{PS02} noted a suspected PMS binary system ([PS02]~125 and
[PS02]~126). An optical spectrum of the unresolved binary shows
$H_\alpha$ in emission and LiI~6708\AA\ in absorption \citep{Chen04},
confirming its youth. We resolve the system as a close pair of ACIS
sources (\#243 \& \#242). Their spectral fits give a rather hard
$kT\sim 2.9$~keV for \#243 and $kT\sim 1.5$~keV for \#242; they share
the same $\log N_H=21.4$ cm$^{-2}$. The light curve of \#242 is
constant while \#243 shows possible but not significant variation
($P_{KS}=0.03$).

{\bf Magnetic Star--} \citet{Bagnulo04} discovered an extraordinarily
strong magnetic field in the very young cluster member NGC 2244-334
(=[OI81] 334; spectral type B3; R.A.=06$^h$32$^m$51.$^s$79,
Dec=+04$^{\circ}$47$^{\prime}$16.1$^{\prime \prime}$ [J2000]), ranking
as the second strongest longitudinal field known among non-degenerate
stars \citep[after HD 215441, or Babcock's star; ][]{Borra78}. We
detected a 10 count X-ray source (\#899, CXOU J063251.79+044715.9) at
its position in the 20 ks observation (it was not covered by the deep
observation).

{\bf Elephant Trunks--} \citet{Schneps80} identified several
spectacular elephant trunk globules in the northwest part of the
Rosette Nebula. Only one of them, a small isolated globule denoted R1,
is in our field of view. One X-ray source (\#169), probably by chance
superposition, is located 6 arcsec away from the bright rim of this
dark globule. No other X-ray/IR source can be found inside the
globule. 

In the southeast quadrant of the nebula towards the RMC, another
molecular pillar is prominent. This region is also highlighted in a
recent Spitzer survey \citep{Balog07}, as the size of the pillar is
comparable to the largest ``pillar-of-creation'' in
M16. \citet{Chen04} noted shocked gas near the pillar, perhaps due to
strong winds from a star nearby that was matched to one of the {\em
ROSAT} sources.  In the vicinity of this pillar, a luminous X-ray
source (\#919, CXOU J063309.61+044624.3) is detected in our
observation, but it is not located at the tip of the elephant
trunk. Its X-ray spectrum can be fit well with $\log N_H=21.2$
cm$^{-2}$ and a hard ($kT=3.3$~keV) plasma. With an unusually high
luminosity of $\log L_{t,c}\sim 32.2$ ergs s$^{-1}$, it is comparable to
the earliest O stars. The light curve is variable, as shown in
Figure~\ref{fig:lightcurve}. The count rate doubles after the first
6~ks and remains in a high state for $\sim$13~ks. Its IR counterpart
is also bright, with a $K$-band magnitude of 9.6 mag that is similar
to the observed B0-B2 stars in the field. Its location in the
color-magnitude diagram also suggests a spectral type of B0-B1. No
$K$-excess is seen.  As an early B-type star, its X-ray variability
can be explained by an unresolved late type companion, although the
lightcurve does not follow the typical PMS fast-rise and slow-decay
phase. A similar transition between high and low states seen in
$\theta ^2$ Ori A is investigated by \citet{Schulz06} and interpreted
as possibly the reconnection from magnetic interactions in a close
binary system.

\section{Summary \label{summary.sec}}

We present a high spatial resolution X-ray study of the NGC 2244
cluster in the Rosette Nebula obtained via deep {\it Chandra}
observations. Our main findings follow:

1. We detect 919 X-ray sources with a limiting X-ray sensitivity of
   $L_{t,c}\sim 1\times 10^{29}$ ergs s$^{-1}$.  Positional
   coincidence matching yields a total of 712 ONIR counterparts. We
   estimate 8\% extragalactic and Galactic contamination. The rest
   of the X-ray sources without ONIR counterparts are likely new
   NGC~2244 members clustered around the massive star HD~46150 or deeply
   embedded in the cloud. The X-ray detected population provides the
   first deep probe of the rich low mass population in this massive
   cluster.

2. The locations of most ACIS sources in the color-magnitude plot
   indicate a large population of 2 Myr old PMS low mass stars ($M
   \la 2 M_{\odot}$) subject to a visual extinction of $1\la A_V \la
   2$ at 1.4 kpc. We derive an overall $K$-excess disk frequency of
   $\sim$6\% for stars with mass $M \ga 0.5M_{\odot}$ using the X-ray
   selected sample, slightly lower than the 10\% $K$-excess disk
   fraction using a FLAMINGOS selected sample. Both fractions are
   significantly lower than the 45\% mid-IR disk fraction in a {\em
   Spitzer} sample that is more sensitive to disks. We emphasize that
   the combination of young stars identified in X-rays (mostly Class
   III stars) and those selected through IR colors will provide the
   best census to date for the young stellar population of this
   region. Three objects have Class I colors.

3. The derived XLF ($L_{h,c}$) for NGC 2244 is compared to the XLFs of
   the ONC, M17, Cep B, and NGC 6357; this exercise indicates that the
   unobscured population in NGC 2244 is 1.2 times larger than that of
   the ONC, or $\sim 1000$ stars detectable in COUP-sensitivity X-ray
   observations. Taking into account the obscured population, the
   total stellar population in NGC 2244 is $\sim 2000$, in good
   agreement with the estimated population from the spatially complete
   2MASS study. The XLF and KLF suggest a normal Salpeter IMF for NGC
   2244; we do not confirm a top heavy IMF reported from earlier
   optical studies.

4. We examine the spatial distribution of the X-ray identified NGC
   2244 cluster members; the stellar surface density map suggests a
   spherical cluster with substructure. We confirm the existence of a
   subcluster around HD 46150 with $\sim 50$ members in a 1~pc region;
   a second small subcluster consisting of a number of late type stars is
   also found.  The O4 star HD 46223 has few companions. The radial
   density profile of NGC 2244 shows a larger relaxed structure around
   the central subcluster. Similar structure is seen in NGC 6357. No
   evidence for significant mass segregation is found in this
   cluster. Altogether we suggest that this 2 Myr cluster is not
   dynamically evolved and has a complex star formation history. Our
   results will strongly constrain models of the cluster formation
   process.

5. We detected all 9 OB stars with spectral types B0.5 or earlier, but
   only 6 out of 14 B stars with spectral types B1--B3 in our field of
   view. X-ray spectra for the massive stars in NGC 2244 all show soft
   emission. We confirm the long-standing
   $\log(L_x/L_{bol})\sim -7$ relation for the NGC 2244 O stars. Large
   scatter around this correlation was found for the B stars.

6. We report X-ray emission detected from a few interesting individual
   objects, including the ionizing source of the optical jet Rosette
   HH1, binary systems, a magnetic star, and a possible X-ray luminous
   uncataloged massive star.

\acknowledgements 

We thank the referee, Scott Wolk, for his time and for many useful
comments that improved this work. We thank Ed Churchwell and Steinn
Sigurdsson for helpful discussions on the cluster structure and
dynamics. We thank Travis Rector and Mark Heyer for kindly providing
the KPNO MOSAIC images of the Rosette Nebula and the CO emission maps
of the Rosette Complex, respectively. This work was supported by {\it
Chandra X-ray Observatory} grants GO1-2008X, GO3-4010X, and
GO6-7006X. FLAMINGOS was designed and constructed by the IR
instrumentation group (PI: R. Elston) at the University of Florida,
Department of Astronomy with support from NSF grant AST97-31180 and
Kitt Peak National Observatory.  The data were collected under the
NOAO Survey Program, ``Towards a Complete Near-Infrared Spectroscopic
Survey of Giant Molecular Clouds'' (PI: E. Lada) and supported by NSF
grants AST97-3367 and AST02-02976 to the University of
Florida. E.A.L. also acknowledges support from NASA LTSA NNG05D66G.
This publication makes use of data products from the Two Micron All
Sky Survey, which is a joint project of the University of
Massachusetts and the Infrared Processing and Analysis
Center/California Institute of Technology, funded by NASA and the
National Science Foundation.  This research has made use of the SIMBAD
database and the VizieR catalogue access tool, operated at CDS,
Strasbourg, France.

{\it Facility}: \facility{CXO (ACIS)}

\clearpage
\onecolumn

\begin{deluxetable}{ccrccccc}
\centering
\tabletypesize{\tiny} \tablewidth{0pt}
\tablecolumns{8}

\tablecaption{ Log of {\it Chandra} Observations
 \label{tbl:obslog}}

\tablehead{
\colhead{Target} &
\colhead{Obs ID} &
\colhead{Start Time} &
\colhead{Exposure Time} &
\multicolumn{2}{c}{Aimpoint} &
\colhead{Roll Angle} &
\colhead{Mode\tablenotemark{a}} \\
\cline{5-6}

\colhead{} &
\colhead{} &
\colhead{(UT)} &
\colhead{(s)} &
\colhead{$\alpha_{\rm J2000}$} &
\colhead{$\delta_{\rm J2000}$} &
\colhead{(deg)} &
\colhead{}
}

\startdata
Rosette Field 1............ & 1874 & 2001 Jan 05 11:53 & 19700 & 06 31 52.85 & +04 55 42.0 & 335.86 & F \\
Rosette Field 2............ & 1875 & 2001 Jan 05 17:46 & 19500 & 06 32 40.84 & +04 42 45.0 & 335.90 & F \\
Rosette Field 3............ & 1876 & 2001 Jan 05 23:28 & 19410 & 06 33 17.15 & +04 34 42.0 & 335.76 & F \\
Rosette Field 4............ & 1877 & 2001 Jan 06 05:10 & 19510 & 06 34 17.34 & +04 27 45.9 & 335.85 & F \\
Rosette Nebula/NGC 2244 & 3750 & 2004 Jan 01 02:20 & 75000 & 06 31 56.45 & +04 56 25.4 & 351.87 & VF \\
NGC 2244 Satellite Cluster & 8454 & 2007 Feb 09 02:25 & 20480 & 06 30 50.40 & +04 59 34.0 & 286.00 & VF \\
\enddata

\tablenotetext{a}{ The observing mode:  F=Faint, VF=Very Faint.}

\tablecomments{ Units of right ascension are hours, minutes, and
seconds; units of declination are degrees, arcminutes, and
arcseconds. Exposure times are the net usable times after various
filtering steps are applied in the data reduction process. The
aimpoints and roll angles are obtained from the satellite aspect
solution before astrometric correction is applied.}

\end{deluxetable}
\begin{deluxetable}{rcrrrrrrrrrrrrrccccc}
\centering \rotate
\tabletypesize{\tiny} \tablewidth{0pt}
\tablecolumns{20}

\tablecaption{ {\it Chandra} Main Catalog:  Basic Source Properties \label{tbl:primary}}

\tablehead{
\multicolumn{2}{c}{Source} &
  &
\multicolumn{4}{c}{Position} &
  &
\multicolumn{5}{c}{Extracted Counts} &
  &
\multicolumn{6}{c}{Characteristics} \\
\cline{1-2} \cline{4-7} \cline{9-13} \cline{15-20}

\colhead{Seq} & \colhead{CXOU J} &
  &
\colhead{$\alpha_{\rm J2000}$} & \colhead{$\delta_{\rm J2000}$} & \colhead{Err} & \colhead{$\theta$} &
  &
\colhead{Net} & \colhead{$\Delta$Net} & \colhead{Bkgd} & \colhead{Net} & \colhead{PSF} &
  &
\colhead{Signif} & \colhead{$\log P_B$} & \colhead{Anom} & \colhead{Var} &\colhead{EffExp} & \colhead{$E_{median}$}  \\

\colhead{\#} & \colhead{} &
  &
\colhead{(deg)} & \colhead{(deg)} & \colhead{(\arcsec)} & \colhead{(\arcmin)} &
  &
\colhead{Full} & \colhead{Full} & \colhead{Full} & \colhead{Hard} &
\colhead{Frac} &
  &
\colhead{} & \colhead{} & \colhead{} & \colhead{} & \colhead{(ks)} & \colhead{(keV)}
 \\

\colhead{(1)} & \colhead{(2)} &
  &
\colhead{(3)} & \colhead{(4)} & \colhead{(5)} & \colhead{(6)} &
  &
\colhead{(7)} & \colhead{(8)} & \colhead{(9)} & \colhead{(10)} & \colhead{(11)} &
  &
\colhead{(12)} & \colhead{(13)} & \colhead{(14)} & \colhead{(15)} & \colhead{(16)} & \colhead{(17)}  }

\startdata
   1 & 063114.36$+$045303.0 &  &   97.809835 &   4.884189 &  0.9 &  9.9 &  &    23.3 &   5.7 &   3.7 &     5.8 & 0.91 &  &   3.7 & $<$-5 & g... & \nodata &   15.0 & 1.4 \\
   2 & 063117.04$+$045228.3 &  &   97.821033 &   4.874535 &  0.7 &  9.5 &  &    32.0 &   6.5 &   4.0 &     1.4 & 0.90 &  &   4.5 & $<$-5 & .... & a &   17.0 & 1.4 \\
   3 & 063118.11$+$045511.8 &  &   97.825464 &   4.919953 &  1.0 &  8.7 &  &    12.4 &   4.4 &   2.6 &     3.5 & 0.90 &  &   2.5 & $<$-5 & .... & a &   16.3 & 1.7 \\
   4 & 063118.29$+$045223.1 &  &   97.826245 &   4.873106 &  1.0 &  9.2 &  &    14.6 &   4.8 &   3.4 &     0.8 & 0.89 &  &   2.7 & $<$-5 & .... & a &   16.8 & 1.2 \\
   5 & 063118.76$+$045207.6 &  &   97.828205 &   4.868779 &  1.1 &  9.2 &  &     9.7 &   4.1 &   3.3 &     1.9 & 0.89 &  &   2.1 & -4.4 & .... & b &   17.0 & 1.1 \\
   8 & 063120.89$+$045003.8 &  &   97.837054 &   4.834396 &  0.2 & 10.7 &  &   599.7 &  25.8 &  33.3 &    32.1 & 0.87 &  &  22.8 & $<$-5 & .... & a &   81.8 & 1.0 \\
 311 & 063152.54$+$050159.1 &  &   97.968929 &   5.033106 &  0.1 &  5.8 &  &   299.3 &  17.8 &   0.7 &     0.6 & 0.43 &  &  16.3 & $<$-5 & .... & a &   85.7 & 0.9 \\
 373 & 063155.51$+$045634.2 &  &   97.981330 &   4.942835 &  0.0 &  0.4 &  &  3588.5 &  60.4 &   0.5 &   121.8 & 0.88 &  &  58.9 & $<$-5 & g... & \nodata &   82.6 & 0.9 \\
 476 & 063200.65$+$045241.2 &  &   98.002718 &   4.878124 &  0.1 &  3.8 &  &   185.9 &  14.2 &   1.1 &    27.3 & 0.89 &  &  12.6 & $<$-5 & .... & a &   90.6 & 1.2 \\
 615 & 063209.32$+$044924.4 &  &   98.038846 &   4.823469 &  0.1 &  7.7 &  &  1685.2 &  41.6 &   4.8 &    17.6 & 0.69 &  &  40.0 & $<$-5 & .... & \nodata &   83.9 & 1.0 \\
 630 & 063210.47$+$045759.6 &  &   98.043639 &   4.966572 &  0.1 &  4.0 &  &   332.6 &  18.8 &   1.4 &     0.3 & 0.89 &  &  17.2 & $<$-5 & .... & a &   88.5 & 0.9 \\
 743 & 063220.77$+$045303.5 &  &   98.086542 &   4.884328 &  0.3 &  7.0 &  &    25.5 &   5.7 &   1.5 &     2.2 & 0.48 &  &   4.1 & $<$-5 & .... & b &   85.0 & 0.9 \\
\enddata

\tablecomments{Table \ref{tbl:primary} is published in its entirety in
the electronic edition of the {\it Astrophysical Journal}. The first
five sources and the interesting sources shown in Figure~1 are listed
here for guidance regarding its form and content and for the convenience
of the reader.}

\tablecomments{{\bf Column 1:} X-ray catalog sequence number, sorted by RA.
{\bf Column 2:} IAU designation.
{\bf Columns 3,4:} Right ascension and declination for epoch J2000.0.
{\bf Column 5:} Estimated random component of position error, $1\sigma$, computed as $\frac{\mbox{standard deviation of PSF inside extraction region}}{\sqrt{\mbox{\# of counts extracted}}}$.
{\bf Column 6:} Off-axis angle.
{\bf Columns 7,8:} Estimated net counts extracted in the total energy band (0.5--8~keV); average of the upper and lower $1\sigma$ errors on column 7.
{\bf Column 9:} Background counts extracted (total band).
{\bf Column 10:} Estimated net counts extracted in the hard energy band (2--8~keV).
{\bf Column 11:} Fraction of the PSF (at 1.497 keV) enclosed within the extraction region.  Note that a reduced PSF fraction (significantly below 90\%) may indicate that the source is in a crowded region.
{\bf Column 12:} Photometric significance computed as $\frac{\mbox{net counts}}{\mbox{upper error on net counts}}$.
{\bf Column 13:} Log probability that extracted counts (total band) are solely from background.  Some sources have $P_B$ values above the 1\% threshold that defines the catalog because local background estimates can rise during the final extraction iteration after sources are removed from the catalog.
{\bf Column 14:}  Source anomalies:  g = fractional time that the source was on a detector (FRACEXPO from {\em mkarf}) is $<0.9$ ; e = source on field edge; p = source piled up; s = source on readout streak.
{\bf Column 15:} Variability characterization based on K-S statistic (total band):  a = no evidence for variability ($0.05<P_{KS}$); b = possibly variable ($0.005<P_{KS}<0.05$); c = definitely variable ($P_{KS}<0.005$).  No value is reported for sources with fewer than 4 counts or for sources in chip gaps or on field edges.
{\bf Column 16:} Effective exposure time: approximate time the source would have to be observed on axis to obtain the reported number of counts.
{\bf Column 17:} Background-corrected median photon energy (total band).}

\end{deluxetable}
\begin{deluxetable}{rcrrrrrrrrrrrrrccccc}
\centering \rotate
\tabletypesize{\tiny} \tablewidth{0pt}
\tablecolumns{20}

\tablecaption{ {\it Chandra} Secondary Catalog:  Tentative Source Properties \label{tbl:tentative}}

\tablehead{
\multicolumn{2}{c}{Source} &
  &
\multicolumn{4}{c}{Position} &
  &
\multicolumn{5}{c}{Extracted Counts} &
  &
\multicolumn{6}{c}{Characteristics} \\
\cline{1-2} \cline{4-7} \cline{9-13} \cline{15-20}

\colhead{Seq} & \colhead{CXOU J} &
  &
\colhead{$\alpha_{\rm J2000}$} & \colhead{$\delta_{\rm J2000}$} & \colhead{Err} & \colhead{$\theta$} &
  &
\colhead{Net} & \colhead{$\Delta$Net} & \colhead{Bkgd} & \colhead{Net} & \colhead{PSF} &
  &
\colhead{Signif} & \colhead{$\log P_B$} & \colhead{Anom} & \colhead{Var} &\colhead{EffExp} & \colhead{$E_{median}$}  \\

\colhead{\#} & \colhead{} &
  &
\colhead{(deg)} & \colhead{(deg)} & \colhead{(\arcsec)} & \colhead{(\arcmin)} &
  &
\colhead{Full} & \colhead{Full} & \colhead{Full} & \colhead{Hard} &
\colhead{Frac} &
  &
\colhead{} & \colhead{} & \colhead{} & \colhead{} & \colhead{(ks)} & \colhead{(keV)}
 \\

\colhead{(1)} & \colhead{(2)} &
  &
\colhead{(3)} & \colhead{(4)} & \colhead{(5)} & \colhead{(6)} &
  &
\colhead{(7)} & \colhead{(8)} & \colhead{(9)} & \colhead{(10)} & \colhead{(11)} &
  &
\colhead{(12)} & \colhead{(13)} & \colhead{(14)} & \colhead{(15)} & \colhead{(16)} & \colhead{(17)}  }

\startdata
  19 & 063124.47$+$045306.2 &  &   97.851975 &   4.885077 &  0.8 &  8.4 &  &     9.0 &   4.9 &  10.0 &     0.1 & 0.90 &  &   1.6 & -2.1 & .... & a &   82.8 & 1.5 \\
  28 & 063126.33$+$045728.8 &  &   97.859728 &   4.958023 &  0.8 &  7.4 &  &     8.0 &   4.4 &   7.0 &     3.0 & 0.90 &  &   1.6 & -2.2 & .... & a &   84.7 & 1.1 \\
  32 & 063127.43$+$045036.4 &  &   97.864312 &   4.843457 &  0.8 &  9.0 &  &    10.3 &   5.5 &  13.7 &     2.8 & 0.90 &  &   1.7 & -2.1 & .... & a &   83.0 & 1.5 \\
  36 & 063128.13$+$045008.0 &  &   97.867223 &   4.835559 &  0.8 &  9.2 &  &    11.7 &   5.8 &  15.3 &     0.0 & 0.90 &  &   1.8 & -2.4 & .... & a &   83.7 & 1.3 \\
  52 & 063130.85$+$044847.3 &  &   97.878575 &   4.813144 &  0.9 &  9.7 &  &    13.4 &   6.8 &  23.6 &     0.1 & 0.89 &  &   1.8 & -2.2 & .... & a &   83.3 & 1.4 \\
  61 & 063132.25$+$050439.0 &  &   97.884410 &   5.077512 &  0.9 & 10.2 &  &    12.3 &   6.8 &  22.7 &     6.9 & 0.90 &  &   1.7 & -2.0 & .... & a &   63.1 & 2.7 \\
  68 & 063133.57$+$045233.3 &  &   97.889886 &   4.875920 &  0.6 &  6.7 &  &     8.2 &   4.1 &   4.8 &     2.9 & 0.89 &  &   1.7 & -2.9 & .... & a &   86.1 & 1.8 \\
  71 & 063134.14$+$044958.5 &  &   97.892258 &   4.832939 &  0.7 &  8.3 &  &    10.6 &   5.3 &  11.4 &     0.7 & 0.90 &  &   1.8 & -2.5 & .... & a &   84.7 & 1.1 \\
  83 & 063135.41$+$045813.0 &  &   97.897576 &   4.970293 &  0.6 &  5.4 &  &     5.8 &   3.4 &   2.2 &     4.5 & 0.90 &  &   1.5 & -2.7 & .... & a &   88.9 & 4.9 \\
  85 & 063135.75$+$050259.6 &  &   97.898985 &   5.049898 &  0.7 &  8.4 &  &    10.7 &   5.3 &  11.3 &     2.5 & 0.90 &  &   1.8 & -2.5 & .... & b &   79.9 & 1.6 \\
\enddata

\tablecomments{Table \ref{tbl:tentative} is published in its entirety in the
electronic edition of the {\it Astrophysical Journal}.  A portion is
shown here for guidance regarding its form and content.}

\tablecomments{{\bf Column 1:} X-ray catalog sequence number, sorted by RA.
{\bf Column 2:} IAU designation.
{\bf Columns 3,4:} Right ascension and declination for epoch J2000.0.
{\bf Column 5:} Estimated random component of position error, $1\sigma$, computed as $\frac{\mbox{standard deviation of PSF inside extraction region}}{\sqrt{\mbox{\# of counts extracted}}}$.
{\bf Column 6:} Off-axis angle.
{\bf Columns 7,8:} Estimated net counts extracted in the total energy band (0.5--8~keV); average of the upper and lower $1\sigma$ errors on column 7.
{\bf Column 9:} Background counts extracted (total band).
{\bf Column 10:} Estimated net counts extracted in the hard energy band (2--8~keV).
{\bf Column 11:} Fraction of the PSF (at 1.497 keV) enclosed within the extraction region.  Note that a reduced PSF fraction (significantly below 90\%) may indicate that the source is in a crowded region.
{\bf Column 12:} Photometric significance computed as $\frac{\mbox{net counts}}{\mbox{upper error on net counts}}$.
{\bf Column 13:} Log probability that extracted counts (total band) are solely from background.  Some sources have $P_B$ values above the 1\% threshold that defines the catalog because local background estimates can rise during the final extraction iteration after sources are removed from the catalog.
{\bf Column 14:}  Source anomalies:  g = fractional time that the source was on a detector (FRACEXPO from {\em mkarf}) is $<0.9$ ; e = source on field edge; p = source piled up; s = source on readout streak.
{\bf Column 15:} Variability characterization based on K-S statistic (total band):  a = no evidence for variability ($0.05<P_{KS}$); b = possibly variable ($0.005<P_{KS}<0.05$); c = definitely variable ($P_{KS}<0.005$).  No value is reported for sources with fewer than 4 counts or for sources in chip gaps or on field edges.
{\bf Column 16:} Effective exposure time: approximate time the source would have to be observed on axis to obtain the reported number of counts.
{\bf Column 17:} Background-corrected median photon energy (total band).}

\end{deluxetable}
\begin{deluxetable}{rcrrrcccrcccccrc}
\centering \rotate
\tabletypesize{\scriptsize} \tablewidth{0pt}
\tablecolumns{16}

\tablecaption{X-ray Spectroscopy for Photometrically Selected Sources:  Thermal Plasma Fits
\label{tbl:thermal_spectroscopy}}

\tablehead{
\multicolumn{4}{c}{Source\tablenotemark{a}} &
  &
\multicolumn{3}{c}{Spectral Fit\tablenotemark{b}} &
  &
\multicolumn{5}{c}{X-ray Luminosities\tablenotemark{c}} &
  &
\colhead{Notes\tablenotemark{d}} \\
\cline{1-4} \cline{6-8} \cline{10-14}

\colhead{Seq} & \colhead{CXOU J} & \colhead{Net} & \colhead{Signif} &
  &
\colhead{$\log N_H$} & \colhead{$kT$} & \colhead{$\log EM$} &
  &
\colhead{$\log L_s$} & \colhead{$\log L_h$} & \colhead{$\log L_{h,c}$} & \colhead{$\log L_t$} & \colhead{$\log L_{t,c}$} &
  &
\colhead{}  \\

\colhead{\#} & \colhead{} & \colhead{Counts} & \colhead{} &
  &
\colhead{(cm$^{-2}$)} & \colhead{(keV)} & \colhead{(cm$^{-3}$)} &
  &
\multicolumn{5}{c}{(ergs s$^{-1}$)} &
  &
\colhead{} \\

\colhead{(1)} & \colhead{(2)} & \colhead{(3)} & \colhead{(4)} &
  &
\colhead{(5)} & \colhead{(6)} & \colhead{(7)} &
  &
\colhead{(8)} & \colhead{(9)} & \colhead{(10)} &\colhead{(11)} & \colhead{(12)} &
  &
\colhead{(13)}
}



\startdata
   1 & 063114.36$+$045303.0 &    23.3 &   3.7 &  &   20.0  & {\em  2.0 } &  53.4  &  & 30.20 & 29.96 & 29.96 & 30.40 & 30.40 &  & \nodata \\
   2 & 063117.04$+$045228.3 &    32.0 &   4.5 &  &  \phantom{{\tiny $+0.5$}} 21.1 {\tiny $+0.5$} & {\em  2.0 } & {\tiny $-0.2$} 53.7 {\tiny $+0.2$} &  & 30.31 & 30.19 & 30.20 & 30.56 & 30.64 &  & \nodata \\
   3 & 063118.11$+$045511.8 &    12.4 &   2.5 &  &  \phantom{{\tiny $+1.3$}} 21.6 {\tiny $+1.3$} & {\tiny $-2.6$} 2.8 \phantom{{\tiny $-2.6$}} &  53.4  &  & 29.78 & 30.03 & 30.06 & 30.23 & 30.39 &  & \nodata \\
   4 & 063118.29$+$045223.1 &    14.6 &   2.7 &  &   21.1  & {\em  2.0 } &  53.3  &  & 29.96 & 29.85 & 29.86 & 30.21 & 30.30 &  & \nodata \\
   5 & 063118.76$+$045207.6 &     9.7 &   2.1 &  &  \phantom{{\tiny $+1.1$}} 20.8 {\tiny $+1.1$} &  0.6  & \phantom{{\tiny $+0.8$}} 52.9 {\tiny $+0.8$} &  & 29.73 & 28.33 & 28.34 & 29.75 & 29.85 &  & \nodata \\
   6 & 063119.27$+$045110.7 &    22.6 &   3.6 &  &   20.1  &  1.9  &  53.4  &  & 30.17 & 29.91 & 29.92 & 30.36 & 30.37 &  & \nodata \\
   7 & 063120.48$+$045023.4 &    98.9 &   8.5 &  &  \phantom{{\tiny $+0.1$}} 21.5 {\tiny $+0.1$} & {\tiny $-0.4$} 1.7 {\tiny $+0.8$} & {\tiny $-0.17$} 53.8 {\tiny $+0.07$} &  & 30.25 & 30.18 & 30.20 & 30.52 & 30.73 &  & \nodata \\
   8 & 063120.89$+$045003.8 &   599.7 &  22.8 &  &   21.4  &  0.6  &  54.4  &  & 31.11 & 29.86 & 29.89 & 31.13 & 31.51 &  & H; HD 46056 \\
   9 & 063120.95$+$045322.7 &    14.5 &   2.6 &  &   21.0  &  3.4  &  53.0  &  & 29.70 & 29.81 & 29.81 & 30.06 & 30.11 &  & \nodata \\
  11 & 063121.46$+$045405.3 &    15.6 &   2.7 &  &   21.4  & {\em  2.0 } &  53.2  &  & 29.71 & 29.70 & 29.72 & 30.01 & 30.16 &  & \nodata \\
\enddata

\tablecomments{Table \ref{tbl:thermal_spectroscopy} is published in its entirety in the
electronic edition of the {\it Astrophysical Journal}.  A portion is
shown here for guidance regarding its form and content.}

\tablenotetext{a}{
For convenience {\bf columns 1--4} reproduce the source identification, net counts, and photometric significance data from Table~\ref{tbl:primary}.
}

\tablenotetext{b}{
All fits used the ``wabs(apec)'' model in {\it XSPEC} and assumed 0.3$Z_{\odot}$ abundances \citep{Imanishi01,Feigelson02}.
{\bf Columns 5 and 6} present the best-fit values for the column density and plasma temperature parameters.
{\bf Column 7} presents the emission measure for the model spectrum, assuming a distance of 1.4~kpc.
{\it Quantities in italics} were frozen in the fit.
Uncertainties represent 90\% confidence intervals.
More significant digits are used for uncertainties $<0.1$ in order to avoid large rounding errors; for consistency, the same number of significant digits is used for both lower and upper uncertainties.
Uncertainties are missing when {\it XSPEC} was unable to compute them or when their values were so large that the parameter is effectively unconstrained.
Fits lacking uncertainties should be considered to be merely a spline to the data to obtain rough estimates of luminosities; actual parameter values are unreliable.
}

\tablenotetext{c}{ X-ray luminosities are presented in {\bf columns 8--12}:  s = soft band (0.5--2 keV); h = hard band (2--8 keV); t = total band (0.5--8 keV).
Absorption-corrected luminosities are subscripted with a $c$; they are omitted when $\log N_H > 22.5$ since the soft band emission is essentially unmeasurable.
}

\tablenotetext{d}{ {\bf 2T} means a two-temperature model was used.
{\bf H} means the fit was performed by hand, usually because the
automated fit yielded non-physical results.  Names of well-known OB
counterparts are listed here for the convenience of the reader.}

\end{deluxetable}

\begin{deluxetable}{rcrrrcccrcccccrc}
\centering \rotate
\tabletypesize{\scriptsize} \tablewidth{0pt}
\tablecolumns{16}

\tablecaption{X-ray Spectroscopy for Photometrically Selected Sources:  Power Law Fits
\label{tbl:powerlaw_spectroscopy}}

\tablehead{
\multicolumn{4}{c}{Source\tablenotemark{a}} &
  &
\multicolumn{3}{c}{Spectral Fit\tablenotemark{b}} &
  &
\multicolumn{5}{c}{X-ray Fluxes\tablenotemark{c}} &
  &
\colhead{Notes} \\
\cline{1-4} \cline{6-8} \cline{10-14}

\colhead{Seq} & \colhead{CXOU J} & \colhead{Net} & \colhead{Signif} &
  &
\colhead{$\log N_H$} & \colhead{$\Gamma$} & \colhead{$\log N_{\Gamma}$} &
  &
\colhead{$\log L_s$} & \colhead{$\log L_h$} & \colhead{$\log L_{h,c}$} & \colhead{$\log L_t$} & \colhead{$\log L_{t,c}$} &
  &
\colhead{}  \\

\colhead{\#} & \colhead{} & \colhead{Counts} & \colhead{} &
  &
\colhead{(cm$^{-2}$)} & \colhead{} & \colhead{} &
  &
\multicolumn{5}{c}{(photons cm$^{-2}$ s$^{-1}$)} &
  &
\colhead{} \\

\colhead{(1)} & \colhead{(2)} & \colhead{(3)} & \colhead{(4)} &
  &
\colhead{(5)} & \colhead{(6)} & \colhead{(7)} &
  &
\colhead{(8)} & \colhead{(9)} &\colhead{(10)} & \colhead{(11)} & \colhead{(12)} &
  &
\colhead{(13)}
}



\startdata
  10 & 063121.28$+$045023.8 &    50.6 &   5.7 &  &   21.5  & {\tiny $-0.9$} 1.4 \phantom{{\tiny $-0.9$}} & {\tiny $-0.5$} -5.6 \phantom{{\tiny $-0.5$}} &  &  29.80 &  30.45 &  30.46 &  30.54 &  30.62 &  & \nodata \\
  15 & 063123.59$+$045141.7 &    60.6 &   6.3 &  &  \phantom{{\tiny $+0.5$}} 21.5 {\tiny $+0.5$} & {\tiny $-0.6$} 1.1 {\tiny $+0.9$} & {\tiny $-0.4$} -5.9 {\tiny $+0.5$} &  &  29.63 &  30.41 &  30.42 &  30.47 &  30.53 &  & \nodata \\
  26 & 063125.51$+$045252.2 &    45.5 &   5.4 &  &   20.0  &  2.0  &  -5.8  &  &  29.86 &  29.73 &  29.73 &  30.10 &  30.11 &  & \nodata \\
  31 & 063127.01$+$050205.0 &    14.3 &   2.3 &  &   22.3  &  1.5  &  -6.0  &  &  28.79 &  29.92 &  29.99 &  29.95 &  30.17 &  & \nodata \\
  84 & 063135.68$+$045322.2 &    26.9 &   4.1 &  &  {\tiny $-0.7$} 22.1 {\tiny $+0.4$} & {\tiny $-1.2$} 2.0 \phantom{{\tiny $-1.2$}} &  -5.7  &  &  29.21 &  29.98 &  30.03 &  30.04 &  30.33 &  & \nodata \\
  89 & 063136.33$+$045251.1 &    46.2 &   5.7 &  &  {\tiny $-0.6$} 21.9 {\tiny $+0.3$} & {\tiny $-0.9$} 1.8 \phantom{{\tiny $-0.9$}} &  -5.6  &  &  29.46 &  30.15 &  30.19 &  30.23 &  30.43 &  & \nodata \\
  91 & 063136.42$+$045602.8 &    12.2 &   2.5 &  &   22.3  &  1.5  &  -6.1  &  &  28.70 &  29.84 &  29.91 &  29.87 &  30.10 &  & \nodata \\
  94 & 063136.49$+$045959.0 &    10.6 &   2.2 &  &  \phantom{{\tiny $+1.0$}} 21.1 {\tiny $+1.0$} &  0.8  &  -6.8  &  &  28.90 &  29.72 &  29.72 &  29.78 &  29.80 &  & \nodata \\
 118 & 063139.24$+$050244.7 &    19.5 &   3.1 &  &  \phantom{{\tiny $+0.3$}} 22.1 {\tiny $+0.3$} &  1.3  &  -6.0  &  &  29.02 &  30.08 &  30.13 &  30.12 &  30.27 &  & \nodata \\
 126 & 063139.92$+$050059.4 &    47.1 &   5.8 &  &  \phantom{{\tiny $+0.5$}} 21.7 {\tiny $+0.5$} & {\tiny $-0.8$} 1.3 \phantom{{\tiny $-0.8$}} & {\tiny $-0.4$} -5.8 \phantom{{\tiny $-0.4$}} &  &  29.54 &  30.31 &  30.33 &  30.38 &  30.47 &  & \nodata \\
\enddata

\tablecomments{Table \ref{tbl:powerlaw_spectroscopy} is published in its entirety in the
electronic edition of the {\it Astrophysical Journal}.  A portion is
shown here for guidance regarding its form and content.}

\tablenotetext{a}{
For convenience {\bf columns 1--4} reproduce the source identification, net counts, and photometric significance data from Table~\ref{tbl:primary}.
}

\tablenotetext{b}{
All fits used the ``wabs(powerlaw)'' model in {\it XSPEC}.
{\bf Columns 5 and 6} present the best-fit values for the column density and power law photon index parameters.
{\bf Column 7} presents the power law normalization for the model spectrum.
{\it Quantities in italics} were frozen in the fit.
Uncertainties represent 90\% confidence intervals.
More significant digits are used for uncertainties $<0.1$ in order to avoid large rounding errors; for consistency, the same number of significant digits is used for both lower and upper uncertainties.
Uncertainties are missing when {\it XSPEC} was unable to compute them or when their values were so large that the parameter is effectively unconstrained.
Fits lacking uncertainties should be considered to merely be a spline to the data to obtain rough estimates of luminosities; actual parameter values are unreliable.
}

\tablenotetext{c}{ X-ray luminosities are presented in {\bf columns 8--12}:  s = soft band (0.5--2 keV); h = hard band (2--8 keV); t = total band (0.5--8 keV).
Absorption-corrected luminosities are subscripted with a $c$; they are omitted when $\log N_H > 22.5$ since the soft band emission is essentially unmeasurable.
}

\end{deluxetable}

\clearpage
\pagestyle{empty}
\begin{deluxetable}{rcccrrrrrrrccccrrrc}
\centering \rotate
\tabletypesize{\tiny} \tablewidth{0pt}
\tablecolumns{19}

\tablecaption{Stellar Counterparts \label{tbl:counterparts}}
\tablehead{
\multicolumn{2}{c}{X-ray Source\tablenotemark{a}} &
  &
\multicolumn{16}{c}{Optical/Infrared Photometry} \\
\cline{1-2} \cline{4-19}
\colhead{Seq} & \colhead{CXOU J} & \colhead{} & \colhead{USNO B1.0} & \colhead{MJD95} & \colhead{PS02} &
\colhead{BC02} & \colhead{U} & \colhead{B} & \colhead{V} & \colhead{R} &
\colhead{I} & \colhead{H$\alpha$} & \colhead{2MASS}
& \colhead{FLAMINGOS} & \colhead{J} & \colhead{H} & \colhead{K} &
\colhead{PhCcFlg} \\
\colhead{\#} & \colhead{} & \colhead{} & \colhead{} & \colhead{} & \colhead{} & \colhead{} & \colhead{(mag)} & \colhead{(mag)} & \colhead{(mag)} &
\colhead{(mag)} & \colhead{(mag)} & \colhead{(mag)} &
\colhead{ID} & \colhead{ID} & \colhead{(mag)} & \colhead{(mag)} & \colhead{(mag)} & \colhead{}  \\
\colhead{(1)} & \colhead{(2)} & \colhead{} & \colhead{(3)} & \colhead{(4)} &
\colhead{(5)} & \colhead{(6)} & \colhead{(7)} &
\colhead{(8)} & \colhead{(9)} & \colhead{(10)} &\colhead{(11)} & \colhead{(12)} &
\colhead{(13)} & \colhead{(14)} & \colhead{(15)} &\colhead{(16)} & \colhead{(17)} & \colhead{(18)}
}
\startdata
   1 & 063114.36+045303.0 & & 0948-0096209 & \nodata & \nodata & \nodata &  \nodata & 19.33 &  \nodata & 17.15 & 17.37 &  \nodata & 06311429+0453032 & \nodata & 14.49 & 13.64 & 13.45 & AAA000 \\
   2 & 063117.04+045228.3 & & 0948-0096228 & \nodata & \nodata & \nodata &  \nodata & 20.75 &  \nodata & 18.32 & 16.59 &  \nodata & 06311696+0452281 & \nodata & 14.94 & 14.17 & 13.85 & AAA000 \\
   3 & 063118.11+045511.8 & & \nodata & \nodata & \nodata & \nodata &  \nodata &  \nodata &  \nodata &  \nodata &  \nodata &  \nodata & \nodata & \nodata &  \nodata &  \nodata &  \nodata & \nodata\nodata \\
   4 & 063118.29+045223.1 & & 0948-0096235 & \nodata & \nodata & \nodata &  \nodata & 19.47 &  \nodata & 16.79 & 15.67 &  \nodata & 06311825+0452234 & \nodata & 14.53 & 13.53 & 13.22 & AAA000 \\
   5 & 063118.76+045207.6 & & 0948-0096241 & 503 & \nodata & 14 & 13.36 & 13.36 & 12.90 & 12.38 & 12.91 & 12.57 & 06311881+0452089 & \nodata & 11.98 & 11.77 & 11.71 & AAA000 \\
   6 & 063119.27+045110.7 & & 0948-0096246 & \nodata & \nodata & \nodata &  \nodata & 19.39 &  \nodata & 16.84 & 15.52 &  \nodata & 06311925+0451121 & \nodata & 14.35 & 13.58 & 13.29 & AAA000 \\
   7 & 063120.48+045023.4 & & \nodata & \nodata & \nodata & \nodata &  \nodata &  \nodata &  \nodata &  \nodata &  \nodata &  \nodata & 06312048+0450239 & \nodata & 13.42 & 12.70 & 12.46 & AAAc00 \\
   8 & 063120.89+045003.8 & & 0948-0096261 & 454 & \nodata & 15 &  7.64 &  8.37 &  8.22 &  8.16 &  8.09 &  8.13 & 06312087+0450038 & 063117+050522 &  7.84 &  7.84 &  7.82 & AAA000 \\
   9 & 063120.95+045322.7 & & 0948-0096264 & \nodata & \nodata & \nodata &  \nodata & 19.50 &  \nodata & 17.52 & 16.43 &  \nodata & 06312100+0453238 & 063120+045323 & 14.93 & 14.14 & 13.97 & AAA000 \\
  10 & 063121.28+045023.8 & & \nodata & \nodata & \nodata & \nodata &  \nodata &  \nodata &  \nodata &  \nodata &  \nodata &  \nodata & \nodata & \nodata &  \nodata &  \nodata &  \nodata & \nodata\nodata \\
\enddata

\tablecomments{Table \ref{tbl:counterparts} with complete notes is
published in its entirety in the electronic edition of the {\it
Astrophysical Journal}.  A portion is shown here for guidance
regarding its form and content.}

\tablenotetext{a}{{\bf Columns 1--2} reproduce the sequence number and
source identification from Table~\ref{tbl:primary} and
Table~\ref{tbl:tentative}. {\bf Columns 3--6} are the catalogs used
for counterparts matching. For convenience, [MJD95]=Massey, Johnson,
\& Degioia-Eastwood (1995), [BC02]=Bergh{\"o}fer \& Christian\ (2002),
[PS02]=Park \& Sung (2002). {\bf Columns 7--12} give available optical
photometry. {\bf Columns 13--17} provide NIR identifications and $JHK$
photometry from FLAMINGOS (\S3.1). {\bf Column 18} lists the 2MASS
photometric quality flags (Cutri et al. 2003). In the note on
individual sources, [OI81]=Ogura \& Ishida (1981), [LR04]=Li \& Rector
(2004), ProbMem=Probability of being a cluster member from proper
motion data: [M82]=Marschall et al.\ (1982), [D06]=Dias et al.\
(2006).  }

\tablenotetext{      5}{ 2\arcsec from [OI81] 76=[BC02] 14 }  
\tablenotetext{      8}{ =HD 46056=BD+04 1291=MWC 808=[OI81] 84 (O8V); FUSE spectrum available}

\end{deluxetable}

\begin{deluxetable}{lcrrcccccrcccc}
\centering
\rotate
\tabletypesize{\tiny} \tablewidth{0pt}
\tablecolumns{13}
\tablecaption{X-ray properties of cataloged OB stars in NGC 2244 \label{tbl:OB.tab}
}
\tablehead{
\multicolumn{5}{c}{Optical/IR Properties} && \multicolumn{7}{c}{X-ray Properties} \\
\cline{1-5} \cline{7-13}
\colhead{Name} & \colhead{SpTy}  &
\colhead{2MASS} &  \colhead{K}  & \colhead{$\log \tilde{L_{bol}}$}
&&  \colhead{Seq} & \colhead{$\Delta \phi$} & \colhead{NetCts} &
\colhead{$\log N_H$} & \colhead{kT} & \colhead{$\log L_h$} & \colhead{$\log L_{t,c}$} \\

&&& \colhead{(mag)} & \colhead{(L$_\odot$)} && \colhead{\#} & \colhead{($\arcsec$)} &&
\colhead{(cm$^{-2}$)} & \colhead{(keV)} & \colhead{(erg s$^{-1}$)} &
\colhead{(erg s$^{-1}$)} \\
\colhead{(1)} & \colhead{(2)} & \colhead{(3)} & \colhead{(4)} & \colhead{(5)} &&
\colhead{(6)} & \colhead{(7)} & \colhead{(8)} &
\colhead{(9)} &\colhead{(10)} & \colhead{(11)} & \colhead{(12)}
}
\startdata
HD 46223 & O4V((f))  &   06320931+0449246               & 6.68  &  5.7  &&  615   & 0.2   &  1685   &  21.6 &   0.3  &  29.71 & 32.38 \\
HD 46150 & O5V((f))  &   06315551+0456343               & 6.44  &  5.5  &&  373   & 0.1   &  3589   &  21.4 &   0.6+0.2  &  30.48 & 32.34 \\
HD 46485 & O7V       &   06335094+0431316               & 7.45 &  5.2  &&  RMC 164  & 0.3   &   310  &  21.6 &   0.3+0.9  &  30.46 & 32.05 \\
HD 46056 & O8V((f))  &    06312087+0450038              & 7.82 &  5.0  &&  8   & 0.3   &   600  &  21.4 &   0.6  &  29.86 & 31.51 \\
HD 46149 & O8.5V((f))  &  06315253+0501591              & 7.25  &  4.9  &&  311   & 0.1   &   299  &  20.0 &   0.7  &  29.62 & 31.04 \\
HD 258691 & O9V((f))  &   06303331+0441276             & 7.93  &  4.8  &&  NFOV   & \nodata   &  \nodata   &  \nodata &   \nodata  &  \nodata & \nodata \\
HD 46202 & O9V((f))  &    06321047+0457597            & 7.72  &  4.8  &&  630   & 0.1   &   333  &  21.2 &   0.3  &  28.75 & 31.11 \\
HD 259238 & B0V  &     06321821+0503216           & 10.28  &  4.5  &&  727   & 1.0   &   36  &  21.8 &   0.5  &  28.74 & 30.65 \\
HD 46106 & B0.2V  &    06313839+0501363            & 7.62  &  4.4  &&  107   & 0.2   &   186  &  21.4 &   0.4+1.4  &  29.81 & 30.94 \\
MJD95 & B0.5V  &    06313708+0445537            & 12.20  &  4.3  &&  NFOV   &  \nodata & \nodata   & \nodata & \nodata   & \nodata & \nodata \\
HD 259135 & B0.5V  &   06320061+0452410           & 8.12  &  4.3  &&  476   & 0.7   &   186  &  20.6 &   1.6  &  30.04 & 30.57 \\
IRAS 06309+0450 & B0.5V  &  06333749+0448470      & 8.64  &  4.3  && NFOV    &  \nodata  & \nodata  & \nodata &  \nodata  &  \nodata & \nodata \\
HD 259012 & B1V  &   06313346+0450396             & 8.79  &  4.0  &&  66   & 0.6   &   275  &  21.1 &   2.5  &  30.75 & 31.12 \\
HD 259105 & B1V  &    06315200+0455573            & 8.95  &  4.0  &&  \nodata   & \nodata   &  \nodata  & \nodata &  \nodata  & $<$28.3 & $<$28.7 \\
BD+04$^{\circ}$1299s & B1III  &     06320613+0452153   & 9.38  &  4.0  &&  \nodata \tablenotemark{a}  &  \nodata   &   \nodata  &  \nodata &  \nodata  &  $<$28.3 & $<$28.7 \\
HD 46484 & B1V  &    06335441+0439446            & 6.86  &  4.0  &&  NFOV   &  \nodata  &  \nodata  &  \nodata &  \nodata  &  \nodata & \nodata \\
BD+05$^{\circ}$1281B & B1.5V  &   06315893+0455398  & 9.74  &  3.7  && 448    & 0.1   &   16  &  21.1 &   2.2  &  29.58 & 30.01 \\
HD 259172 & B2V  &   06320259+0505086            & 10.08  &  3.5  &&  NFOV   & \nodata   & \nodata &  \nodata &  \nodata  & \nodata & \nodata \\
OI81 345 & B2  &     06330656+0506034           & 11.20  &  3.5  &&  NFOV   & \nodata   &  \nodata  & \nodata &  \nodata  &  \nodata & \nodata \\
BD+04$^{\circ}$1295p & B2.5V  &   06313146+0450596  & 10.24  &  3.4  && 53  & 0.5   &   28  &  21.5 &   2.0  &  29.57 & 30.03 \\
OI81 130 & B2.5V  &   06314789+0454181             & 10.87  &  3.4  && \nodata    & \nodata   &  \nodata  &  \nodata &  \nodata  &  $<$28.3 & $<$28.7 \\
OI81 190 & B2.5Vn  &   06315891+0456162            & 10.54  &  3.4  && \nodata    & \nodata   & \nodata  & \nodata &  \nodata  & $<$28.3 & $<$28.7 \\
OI81 172 & B2.5V  &    06320984+0502134            & 10.43  &  3.4  && \nodata    & \nodata  &\nodata  & \nodata & \nodata  & $<$28.3 & $<$28.7\\
OI81 274 & B2.5V  &   06322424+0447037           & 10.58  &  3.4  &&  \nodata   & \nodata   &  \nodata  &  \nodata &   \nodata  &  $<$28.3 & $<$28.7 \\
OI81 392 & B2.5V  &   06335056+0501376           & 9.99  &  3.4  &&  NFOV   & \nodata   &  \nodata  &  \nodata &  \nodata  & \nodata  & \nodata \\
OI81 194 & B3  &    06321548+0455203            & 10.96  &  3.2  && 697    & 0.1   &   62  &  21.2 &   3.6  &  30.16 & 30.45 \\
MJD95 & B3V  &     06322249+0455342           & 13.48  &  3.2  &&  \nodata   & \nodata  & \nodata  & \nodata & \nodata  & $<$28.3 & $<$28.7 \\
HD 259268 & B3  &  06322304+0502457         & 10.33  &  3.2  &&  \nodata   & \nodata & \nodata  & \nodata &  \nodata  & $<$28.4 & $<$28.8\\
HD 259300 & B3Vp  &    06322939+0456560            & 9.34  &  3.2  &&  815   & 0.3   &   60  &  21.5 &   1.4  &  29.88 & 30.54 \\
OI81 334 & B3  &   06325179+0447161             & 11.53  &  3.2  &&  899   & 0.2   &   10  &  20.9 &   1.4  &  29.33 & 29.98 \\
MJD95 & B3V  &     06331016+0459499          & 12.39  &  3.2  &&  NFOV   & \nodata  & \nodata  &  \nodata &  \nodata  & \nodata & \nodata \\
\enddata

\tablecomments{ {\bf Column 1:} This list is obtained from Appendix A
of TFM03 which gives optical cross-identifications, positions, and
spectral types. The stars are listed first in order of decreasing
mass, and then by right ascension. OI=~\citet{OI81}.  {\bf Column 2:}
Spectral types are from Ogura \& Ishida (1981) and Massey et
al.(1995).\\ {\bf Columns 3--4:} Source numbers and $K$-band
magnitudes are from the 2MASS All-sky Point Sources Catalog.\\ {\bf Column
5:} Bolometric luminosities are estimated from calibrations of
$L_{bol}$ with spectral type, \citet{Martins05} for O3--O9.5 stars and
\citet{deJager87} for B stars.  No use is made of available
photometry. \\ {\bf Column 6:} {\em Chandra} source number, from
Table~\ref{tbl:primary}. NFOV=object is not covered in the
FOV. ...=non detection. HD 46485 is observed in Rosette Field 4 (RMC
source \#164, Paper II).\\ {\bf Column 7:} Offset (in arcseconds)
between the {\em Chandra} and 2MASS sources.  \\ {\bf Columns 8--12:}
X-ray properties from Table~\ref{tbl:thermal_spectroscopy}: extracted
counts after background subtraction; column density and plasma energy
from fits to the ACIS spectra (: in $\log N_H$ values are approximated
from median energy \citep{Feigelson05}; : in $kT$ are assumed);
observed hard band luminosity ($2-8$ keV); inferred total band
luminosity corrected for absorption ($0.5-8$~keV).  Upper limits to
luminosities are estimated using the faintest sources in
Table~\ref{tbl:thermal_spectroscopy} and scaled with the corresponding
exposure time.}

\tablenotetext{a}{BD+04$^{\circ}$1299s was reported as a detection in
TFM03 (20ks observation). However the deep observation resolved this
source into two X-ray sources. Both are separated by $\sim$2\arcsec
from the optical position. Therefore we do not report
BD+04$^{\circ}$1299s as a detection here.}

\end{deluxetable}

\clearpage
\pagestyle{plaintop}
\clearpage
\centerline{\includegraphics[width=0.62\textwidth]{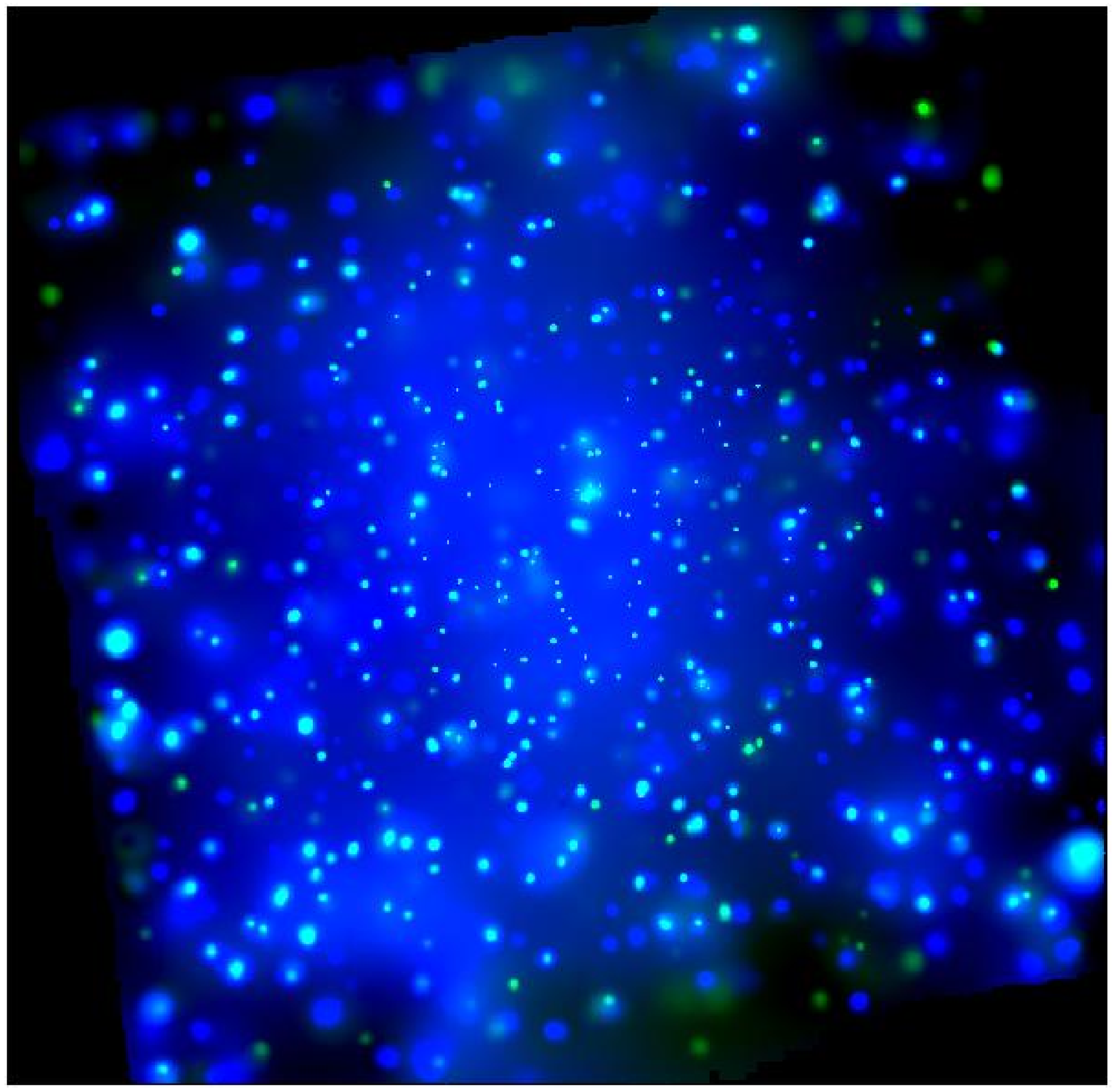}}
\centerline{\includegraphics[width=0.62\textwidth]{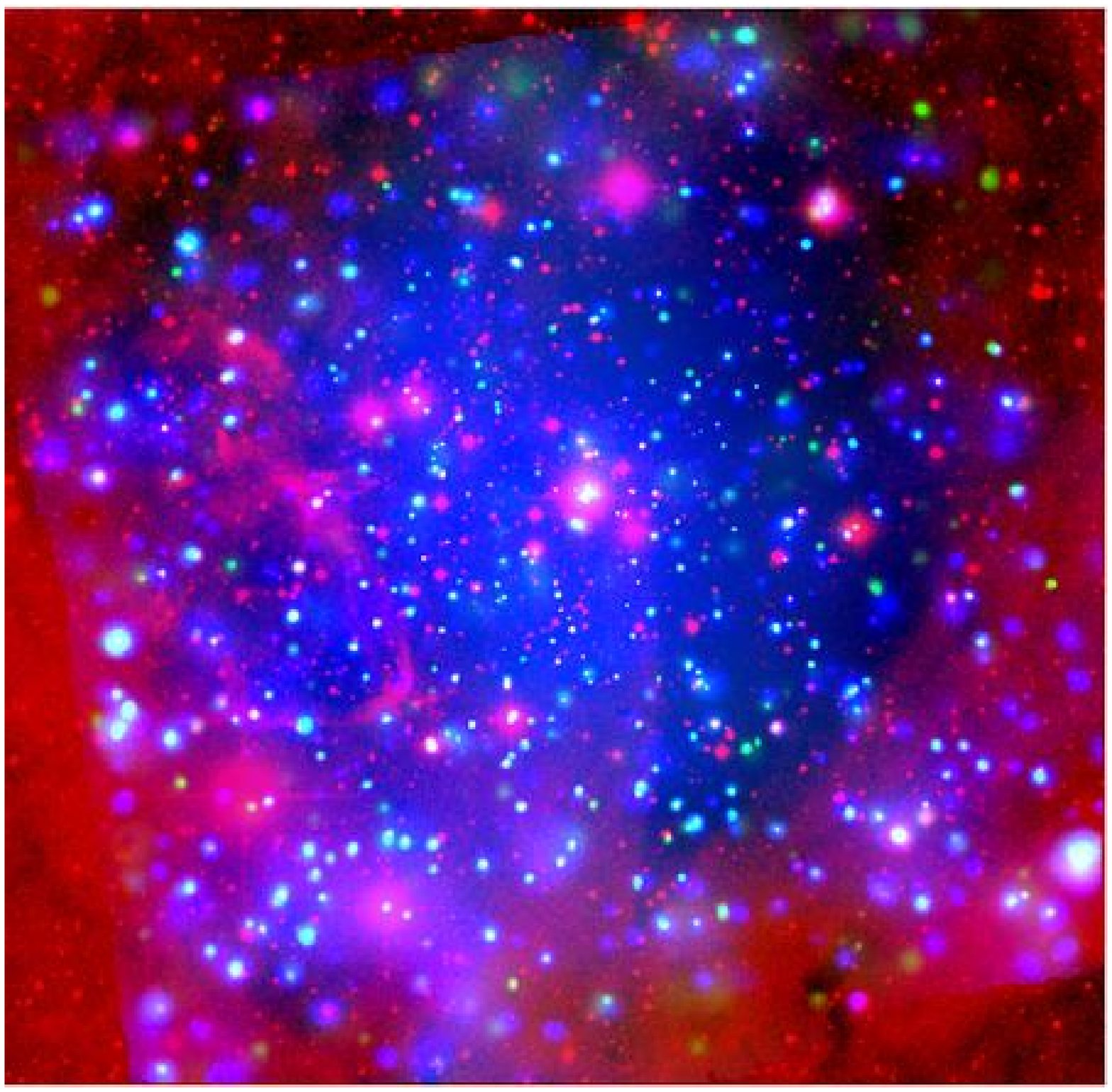}}
\centerline{Fig. 2. --- Continued.}
\clearpage
\clearpage

\clearpage

\begin{figure}
\centering
\plotone{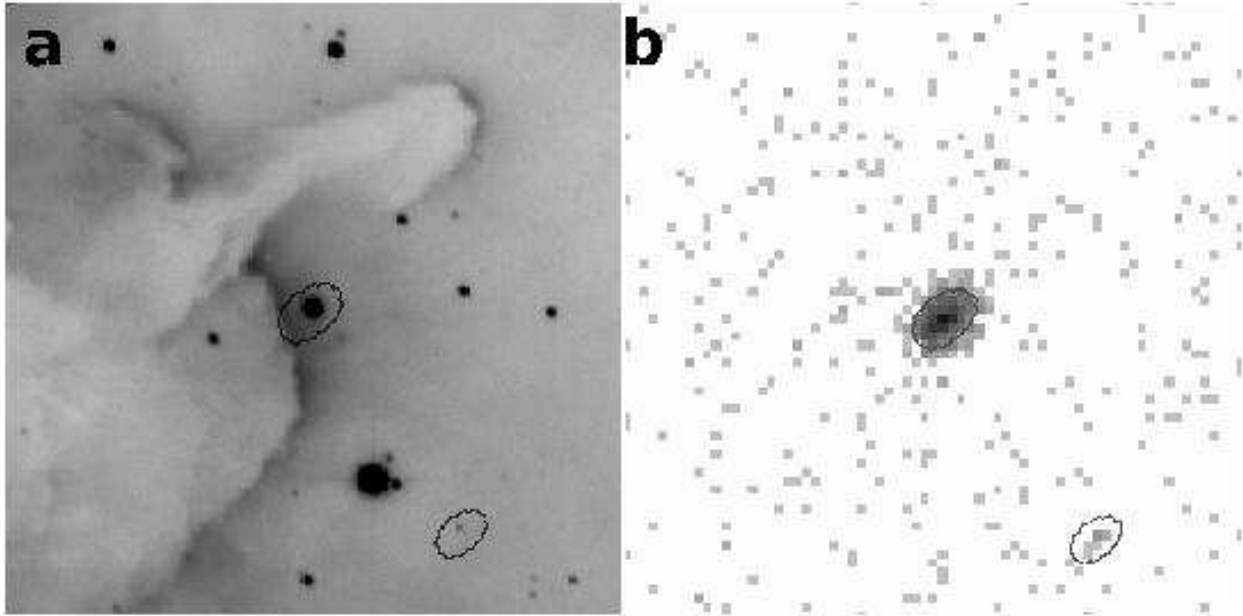}
\includegraphics[width=0.72\textwidth,angle=-90]{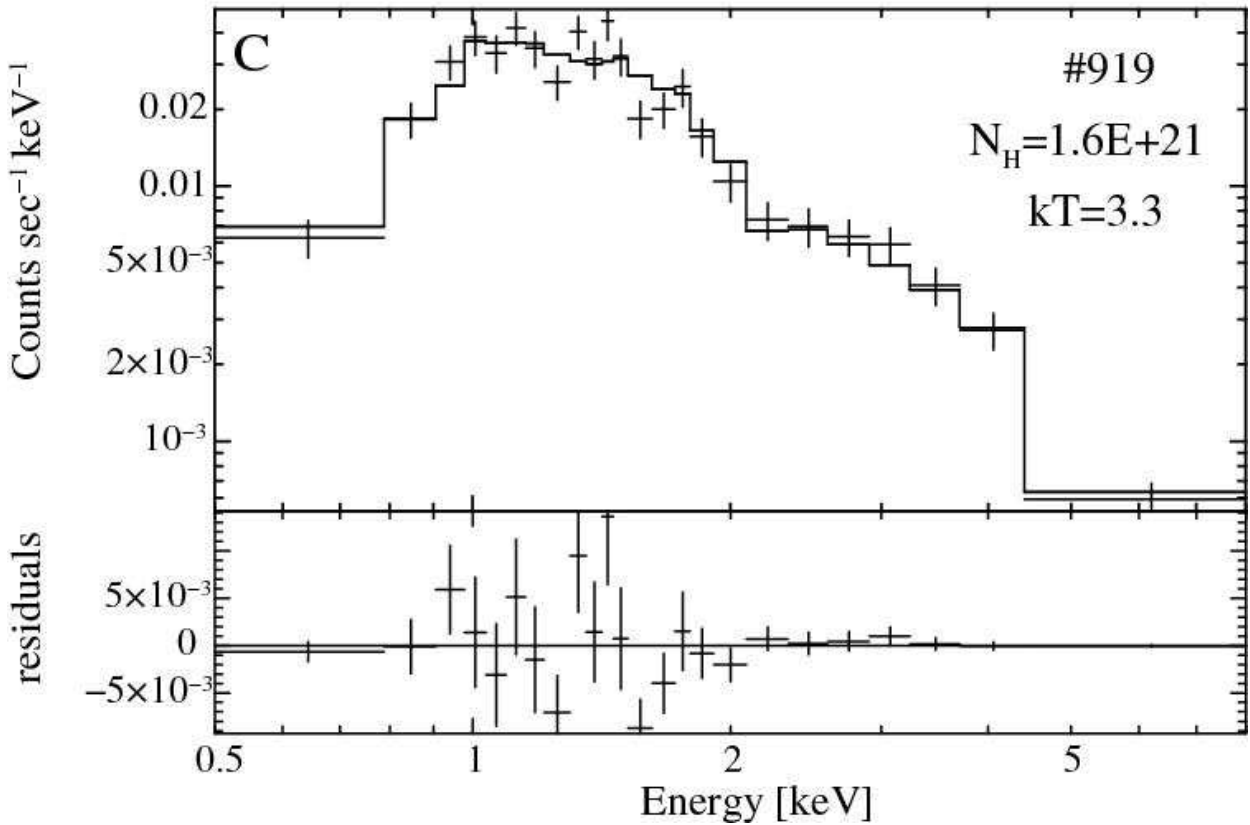}
\caption{(a): $H_{\alpha}$ image of the neighborhood of ACIS source \#919, near a
molecular pillar. (b): The X-ray image of the same region. The
2.5\arcmin $\times$ 2.5\arcmin\ images are both centered at the X-ray
bright star \#919. (c): The spectral fit to the X-ray spectrum of
source \#919 with $\log N_H=21.2$ and a hard ($kT=3.3$~keV)
plasma.\label{fig:919}}
\end{figure}
\begin{figure}
\centering
\plotone{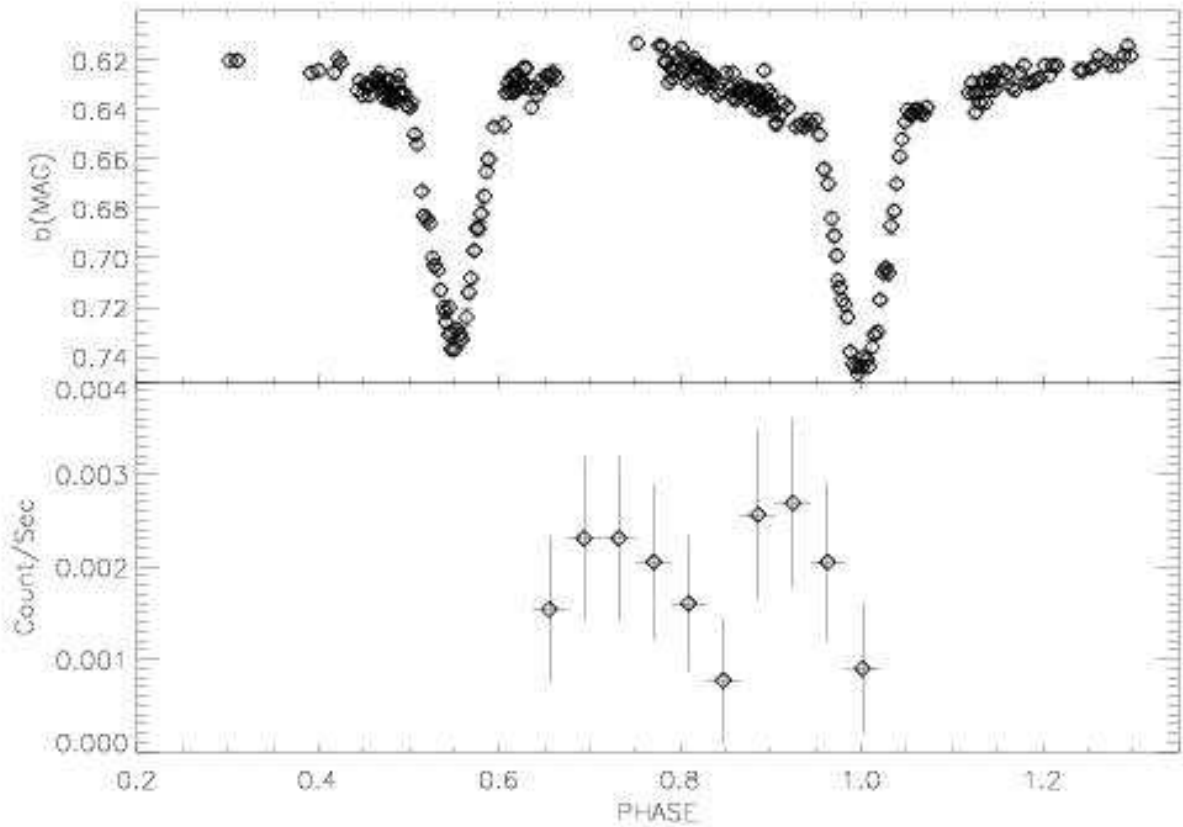}
\caption{X-ray light curve of the eclipsing binary V578 Mon (lower
panel), together with the optical light curve ($b$-band photometry
with 1$\sigma$ error-bars taken from Hensberge et al. 2000) in orbital
phases (upper panel). The orbital phase is calculated using the eclipse
ephemeris reported in \citet{Hensberge00}. \label{fig:eclipse}}
\end{figure}

\begin{thebibliography}{}

\bibitem[Adams et al.(2001)]{Adams01} Adams, J.~D., Stauffer, 
J.~R., Monet, D.~G., Skrutskie, M.~F., \& Beichman, C.~A.\ 2001, \aj, 121, 
2053 

\bibitem[Albacete Colombo et al.(2007)]{Albacete-Colombo07} Albacete
Colombo, J.~F., Flaccomio, E., Micela, G., Sciortino, S., \& Damiani,
F.\ 2007, \aap, 464, 211

\bibitem[Arnaud(1996)]{Arnaud96}
Arnaud, K.~A.\ 1996, ASP Conf.~Ser.~101: Astronomical Data
Analysis Software and Systems V, 5, 17

\bibitem[Babel \& Montmerle(1997a)]{Babel97a} 
Babel, J., \& Montmerle, T.\ 1997, \aap, 323, 121 

\bibitem[Babel \& Montmerle(1997b)]{Babel97b}
Babel, J., \& Montmerle, T.\ 1997, \apjl, 485, L29 

\bibitem[Bagnulo et al.(2004)]{Bagnulo04} Bagnulo, S., Hensberge,
H., Landstreet, J.~D., Szeifert, T., \& Wade, G.~A.\ 2004, \aap, 416, 1149

\bibitem[Balog et al.(2006)]{Balog06} Balog, Z., Rieke, G.~H., 
Su, K.~Y.~L., Muzerolle, J., \& Young, E.~T.\ 2006, \apjl, 650, L83 

\bibitem[Balog et al.(2007)]{Balog07} Balog, Z., Muzerolle, J., Rieke,
G.~H., Su, K.~Y.~L., \& Young, E.~T.\ 2007, ApJ in press,
astro-ph/0701741

\bibitem[Bergh{\"o}fer \& Schmitt(1994)]{Berghoefer94} Bergh{\"o}fer, T.~W., 
\& Schmitt, J.~H.~M.~M.\ 1994, \aap, 292, L5 

\bibitem[Bergh{\"o}fer et al.(1996)]{Berghoefer96} Bergh{\"o}fer, T.~W., 
Schmitt, J.~H.~M.~M., \& Cassinelli, J.~P.\ 1996, \aaps, 118, 481 

\bibitem[Bergh{\"o}fer \& Christian(2002)]{BC02}
Bergh{\"o}fer, T.~W., \& Christian, D.~J.\ 2002, \aap, 384, 890

\bibitem[Bergh{\"o}fer et al.(1997)]{Berghoefer97}
Bergh{\"o}fer, T.~W., Schmitt, J.~H.~M.~M., Danner, R., \& Cassinelli, J.~P.\ 1997, \aap, 322, 167

\bibitem[Binney \& Tremaine(1987)]{Binney87} Binney, J., \& 
Tremaine, S.\ 1987, Galactic Dynamics, Princeton University Press, 1987

\bibitem[Blitz \& Thaddeus(1980)]{Blitz80} Blitz, L., \&
Thaddeus, P.\ 1980, \apj, 241, 676

\bibitem[Bonatto et al.(2006)]{Bonatto06} Bonatto, C., Santos, 
J.~F.~C., Jr., \& Bica, E.\ 2006, \aap, 445, 567

\bibitem[Bonnell \& Davies(1998)]{Bonnell98} Bonnell, I.~A., \& 
Davies, M.~B.\ 1998, \mnras, 295, 691 

\bibitem[Borra \& Landstreet(1978)]{Borra78} Borra, E.~F., \& 
Landstreet, J.~D.\ 1978, \apj, 222, 226 

\bibitem[Brandt et al.(2001)]{Brandt01}
Brandt, W.~N., et al.\ 2001, \aj, 122, 2810

\bibitem[Broos et al.(2002)]{Broos02} Broos, P.~S., Townsley, L.~K.,
Getman, K.~V., \& Bauer, F.~E. 2002, ACIS Extract, An ACIS Point
Source Extraction Package (University Park: Pennsylvania State Univ.)

\bibitem[Broos et al.(2007)]{Broos07} Broos, P.~S., Feigelson, E.~D.,
Townsley, L.~K., Getman, K.~V., Wang, J.~F., Garmire, G.~P., Jiang,
Z., \& Tsuboi, Y.\ 2007, \apjs, 169, 353

\bibitem[Burningham et al.(2005)]{Burningham05} Burningham, B., 
Naylor, T., Littlefair, S.~P., \& Jeffries, R.~D.\ 2005, \mnras, 363, 1389 

\bibitem[Cash(1979)]{Cash79} Cash, W.\ 1979, \apj, 228, 939

\bibitem[Carpenter et al.(1997)]{Carpenter97} Carpenter, J.~M., 
Meyer, M.~R., Dougados, C., Strom, S.~E., \& Hillenbrand, L.~A.\ 1997, \aj, 
114, 198

\bibitem[Chen et al.(2004)]{Chen04} Chen, W.~P., Chiang,
P.~S., \& Li, J.~Z.\ 2004, Chinese Journal of Astronony and Astrophysics,
4, 153

\bibitem[Celnik(1985)]{Celnik85} Celnik, W.~E.\ 1985, \aap, 144, 171

\bibitem[Celnik(1986)]{Celnik86} Celnik, W.~E.\ 1986, \aap, 160, 287 

\bibitem[Chen et al.(2007)]{Chen07} Chen, L., de Grijs, R., \& 
Zhao, J.~L.\ 2007, \aj, 134, 1368 

\bibitem[Chini et al.(2004)]{Chini04} Chini, R., Hoffmeister, 
V., Kimeswenger, S., Nielbock, M., N{\"u}rnberger, D., Schmidtobreick, L., 
\& Sterzik, M.\ 2004, \nat, 429, 155 

\bibitem[Chini et al.(2005)]{Chini05} Chini, R., Hoffmeister, 
V.~H., Nielbock, M., Scheyda, C.~M., N{\"u}rnberger, D., Feigelson, E.~D., 
Getman, K., \& Townsley, L.~K.\ 2005, Massive Star Birth: A Crossroads of 
Astrophysics, 227, 145 

\bibitem[Chlebowski et al.(1989)]{Chlebowski89}
Chlebowski, T., Harnden, F.~R., \& Sciortino, S.\ 1989, \apj, 341, 427

\bibitem[Clarke et al.(2000)]{Clarke00} Clarke, C.~J., Bonnell, 
I.~A., \& Hillenbrand, L.~A.\ 2000, Protostars and Planets IV, 151 

\bibitem[Comeron \& Pasquali(2007)]{Comeron07} Comeron, F., \& 
Pasquali, A.\ 2007, astro-ph/0704.0676

\bibitem[Corcoran et al.(1994)]{Corcoran94}
Corcoran, M.~F., et al.\ 1994, \apjl, 436, L95

\bibitem[Cox et al.(1990)]{Cox90}
Cox, P., Deharveng, L., \& Leene, A.\ 1990, \aap, 230, 181

\bibitem[Cutri et al.(2003)]{Cutri03} Cutri, R.~M., et al.\
2003, The IRSA 2MASS All-Sky Point Source Catalog, NASA/IPAC Infrared
Science Archive.~http://irsa.ipac.caltech.edu/applications/Gator/

\bibitem[Damiani et al.(1994)]{Damiani94} Damiani, F., Micela,
G., Sciortino, S., \& Harnden, F.~R., Jr.\ 1994, \apj, 436, 807

\bibitem[de Jager \& Nieuwenhuijzen(1987)]{deJager87} de Jager, 
C., \& Nieuwenhuijzen, H.\ 1987, \aap, 177, 217 

\bibitem[Dias et al.(2006)]{Dias06}
Dias, W.~S., Assafin, M., Fl{\'o}rio, V., Alessi, B.~S., \& L{\'{\i}}bero, V.\ 2006, \aap, 446, 949

\bibitem[Dickey \& Lockman(1990)]{Dickey90}
Dickey, J.~M., \& Lockman, F.~J.\ 1990, \araa, 28, 215

\bibitem[Ebeling et al.(2006)]{Ebeling06}
Ebeling, H., White, D.~A., \& Rangarajan, F.~V.~N.\ 2006, \mnras, 368, 65

\bibitem[Elston et al.(2003)]{Elston03} Elston, R., Raines, 
S.~N., Hanna, K.~T., Hon, D.~B., Julian, J., Horrobin, M., Harmer, 
C.~F.~W., \& Epps, H.~W.\ 2003, \procspie, 4841, 1611

\bibitem[Evans et al.(2004)]{Evans04}
Evans, N.~R., Schlegel, E.~M., Waldron, W.~L., Seward, F.~D., Krauss, M.~I., Nichols, J., \& Wolk, S.~J.\ 2004, \apj, 612, 1065

\bibitem[Ezoe et al.(2006)]{Ezoe06} Ezoe, Y., Kokubun, M.,
Makishima, K., Sekimoto, Y., \& Matsuzaki, K.\ 2006, \apj, 638, 860

\bibitem[Favata \& Micela(2003)]{Favata03}
Favata, F., \& Micela, G.\ 2003, Space Science Reviews, 108, 577

\bibitem[Favata et al.(2005)]{Favata05}
Favata, F., Flaccomio, E., Reale, F., Micela, G., Sciortino, S., Shang, H., Stassun, K.~G., \&
Feigelson, E.~D.\ 2005, \apjs, 160, 469

\bibitem[Feigelson \& Montmerle(1999)]{Feigelson99} Feigelson,
E.~D., \& Montmerle, T.\ 1999, \araa, 37, 363

\bibitem[Feigelson et al.(2002)]{Feigelson02} Feigelson, E.~D., 
Broos, P., Gaffney, J.~A., III, Garmire, G., Hillenbrand, L.~A., Pravdo, 
S.~H., Townsley, L., \& Tsuboi, Y.\ 2002, \apj, 574, 258 

\bibitem[Feigelson et al.(2005)]{Feigelson05}
Feigelson, E.~D., et al.\ 2005, \apjs, 160, 379

\bibitem[Feigelson et al.(2007)]{Feigelson07} Feigelson, E., Townsley,
L., Gudel, M., \& Stassun, K.\ 2007, Protostars \& Planets V,
eds. Reipurth et al., University of Arizona Press, Tucson, 313

\bibitem[Freeman et al.(2002)]{Freeman02}
Freeman, P.~E., Kashyap, V., Rosner, R., \& Lamb, D.~Q.\ 2002, \apjs, 138, 185

\bibitem[Froebrich et al.(2005)]{Froebrich05}
Froebrich, D., Scholz, A., Eisl{\"o}ffel, J., \& Murphy, G.~C.\ 2005, \aap, 432, 575

\bibitem[Gagn{\'e} et al.(2005)]{Gagne05} 
Gagn{\'e}, M., Oksala, M.~E., Cohen, D.~H., Tonnesen, S.~K., ud-Doula, A., Owocki, S.~P., Townsend, R.~H.~D., \& MacFarlane, J.~J.\ 2005, \apj, 628, 986 

\bibitem[Getman et al.(2005a)]{Getman05a}
Getman, K.~V., Feigelson, E.~D., Grosso, N., McCaughrean, M.~J., Micela, G., Broos, P., Garmire, G., \& Townsley, L.\ 2005a, \apjs, 160, 353

\bibitem[Getman et al.(2005b)]{Getman05b} Getman, K.~V., et al.\
2005b, \apjs, 160, 319

\bibitem[Getman et al.(2006)]{Getman06}
Getman, K.~V., et~al., 2006, \apjs, 163, 306

\bibitem[Getman et al.(2007)]{Getman07} Getman, K.~V., 
Feigelson, E.~D., Garmire, G., Broos, P., \& Wang, J.\ 2007, \apj, 654, 316 

\bibitem[Gregorio-Hetem et al.(1998)]{Gregorio98} Gregorio-Hetem, J.,
Montmerle, T., Casanova, S., \& Feigelson, E.~D.\ 1998, \aap, 331, 193

\bibitem[Grosso et al.(2004)]{Grosso04}
Grosso, N., Montmerle, T., Feigelson, E.~D., \& Forbes, T.~G.\ 2004, \aap, 419, 653

\bibitem[G{\" u}del et al.(1998)]{Gudel98}
G{\" u}del, M., Guinan, E.~F., \& Skinner, S.~L.\ 1998, ASP Conf.~Ser.~154: Cool Stars, Stellar Systems, and the Sun, 154, 1041

\bibitem[Guarcello et al.(2007)]{Guarcello07} Guarcello, M.~G.,
Prisinzano, L., Micela, G., Damiani, F., Peres, G., \& Sciortino, S.\
2007, \aap, 462, 245

\bibitem[Haisch et al.(2001a)]{Haisch01a} Haisch, K.~E., Jr., 
Lada, E.~A., Pi{\~n}a, R.~K., Telesco, C.~M., \& Lada, C.~J.\ 2001, \aj, 
121, 1512 

\bibitem[Haisch et al.(2001b)]{Haisch01b} Haisch, K.~E., Jr., Lada,
E.~A., \& Lada, C.~J.\ 2001, \aj, 121, 2065

\bibitem[Hamaguchi et al.(2005)]{Hamaguchi05} Hamaguchi, K.,
Yamauchi, S., \& Koyama, K.\ 2005, \apj, 618, 360

\bibitem[Harnden et al.(1979)]{Harnden79} Harnden, F.~R., Jr., et 
al.\ 1979, \apjl, 234, L51 

\bibitem[Harries \& Hilditch(1998)]{Harries98} Harries, T.~J., \& 
Hilditch, R.~W.\ 1998, ASP Conf.~Ser.~131: Properties of Hot Luminous 
Stars, 131, 401 

\bibitem[Hartmann \& Burkert(2007)]{Hartmann07} Hartmann, L., \& 
Burkert, A.\ 2007, \apj, 654, 988 

\bibitem[Hensberge et al.(2000)]{Hensberge00} Hensberge, H.,
Pavlovski, K., \& Verschueren, W.\ 2000, \aap, 358, 553

\bibitem[Hillenbrand et al.(1993)]{Hillenbrand93} Hillenbrand, L.~A., 
Massey, P., Strom, S.~E., \& Merrill, K.~M.\ 1993, \aj, 106, 1906 

\bibitem[Hillenbrand \& Hartmann(1998)]{Hillenbrand98} Hillenbrand, 
L.~A., \& Hartmann, L.~W.\ 1998, \apj, 492, 540 

\bibitem[Heyer et al.(2006)]{Heyer06} Heyer, M.~H., Williams, 
J.~P., \& Brunt, C.~M.\ 2006, \apj, 643, 956 

\bibitem[Hollenbach et al.(2000)]{Hollenbach00} Hollenbach, D.~J., 
Yorke, H.~W., \& Johnstone, D.\ 2000, Protostars and Planets IV, 401 

\bibitem[H{\"u}nsch et al.(1999)]{Huensch99} H{\"u}nsch, M.,
Schmitt, J.~H.~M.~M., Sterzik, M.~F., \& Voges, W.\ 1999, \aaps, 135, 319

\bibitem[Isobe et al.(1986)]{Isobe86} Isobe, T., Feigelson, E.~D., \&
Nelson, P.~I.\ 1986, \apj, 306, 490

\bibitem[Imanishi et al.(2001)]{Imanishi01}
Imanishi, K., Koyama, K., \& Tsuboi, Y.\ 2001, \apj, 557, 747

\bibitem[Jiang et al.(2002)]{Jiang02} Jiang, Z., et al.\ 2002, 
\apj, 577, 245 

\bibitem[Johnstone et al.(1998)]{Johnstone98} Johnstone, D.,
Hollenbach, D., \& Bally, J.\ 1998, \apj, 499, 758

\bibitem[Kenyon et al.(1993)]{Kenyon93} Kenyon, S.~J., Calvet, 
N., \& Hartmann, L.\ 1993, \apj, 414, 676 

\bibitem[King(1962)]{King62} King, I.\ 1962, \aj, 67, 471 

\bibitem[Kroupa(2002)]{Kroupa02} Kroupa, P.\ 2002, Science, 295, 
82 

\bibitem[Kroupa(2004)]{Kroupa04} Kroupa, P.\ 2004, New Astronomy 
Review, 48, 47 

\bibitem[Kharchenko et al.(2005)]{Kharchenko05} Kharchenko, N.~V.,
Piskunov, A.~E., R{\"o}ser, S., Schilbach, E., \& Scholz, R.-D.\ 2005,
\aap, 438, 1163

\bibitem[Kouwenhoven et al.(2007)]{Kouwenhoven07} Kouwenhoven, 
M.~B.~N., Brown, A.~G.~A., \& Kaper, L.\ 2007, \aap, 464, 581 

\bibitem[Lada(1991)]{Lada91} Lada, C.~J.\ 1991, NATO ASIC 
Proc.~342: The Physics of Star Formation and Early Stellar Evolution, 329 

\bibitem[Lada(1992)]{Lada92} Lada, E.~A.\ 1992, \apjl, 393, L25


\bibitem[Lada \& Lada(1995)]{Lada95} Lada, E.~A., \& Lada, C.~J.\ 1995, \aj, 109, 1682

\bibitem[Lada et al.(2000)]{Lada00} Lada, C.~J., Muench, A.~A.,
Haisch, K.~E., Jr., Lada, E.~A., Alves, J.~F., Tollestrup, E.~V., \&
Willner, S.~P.\ 2000, \aj, 120, 3162

\bibitem[Lada \& Lada(2003)]{Lada03} Lada, C.~J., \& Lada, 
E.~A.\ 2003, \araa, 41, 57 

\bibitem[Lada et al.(2004)]{Lada04} Lada, C.~J., Muench, 
A.~A., Lada, E.~A., \& Alves, J.~F.\ 2004, \aj, 128, 1254 

\bibitem[Lada et al.(2006)]{Lada06} Lada, C.~J., et al.\ 2006, 
\aj, 131, 1574 

\bibitem[Leahy(1985)]{Leahy85} Leahy, D.~A.\ 1985, \mnras, 217, 69 

\bibitem[Li et al.(2002)]{Li02} Li, J.~Z., Wu, C.~H., Chen, W.~P.,
Rector, T., Chu, Y.~H., \& Ip, W.~H.\ 2002, \aj, 123, 2590

\bibitem[Li(2003)]{Li03} Li, J.-Z.\ 2003, Chinese Journal of
Astronomy and Astrophysics, 3, 495

\bibitem[Li \& Rector(2004)]{Li04} Li, J.~Z., \& Rector,
T.~A.\ 2004, \apjl, 600, L67

\bibitem[Li(2005)]{Li05} Li, J.~Z.\ 2005, \apj, 625, 242

\bibitem[Li \& Smith(2005a)]{Li05a} Li, J.~Z., \& Smith, M.\ 
2005, \aj, 130, 721 

\bibitem[Li \& Smith(2005b)]{Li05b} Li, J.~Z., \& Smith, 
M.~D.\ 2005, \apj, 620, 816 

\bibitem[Li et al.(2006)]{Li06} Li, J.~Z., Chu, Y.~H., \&
Gruendl, R.~A.\ 2006, astro-ph/0602456

\bibitem[Linsky et al.(2007)]{Linsky07} Linsky, J.~L., 
Gagn{\'e}, M., Mytyk, A., McCaughrean, M., \& Andersen, M.\ 2007, \apj, 
654, 347 

\bibitem[Lucy(1974)]{Lucy74}
Lucy, L.~B.\ 1974, \aj, 79, 745

\bibitem[Lucy \& White(1980)]{Lucy80}
Lucy, L.~B., \& White, R.~L.\ 1980, \apj, 241, 300

\bibitem[Marschall et al.(1982)]{Marschall82}
Marschall, L.~A., van Altena, W.~F., \& Chiu, L.-T.~G.\ 1982, \aj, 87, 1497

\bibitem[Martins et al.(2005)]{Martins05} Martins, F., Schaerer, 
D., \& Hillier, D.~J.\ 2005, SF2A-2005: Semaine de l'Astrophysique Francaise, 633 

\bibitem[Massey et al.(1995)]{MJD95} Massey, P., Johnson,
K.~E., \& Degioia-Eastwood, K.\ 1995, \apj, 454, 151

\bibitem[Mathews(1966)]{Mathews66} Mathews, W.~G.\ 1966, \apj,
144, 206

\bibitem[McMillan et al.(2007)]{McMillan07} McMillan, S.~L.~W., 
Vesperini, E., \& Portegies Zwart, S.~F.\ 2007, \apjl, 655, L45 

\bibitem[Meaburn \& Walsh(1986)]{Meaburn86} Meaburn, J., \& Walsh,
J.~R.\ 1986, \mnras, 220, 745

\bibitem[Meaburn et al.(2005)]{Meaburn05}
Meaburn, J., L{\'o}pez, J.~A., Richer, M.~G., Riesgo, H., \& Dyson, J.~E.\ 2005, \aj, 130, 730

\bibitem[Monet et al.(2003)]{Monet03}
Monet, D.~G., et al.\ 2003, \aj, 125, 984

\bibitem[Moretti et al.(2003)]{Moretti03}
Moretti, A., Campana, S., Lazzati, D., \& Tagliaferri, G.\ 2003, \apj, 588, 696

\bibitem[Mori et al.(2001)]{Mori01}
Mori, K., Tsunemi, H., Miyata, E.,
Baluta, C.~J., Burrows, D.~N., Garmire, G.~P., \& Chartas, G.\ 2001,
ASP Conf.~Ser.~251: New Century of X-ray Astronomy, 251, 576

\bibitem[Muench et al.(2002)]{Muench02} Muench, A.~A., Lada, 
E.~A., Lada, C.~J., \& Alves, J.\ 2002, \apj, 573, 366 

\bibitem[Muench et al.(2007)]{Muench07} Muench, A.~A., Lada, 
C.~J., Luhman, K.~L., Muzerolle, J., \& Young, E.\ 2007, \aj, 134, 411 

\bibitem[Muno et al.(2006)]{Muno06} Muno, M.~P., Bauer, F.~E., 
Bandyopadhyay, R.~M., \& Wang, Q.~D.\ 2006, \apjs, 165, 173 

\bibitem[Muzerolle et al.(1998)]{Muzerolle98} Muzerolle, J.,
Hartmann, L., \& Calvet, N.\ 1998, \aj, 116, 455

\bibitem[Muzerolle et al.(2001)]{Muzerolle01} Muzerolle, J.,
Calvet, N., \& Hartmann, L.\ 2001, \apj, 550, 944

\bibitem[Natta et al.(2000)]{Natta00} Natta, A., Grinin, V., \& 
Mannings, V.\ 2000, Protostars and Planets IV, 559 

\bibitem[Nielbock et al.(2001)]{Nielbock01} Nielbock, M., Chini, 
R., J{\"u}tte, M., \& Manthey, E.\ 2001, \aap, 377, 273 

\bibitem[O'Dell \& Wong(1996)]{Odell96} O'Dell, C.~R., \& Wong, 
K.\ 1996, \aj, 111, 846 

\bibitem[Ogura \& Ishida(1981)]{OI81} Ogura, K., \& Ishida, 
K.\ 1981, \pasj, 33, 149 

\bibitem[Owocki et al.(1988)]{Owocki88}
Owocki, S.~P., Castor, J.~I., \& Rybicki, G.~B.\ 1988, \apj, 335, 914

\bibitem[Owocki \& Cohen(1999)]{Owocki99}
Owocki, S.~P., \& Cohen, D.~H.\ 1999, \apj, 520, 833

\bibitem[Pallavicini et al.(1981)]{Pallavicini81} Pallavicini, R., 
Golub, L., Rosner, R., Vaiana, G.~S., Ayres, T., \& Linsky, J.~L.\ 1981, 
\apj, 248, 279 

\bibitem[Park \& Sung(2002)]{PS02} Park, B.-G., \& Sung, H.\
2002, \aj, 123, 892

\bibitem[Perez(1991)]{Perez91}
Perez, M.~R.\ 1991, Revista Mexicana de Astronomia y Astrofisica, 22, 99

\bibitem[Pflamm-Altenburg \& Kroupa(2006)]{Pflamm06} 
Pflamm-Altenburg, J., \& Kroupa, P.\ 2006, \mnras, 373, 295 

\bibitem[Phelps \& Lada(1997)]{Phelps97} Phelps, R.~L., \& Lada,
E.~A.\ 1997, \apj, 477, 176

\bibitem[Pizzolato et al.(2000)]{Pizzolato00} Pizzolato, N.,
Maggio, A., \& Sciortino, S.\ 2000, \aap, 361, 614

\bibitem[Pollock et al.(2005)]{Pollock05} Pollock, A.~M.~T., 
Corcoran, M.~F., Stevens, I.~R., \& Williams, P.~M.\ 2005, \apj, 629, 482 

\bibitem[Pozzo et al.(2000)]{Pozzo00} Pozzo, M., Jeffries,
R.~D., Naylor, T., Totten, E.~J., Harmer, S., \& Kenyon, M.\ 2000, \mnras,
313, L23

\bibitem[Preibisch \& Feigelson(2005)]{Preibisch05} Preibisch, T., 
\& Feigelson, E.~D.\ 2005, \apjs, 160, 390 

\bibitem[Preibisch et al.(2005)]{Preibisch05a} Preibisch, T., et
al.\ 2005, \apjs, 160, 401

\bibitem[Rauw et al.(2002)]{Rauw02} Rauw, G., Naz{\'e}, Y., 
Gosset, E., Stevens, I.~R., Blomme, R., Corcoran, M.~F., Pittard, J.~M., \& 
Runacres, M.~C.\ 2002, \aap, 395, 499 

\bibitem[Rho et al.(2004)]{Rho04} Rho, J., Ram{\'{\i}}rez, 
S.~V., Corcoran, M.~F., Hamaguchi, K., \& Lefloch, B.\ 2004, \apj, 607, 904 

\bibitem[Robin et al.(2003)]{Robin03}
Robin, A.~C., Reyl{\'e}, C., Derri{\`e}re, S., \& Picaud, S.\ 2003, \aap, 409, 523

\bibitem[Roeser \& Bastian(1988)]{Roeser88} Roeser, S., \& 
Bastian, U.\ 1988, \aaps, 74, 449 

\bibitem[Rom\'an-Z\'u\~niga et al.(2005)]{Roman05}
Rom{\'a}n-Z{\'u}{\~n}iga, C.~G., Lada, E.~A., \& Williams, J.~P.\ 2005,
Revista Mexicana de Astronomia y Astrofisica Conference Series, 24, 66

\bibitem[Rom\'an-Z\'u\~niga(2006)]{Roman06}
Rom{\'a}n-Z{\'u}{\~n}iga, C.~G., PhD Thesis, University of Florida 2006

\bibitem[Rom\'an-Z\'u\~niga \& Lada (2007)]{RL07} Rom\'an-Z\'u\~niga,
C. \& Lada, E., 2007, Star Formation Handbook, eds. Reipurth, B. (RL07)

\bibitem[Rom\'an-Z\'u\~niga et al.(2007)]{Roman07} Rom\'an-Z\'u\~niga,
C., Elston R., Ferreira, B., \& Lada, E.,\ 2007, ApJ in press,
astro-ph/0709.3004

\bibitem[Sana et al.(2006)]{Sana06} Sana, H., Rauw, G., Naz{\'e}, Y.,
Gosset, E., \& Vreux, J.-M.\ 2006, \mnras, 372, 661

\bibitem[Schilbach et al.(2006)]{Schilbach06} Schilbach, E., 
Kharchenko, N.~V., Piskunov, A.~E., R{\"o}ser, S., \& Scholz, R.-D.\ 2006, 
\aap, 456, 523 

\bibitem[Schmitt et al.(1985)]{Schmitt85} Schmitt, J.~H.~M.~M., 
Golub, L., Harnden, F.~R., Jr., Maxson, C.~W., Rosner, R., \& Vaiana, 
G.~S.\ 1985, \apj, 290, 307 

\bibitem[Schmitt et al.(1995)]{Schmitt95} Schmitt, J.~H.~M.~M.,
Fleming, T.~A., \& Giampapa, M.~S.\ 1995, \apj, 450, 392

\bibitem[Schmitt(1997)]{Schmitt97} Schmitt, J.~H.~M.~M.\ 1997,
\aap, 318, 215

\bibitem[Schmitt \& Favata(1999)]{Schmitt99} Schmitt, J.~H.~M.~M., \&
Favata, F.\ 1999, \nat, 401, 44

\bibitem[Schneps et al.(1980)]{Schneps80} Schneps, M.~H., Ho,
P.~T.~P., \& Barrett, A.~H.\ 1980, \apj, 240, 84

\bibitem[Schulz et al.(2003)]{Schulz03} Schulz, N.~S., 
Canizares, C., Huenemoerder, D., \& Tibbets, K.\ 2003, \apj, 595, 365 

\bibitem[Schulz et al.(2006)]{Schulz06} Schulz, N.~S., Testa, P.,
Huenemoerder, D.~P., Ishibashi, K., \& Canizares, C.~R.\ 2006, \apj,
653, 636

\bibitem[Siess et al.(2000)]{Siess00} Siess, L., Dufour, E., \&
Forestini, M.\ 2000, \aap, 358, 593

\bibitem[Sharpless(1954)]{Sharpless54}
Sharpless, S.\ 1954, \apj, 119, 334

\bibitem[Skinner et al.(2005)]{Skinner05} Skinner, S.~L., Zhekov,
S.~A., Palla, F., \& Barbosa, C.~L.~D.~R.\ 2005, \mnras, 361, 191

\bibitem[Skinner et al.(2006)]{Skinner06} Skinner, S., G{\"u}del, 
M., Schmutz, W., \& Zhekov, S.\ 2006, \apss, 304, 97 

\bibitem[Smith et al.(2001)]{Smith01} Smith, R.~K., Brickhouse, N.~S.,
Liedahl, D.~A., \& Raymond, J.~C.\ 2001, \apjl, 556, L91

\bibitem[Stassun et al.(2006)]{Stassun06} Stassun, K.~G., van den 
Berg, M., Feigelson, E., \& Flaccomio, E.\ 2006, \apj, 649, 914 

\bibitem[Stelzer et al.(2003)]{Stelzer03} Stelzer, B., 
Hu{\'e}lamo, N., Hubrig, S., Zinnecker, H., \& Micela, G.\ 2003, \aap, 407, 
1067 

\bibitem[Stelzer et al.(2005)]{Stelzer05} Stelzer, B., Flaccomio,
E., Montmerle, T., Micela, G., Sciortino, S., Favata, F., Preibisch, T., \&
Feigelson, E.~D.\ 2005, \apjs, 160, 557

\bibitem[Stelzer et al.(2006)]{Stelzer06a} Stelzer, B.,
Hu{\'e}lamo, N., Micela, G., \& Hubrig, S.\ 2006, \aap, 452, 1001

\bibitem[Stelzer et al.(2006)]{Stelzer06b} Stelzer, B., Micela, 
G., Hamaguchi, K., \& Schmitt, J.~H.~M.~M.\ 2006, \aap, 457, 223 

\bibitem[Strom et al.(1995)]{Strom95} Strom, K.~M., Kepner, J., 
\& Strom, S.~E.\ 1995, \apj, 438, 813

\bibitem[Telleschi et al.(2007)]{Telleschi07} Telleschi, A., Guedel,
M., Briggs, K.~R., Audard, M., \& Palla, F.\ 2007, \aap in press,
astro-ph/0612338

\bibitem[Throop \& Bally(2005)]{Throop05} Throop, H.~B., \&
Bally, J.\ 2005, \apjl, 623, L149

\bibitem[Townsley et al.(2002)]{Townsley02}
Townsley, L.~K., Broos, P.~S., Chartas, G., Moskalenko, E., Nousek,
J.~A., \& Pavlov, G.~G.\ 2002, Nuclear Instruments and Methods in
Physics Research A, 486, 716

\bibitem[Townsley et al.(2003)]{TFM03} Townsley, L.~K., Feigelson,
E.~D., Montmerle, T., Broos, P.~S., Chu, Y.-H., \& Garmire, G.~P.\
2003, \apj, 593, 874

\bibitem[Townsley et al.(2006a)]{Townsley06} Townsley, L.~K., Broos,
P.~S., Feigelson, E.~D., Brandl, B.~R., Chu, Y.-H., Garmire, G.~P., \&
Pavlov, G.~G.\ 2006, \aj, 131, 2140

\bibitem[Townsley et al.(2006b)]{Townsley06b} Townsley, L.~K., Broos,
P.~S., Feigelson, E.~D., Garmire, G.~P., \& Getman, K.~V.\ 2006, \aj,
131, 2164

\bibitem[ud-Doula \& Owocki(2002)]{ud-Doula02} ud-Doula, A., \& 
Owocki, S.~P.\ 2002, \apj, 576, 413 

\bibitem[Verschueren(1991)]{Verschueren91} Verschueren, W.\ 1991, 
Ph.D.~Thesis, Vrije Universiteit Brussel, Belgium  

\bibitem[Vuong et al.(2003)]{Vuong03}
Vuong, M.~H., Montmerle, T., Grosso, N., Feigelson, E.~D., Verstraete, L., \& Ozawa, H.\ 2003, \aap, 408, 581

\bibitem[Walborn et al.(2002)]{Walborn02} Walborn, N.~R., et al.\
2002, \aj, 123, 2754

\bibitem[Waldron et al.(2004)]{Waldron04}
Waldron, W.~L., Cassinelli, J.~P., Miller, N.~A., MacFarlane, J.~J., \& Reiter, J.~C.\
2004, \apj, 616, 542

\bibitem[Q. Wang et al.(2006)]{Wangqd06} Wang, Q.~D., Dong, H., \& Lang,
C.\ 2006, \mnras, 371, 38

\bibitem[Wang et al.(2007)]{Wang07} Wang, J.~F., Townsley, L.~K.,
Feigelson, E.~D., Getman, K.~V., Broos, P.~S., Garmire, G.~P., \&
Tsujimoto, M.\ 2007, \apjs, 168, 100

\bibitem[Weaver et al.(1977)]{Weaver77} Weaver, R., McCray, R., 
Castor, J., Shapiro, P., \& Moore, R.\ 1977, \apj, 218, 377 

\bibitem[Williams et al.(1995)]{Williams95} Williams, J.~P.,
Blitz, L., \& Stark, A.~A.\ 1995, \apj, 451, 252

\bibitem[Williams \& Blitz(1998)]{Williams98} Williams, J.~P., \&
Blitz, L.\ 1998, \apj, 494, 657

\bibitem[Winston et al.(2007)]{Winston07} Winston, E., et al.\ 
2007, ApJ, 669, 493 

\bibitem[Wolk et al.(2006)]{Wolk06} Wolk, S.~J., Spitzbart, 
B.~D., Bourke, T.~L., \& Alves, J.\ 2006, \aj, 132, 1100 

\bibitem[Zacharias et al.(2004)]{Zacharias04} Zacharias, N., Monet, 
D.~G., Levine, S.~E., Urban, S.~E., Gaume, R., \& Wycoff, G.~L.\ 2004, 
Bulletin of the American Astronomical Society, 36, 1418 

\bibitem[Zinnecker \& Preibisch(1994)]{Zinnecker94} 
Zinnecker, H., \& Preibisch, T.\ 1994, \aap, 292, 152 

\end{thebibliography}
\end{document}